\documentclass[twoside,12pt]{article}
\usepackage{epsfig}

\usepackage{amssymb}
\usepackage{amsmath}
\usepackage{mathtools}
\usepackage{mathrsfs}
\usepackage{gensymb}
\usepackage{setspace}

\newcommand{\cf}{\textit{cf.}}
\newcommand{\ie}{\textit{i.e.}}
\newcommand{\eg}{\textit{e.g.}}
\newcommand{\ms}{{\rm ms}}
\newcommand{\km}{{\rm km}}
\newcommand{\Hz}{{\rm Hz}}
\newcommand{\kHz}{{\rm kHz}}
\newcommand{\Gpc}{{\rm Gpc}}
\newcommand{\Mpc}{{\rm Mpc}}

\newcommand{\MeV}{{\rm MeV}}
\newcommand{\Msun}{M_\odot}

\newcommand{\be}{\begin{equation}}
\newcommand{\ee}{\end{equation}}
\newcommand{\bea}{\begin{eqnarray}}
\newcommand{\eea}{\end{eqnarray}}

\begin{document}

\title{ \vspace{1cm} Gravitational waves from neutron star mergers and their relation to the nuclear equation of state}
\author{L.\ Baiotti,$^{1}$ \\
\\
$^1$Department of Earth and Space Science, Graduate School of Science,
\\ Osaka University, Japan\\
}

\maketitle
\begin{abstract}

In this article, I introduce ideas and techniques to extract information
about the equation of state of matter at very high densities from
gravitational waves emitted before, during and after the merger of binary
neutron stars. I also review current work and results on the actual use
of the first gravitational-wave observation of a neutron-star merger to
set constraints on properties of such equation of state. In passing, I
also touch on the possibility that what we observe in gravitational waves
are not neutron stars, but something more exotic. In order to make this
article more accessible, I also review the dynamics and
gravitational-wave emission of neutron-star mergers in general, with
focus on numerical simulations and on which representations of the
equation of state are used for studies on binary systems.
\end{abstract}
\tableofcontents

\section{Opening words}
\label{opening}

For decades hope has grown that measurements of gravitational waves
(GWs), and related electromagnetic radiation, from mergers of binaries
composed of compact stars\footnote{I have used the term {\it compact
    stars}, because it does not contain any hints on their composition:
  It just differentiates them from black holes or less dense
  stars. Indeed, what is inside compact stars like those in the first GW
  detection (GW170817) of objects thought not to be black holes
  \cite{Abbott2017} (but see \cite{Hinderer2018} for the possibility that
  GW170817 was a merger of a compact star with a black hole) is still
  largely unknown and it is indeed one of the scopes of this article to
  describe the status of research about the composition of compact stars
  (see Ref. \cite{Baym2017} for a recent review), in view of GW
  observations. However, in most of the literature the term {\it neutron
    stars} is employed, without assuming that they are made only or
  mostly of neutrons. Exceptions to this routine use are works that
  explicitly study non-nucleonic equations of state for compact stars
  (see Sects. \ref{postmerger phase transitions}, \ref{quark stars} and
  \ref{hybrid stars}). These works often distinguish compact stars into
  neutron stars, quark stars, strange stars (quark stars made of strange
  quark matter), hybrid stars (with a core of quark matter and an outer
  region of nucleonic matter). Furthermore, the term compact stars may be
  used to refer to objects that are not made of ordinary matter, such as
  boson stars (see Sect. \ref{other-compact-objects}). Having presented
  these notes of caution on nomenclature, in the rest of this article, I
  will mostly use the term {\it neutron stars} to refer to compact stars
  made of ordinary matter in general. Only where specification may be
  necessary, or at least useful, I will employ the term compact stars or
  one with further particularization.} may set more stringent constraints
on the properties of the equation of state (EoS) of matter at densities
higher and much higher than nuclear density. Such hope has become reality
since August 2017, when the first measurement of gravitational radiation
interpreted as the coalescence of a binary--neutron-star (BNS) merger has
been recorded as GW170817 \cite{Abbott2017}. There have been numerous
follow-up observations (triggered by the GW detection) of electromagnetic
radiation \cite{Abbott2017b, Abbott2017c, Alexander2017, Arcavi2017,
  Arcavi2017a, Arcavi2018, Chornock2017, Coulter2017, Covino2017,
  Cowperthwaite2017, Drout2017, Evans2017, Goldstein2017, Hallinan2017,
  Kasliwal2017, Margutti2017, Murguia-Berthier2017, Nicholl2017,
  Pian2017, Savchenko2017, Smartt2017, Soares-Santos2017, Tanaka2017,
  Tanvir2017, Troja2017} from what are thought to be the material ejecta
of the merger. While concluding this review, the third observational run
of the LIGO-Virgo Collaborations has started and the observation of some
candidate BNS waveforms has been announced. However, details have not
been released and therefore for analysis purposes GW170817 is currently
the only observation of the kind. And this one observation has indeed
already been used to gain knowledge\footnote{There is, naturally, much
  excitement about this first detection, but several experts in the field
  think that GW170817 has actually not provided new insights about the
  EoS that cannot be obtained from knowledge already available from
  nuclear physics theory and experiments, and from preceding
  astrophysical observations. However, to my knowledge, statements of
  this kind have appeared only in a few articles \cite{Zhang2018,
    Lim2018, Tews2019}. See the rest of this review for more details.}
on the EoS, in addition to several other findings, such as more stringent
limits on the difference between the speed of light and that of gravity
\cite{Abbott2017d}, on the equivalence principle (through Shapiro-delay
\cite{Shapiro1964} measurements), on Lorentz invariance
\cite{Abbott2017d} and on cosmologically modified gravity
\cite{Baker2017, Creminelli2017, Ezquiaga2017, Sakstein2017, deRham2018,
  Creminelli2018, Creminelli2019}.

In this review I assume that the reader is familiar with nuclear
EoSs. Therefore, I will not enter into details of the various proposals
for describing and computing the EoS from fundamental principles with
various levels of approximations, or the details of any of the results
of such computations. For reviews on the current status of research in
computations of EoS with NSs in mind see, \eg,
Refs. \cite{Lattimer2012rev, Baym2017}.

As is well known, a large number of experiments carried out in
laboratories on Earth \cite{Danielewicz2002, Steiner2005, Li2008,
  Abrahamyan2012, Horowitz2012, Tsang2012a, Lattimer2016, Watts2016,
  Ozel2016, Oertel2017}, especially heavy-ion reactions with radioactive
beams, constrain the EoS of matter in various ways and to various degrees
up to around nuclear density. Considerable attempts have been made to
constrain the EoS at densities higher than the nuclear density,
but it has turned out to be difficult to put stringent constraints on the
EoS parameters \cite{Danielewicz2002, Tsang2004, Fuchs2006, Lynch2009,
  Tsang2009, Kortelainen2010, Tamii2011, Tsang2012a, Brown2013,
  Lefevre2016, Russotto2016}.

Astrophysical observations (like those involving binary pulsars and x-ray
binaries \cite{Ozel2016, Watts2016, Miller2016a, Gendreau2016,
  Arzoumanian2018, Degenaar2018, Watts2019}) have been giving additional
constraints on the EoS of ultra-high density matter, by trying to measure
the masses and radii of neutron stars (NSs). Further information on the
EoS of ultra-high density matter can be obtained through BNS mergers
observations by determining (see further below for definitions): (i) the
tidal deformability (sometimes also called tidal polarizability)
$\Lambda$ for a NS of a given mass; (ii) the amount of material ejected
from the merger (ejecta), which gives rise to a macronova and to radio
emissions (see, \eg, Ref. \cite{Metzger2012}); (iii) the maximum mass
$M_{\rm max}$ of a non-rotating compact star that is stable against
collapse to a black hole.

The tidal deformability $\Lambda$ is an EoS-dependent, dimensionless
function of the NS mass that correlates with the pressure gradients
inside the star, namely with the stiffness of the EoS. Tidal deformations
cause a phase shift (the phasing of the GW signal is significantly more
important for parameter estimation than its amplitude \cite{Favata2014})
in the waveform relative to the merger of point-particles, which is
measurable with existing GW detectors \cite{Flanagan08, Hinderer08,
  Read:2009b, Read2013}. More details will be given in
Sect. \ref{premerger-basic}.

A macronova \cite{Li1998} is an astronomical source of electromagnetic
radiation about one to three orders of magnitudes brighter than a regular
nova\footnote{Novae are caused by hydrogen fusion explosions on a white
  dwarf accreting from a larger companion star.}, hence the name {\it
  macronova} \cite{Kulkarni2005_macronova-term} or {\it kilonova}
\cite{Metzger:2010} (the term {\it mergernova} has also been recently
used \cite{Li2018a}). In the standard scenario, macronovae shine hours to
days after the merger in ultra-violet, optical, and infrared bands and
their power source is thought to be, with support by the observation of the
macronova associated with GW170817 \cite{Arcavi2017, Pian2017,
  Smartt2017}, the radioactive decay of $r$-process elements
\cite{Li1998, Kulkarni2005_macronova-term, Metzger:2010} (however, see
also \cite{Kisaka2015} for an alternative explanation). Macronovae are
particularly promising electromagnetic counterparts to BNS mergers
because their emission is relatively isotropic, contrary to gamma-ray
bursts, which are thought to be highly beamed and thus observable only in
some cases. In addition to the macronova associated with GW170817, named
also AT2017gfo, a few other candidates have been identified (without a
GW counterpart, of course) \cite{kilonovacatalogueweb,
  Guillochon2017, Perley2009, Berger2013, Tanvir2013, Yang2015, Jin2015,
  Jin2016}.

Observational constraints on the quantities mentioned above (tidal
deformability, maximum mass, amount of ejecta) can be obtained from BNS
mergers through (i) GWs from the late inspiral, (ii) GWs from post-merger
oscillations and (iii) electromagnetic emissions (macronova and radio
emission).

The rest of this article is organized as follows. In Sect. \ref{intro}, I
will review the dynamics and GW emission of BNS mergers in general, with
focus on numerical simulations and on which representations of the EoS
are used for studies on BNS systems. Then, in Sect. \ref{premerger}, I
will introduce ideas and techniques to find information about the EoS
from GWs emitted before the merger of the NSs (coalescence or inspiral
phase), while in Section \ref{postmerger} I will focus on analyses of
gravitational radiation emitted during and after the merger (merger and
post-merger phases) and in Sect. \ref{pre+post merger} on ideas for
combining pre- and post-merger waveforms. Finally, before concluding, in
Sect. \ref{constraints} I will review current work and results on the
actual use of the one GW observation of BNS merger currently available to
set constraints on EoS properties. In passing, in Sect. \ref{exotic
  binaries}, I will also touch on the possibility that what we observe in
GWs are not NSs, but something more exotic.

\section{Introduction}
\label{intro}

For the convenience of the reader, I present here some of the generalities
of BNS systems and of popular ways used to advance our
knowledge on them, in particular on their EoS.

\subsection{Status of numerical-relativity simulations for binary
  neutron-star mergers}
\label{simulations}

In order to help and interpret observations, we need solutions of the
general-relativistic equations describing spacetime\footnote{Except for
  some side remarks and Sect. \ref{exotic binaries}, in this review I
  will assume that the theory of gravity is general relativity, because
  the prospects for deviations from general relativity in stellar-mass
  objects (see Refs. \cite{Berti2015, Barack2019} for reviews) are
  severely limited, also by observations of GWs from mergers of binary
  black holes \cite{Yunes2016} and NSs \cite{Abbott2017d}, and because
  numerical simulations of BNS mergers in alternative theories of gravity
  are very few \cite{Barausse2013, Shibata2014, Palenzuela2014,
    Sampson2014, Taniguchi2015, Ponce2014a, Bezares2017, Palenzuela2017,
    Sagunski2018, Bezares2018}.}, matter and radiation (in particular,
magnetic fields). As everyone knows, analytic solutions of astrophysical
relevance for BNS systems are not available, therefore the field of
numerical relativity -- the science of simulating (solving)
general-relativistic dynamics on computers -- has seen lots of efforts
being put into it and a reasonable amount of reliable results. Numerical
relativity is now mature, but at the beginning, a couple of decades ago,
it has struggled to get decent results because straightforward
discretization of the Einstein equations just does not work. In addition
to the standard problems of any numerical simulation, there are multiple
reasons for the increased difficulty inherent to general-relativistic
simulations: (i) The formulation of the equations is not self-evident;
\eg~time is not “simply” defined and very careful variables definitions
are needed to obtain a system that is strongly hyperbolic; (ii) Physical
singularities may be present and need special treatment; (iii) While not
carrying physical information, gauges play an important role in numerical
stability, for example in countering grid stretching.  For explanations
on all these issues, well-written textbooks are nowadays available
\cite{Alcubierre:2008, Baumgarte2010, Gourgoulhon2012,
  Rezzolla_book:2013, Shibata_book:2016}.

Despite such difficulties, nowadays the number of research groups with
their own independent codes capable of performing (at least in some
respects) state-of-the-art BNS simulations is of order ten. The current
status of the capabilities of codes used for simulating BNS can be
summarized as follows.

(i) All codes can robustly compute the matter and spacetime
dynamics (including long-term evolutions of the formed black hole and
accretion disc), even if improvements are being constantly made. The
selection of appropriate gauges and the extraction of GW signals from the
dynamics is nowadays routinely done by everyone in the field. The EoSs
used in simulations are based on published work from specialists in
nuclear theory and so on and are either piecewise polytropes\footnote{See
  Sect. \ref{eos-parameterization}.}  \cite{Read:2009a} (plus a thermal
part, at times) or tabulated \cite{Neilsen2014, Palenzuela2015,
  Radice2016, Endrizzi2018}. Also the first general-relativistic
simulations of merging NSs including quarks at finite temperatures have
been performed recently \cite{Most2018b, Bauswein2019}. See
Sect. \ref{postmerger phase transitions} for more details on these works
on phase transitions.

(ii) In addition to the theme of this review, namely the linking of GW
observations to physical properties of the emitting system and in
particular the EoS, there is a lot of ongoing work and some robust
results also on heavy-element production \cite{Bauswein2013b,
  Rosswog2013, Piran2013, Rosswog2014a, Grossman2014, Perego2014,
  Wanajo2014, Metzger2015, Just2015, Sekiguchi2015, Foucart2015,
  Palenzuela2015, Radice2016, Just2016, Lehner2016, Sekiguchi2016,
  Hotokezaka2016b, Barnes2016, Bovard2017} and macronovae
\cite{Bauswein2013b, Tanaka2013, Metzger2014, Wanajo2014, Metzger2015,
  Radice2016, Perego2017, Shibata2017c, Hotokezaka2018, Tanaka2018}, on
improved initial data, including better computations of BNS with
non-negligibly spinning stars \cite{Marronetti03, Baumgarte:2009,
  Tichy11, Tichy12, East2012d, Kastaun2013, Tsatsin2013, Bernuzzi2013,
  Kastaun2014, Tsokaros2015, Tacik15, Dietrich:2015b, Kastaun2017,
  Dietrich2018a, Tsokaros2018, Ruiz2018, Ruiz2019} and BNS with
eccentricity \cite{East2012c, Gold2012, Rosswog2013, Radice2016,
  East2016a, Chaurasia2018, Yang2018, Papenfort2018}, and on the
treatment of physical viscosity \cite{Radice2017, Alford2017,
  Shibata:2017b, Kiuchi2017, Zappa2018}.

(iii) Some other issues, instead, are still very open, in particular the
effects of magnetic fields and of radiation transport, especially
neutrino radiation transport. Simulating magnetic fields is challenging
because of physical instabilities that require very high resolutions to
be resolved and because of limitations in the modelling of
electromagnetic interactions. Most simulations, in fact, are carried out
in the magnetohydrodynamics approximation, which does not capture all the
physical effects, like upper limits to the growth of
instabilities. Resistive-magnetohydrodynamics simulations exist in small
numbers \cite{Palenzuela2013b, Ponce2014, Dionysopoulou2015}, but they
are limited by our lack of knowledge about the resistivity of matter in
and around NSs. The open problems with magnetic fields in BNS mergers
apply especially to the post-merger phase, where magnetic fields may have
a huge importance for the dynamics itself, for the ejecta, and for
electromagnetic emissions from the vicinity of the merged object (like
those thought to produce short gamma-ray bursts). Before the merger,
magnetic fields are not relevant for the global dynamics
\cite{Giacomazzo:2009mp}, but may produce observable electromagnetic
radiation, as found in works employing resistive magnetohydrodynamics
\cite{Palenzuela2013b, Ponce2014}. Effects of neutrinos and in general
of radiation transport are also important for the production of ejecta
and direct electromagnetic emissions and advances are progressively being
made \cite{Galeazzi2013, Neilsen2014, Palenzuela2015, Sekiguchi2016,
  Alford2017, Bovard2017, Kyutoku2018}.

\begin{figure*}[ht]
\begin{center}
   \includegraphics[width=0.3\textwidth]{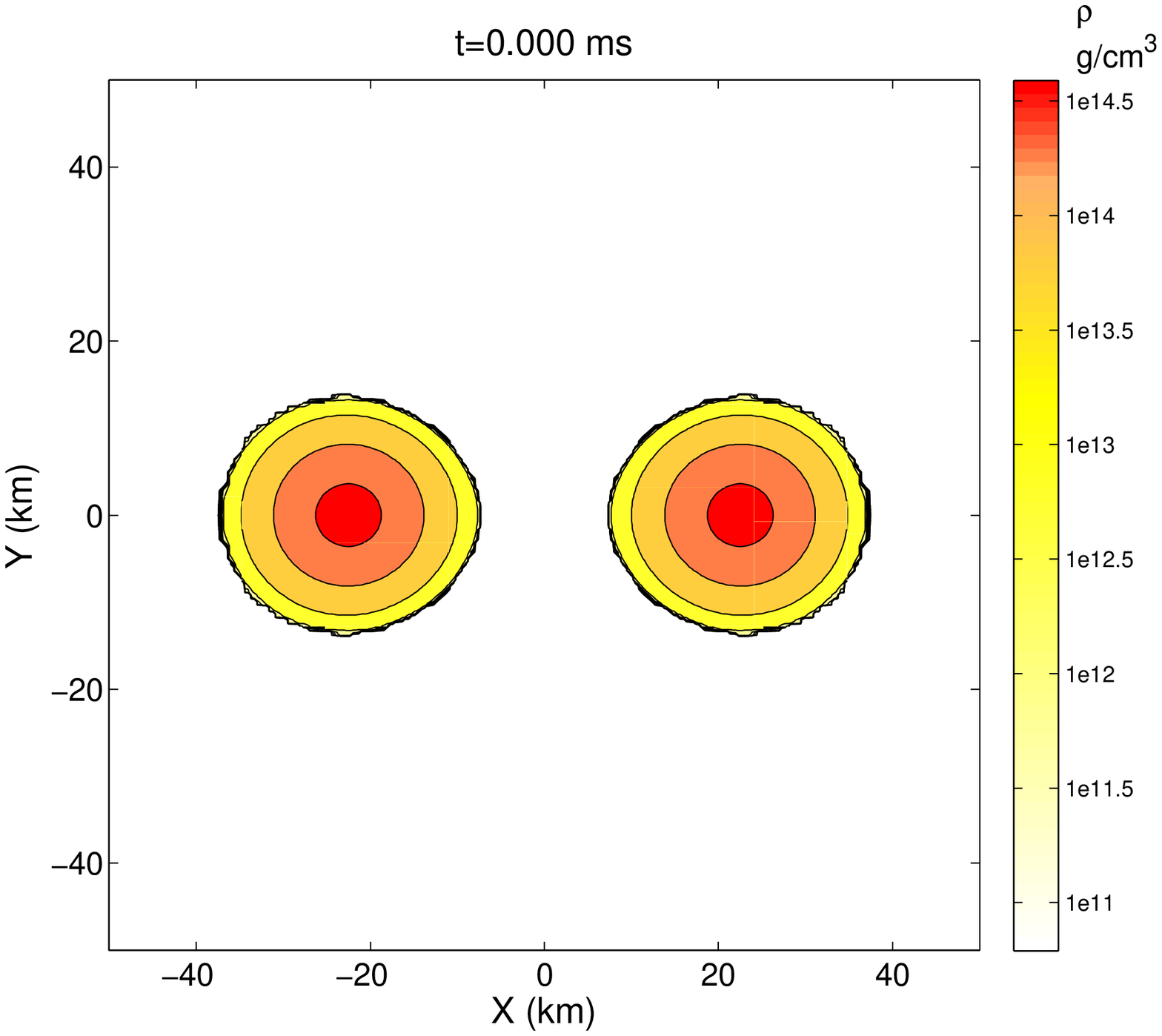} 
   \hskip 0.2cm             
   \includegraphics[width=0.3\textwidth]{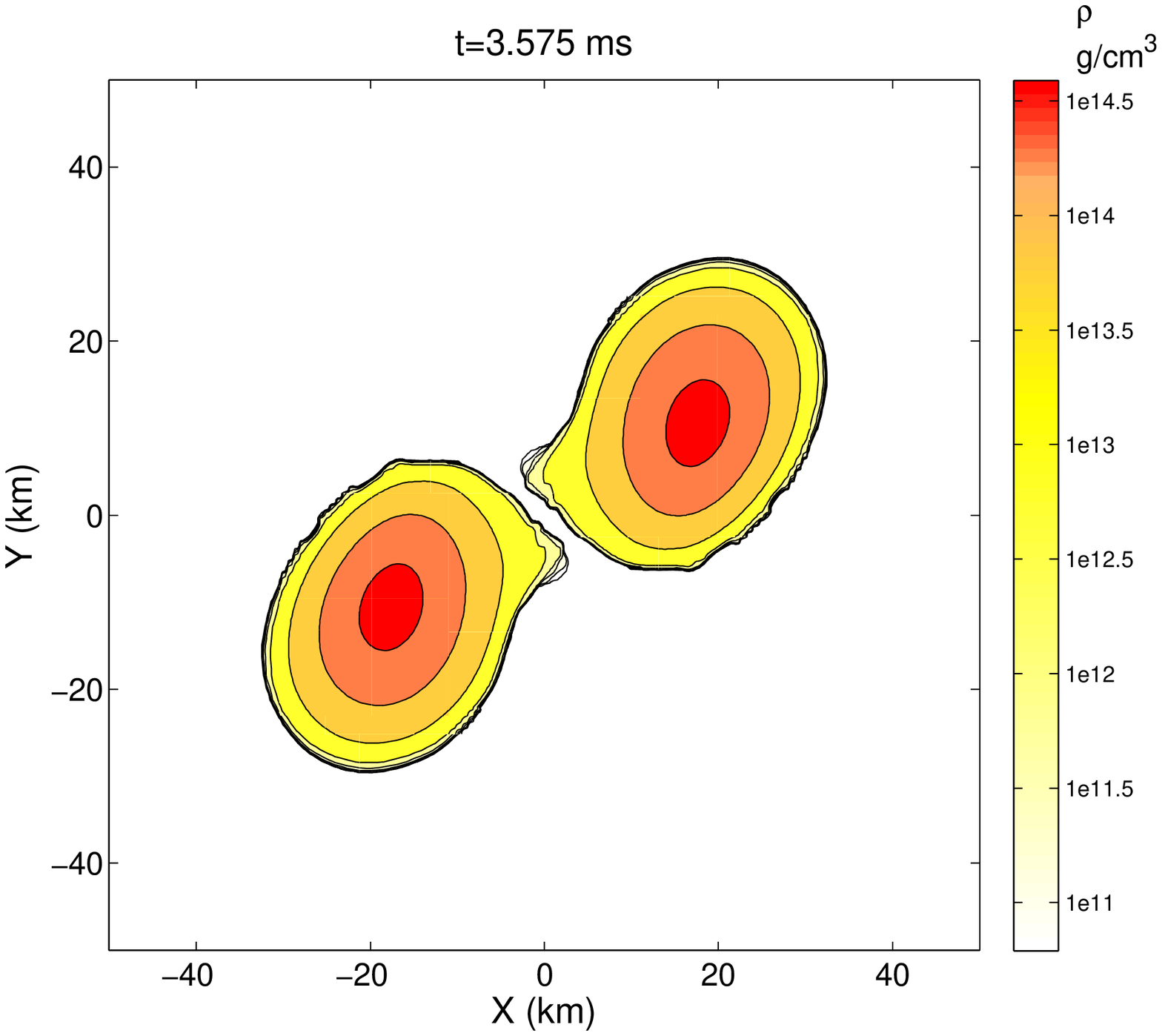} 
   \includegraphics[width=0.3\textwidth]{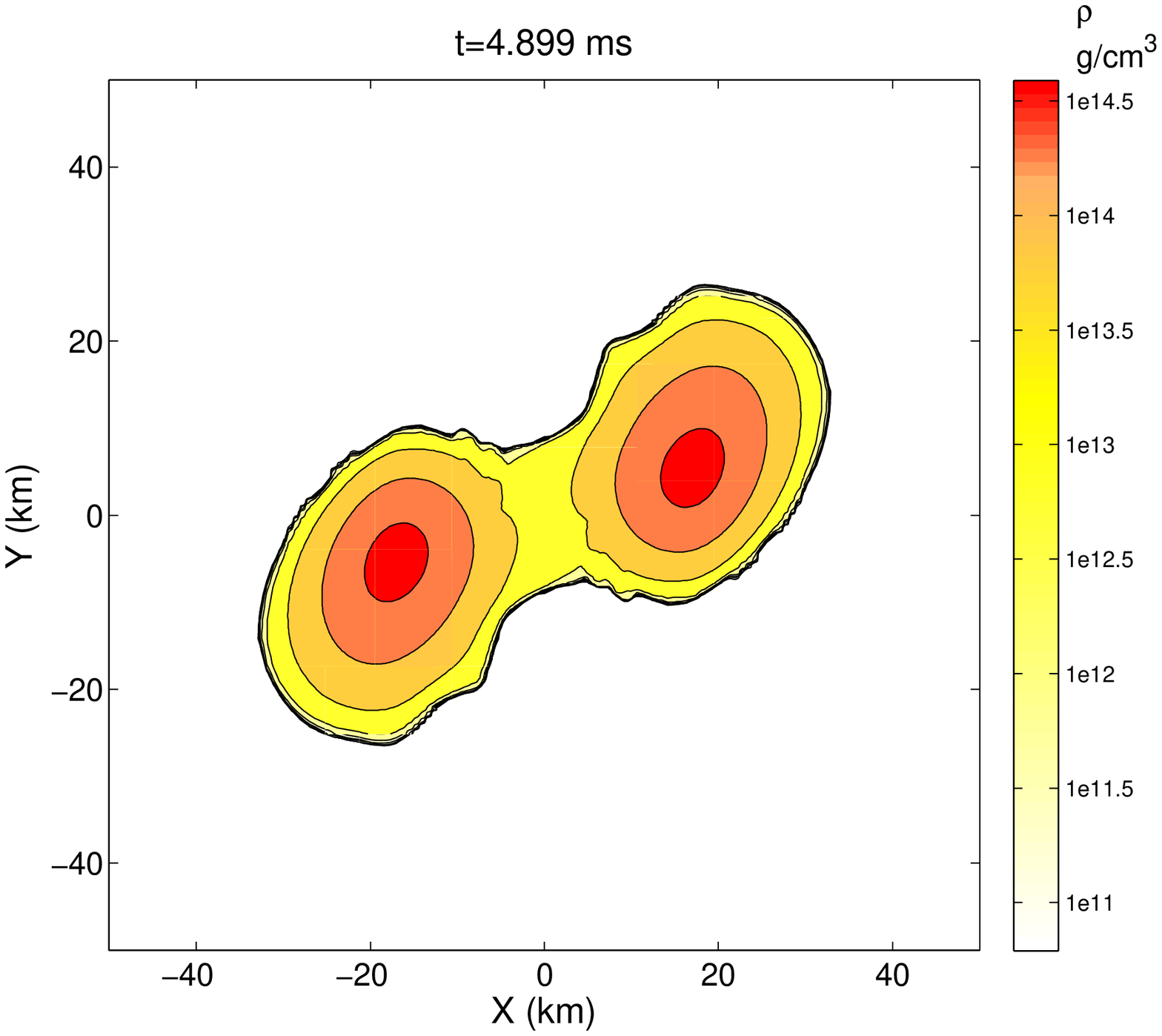} 
   \hskip 0.2cm             
   \includegraphics[width=0.3\textwidth]{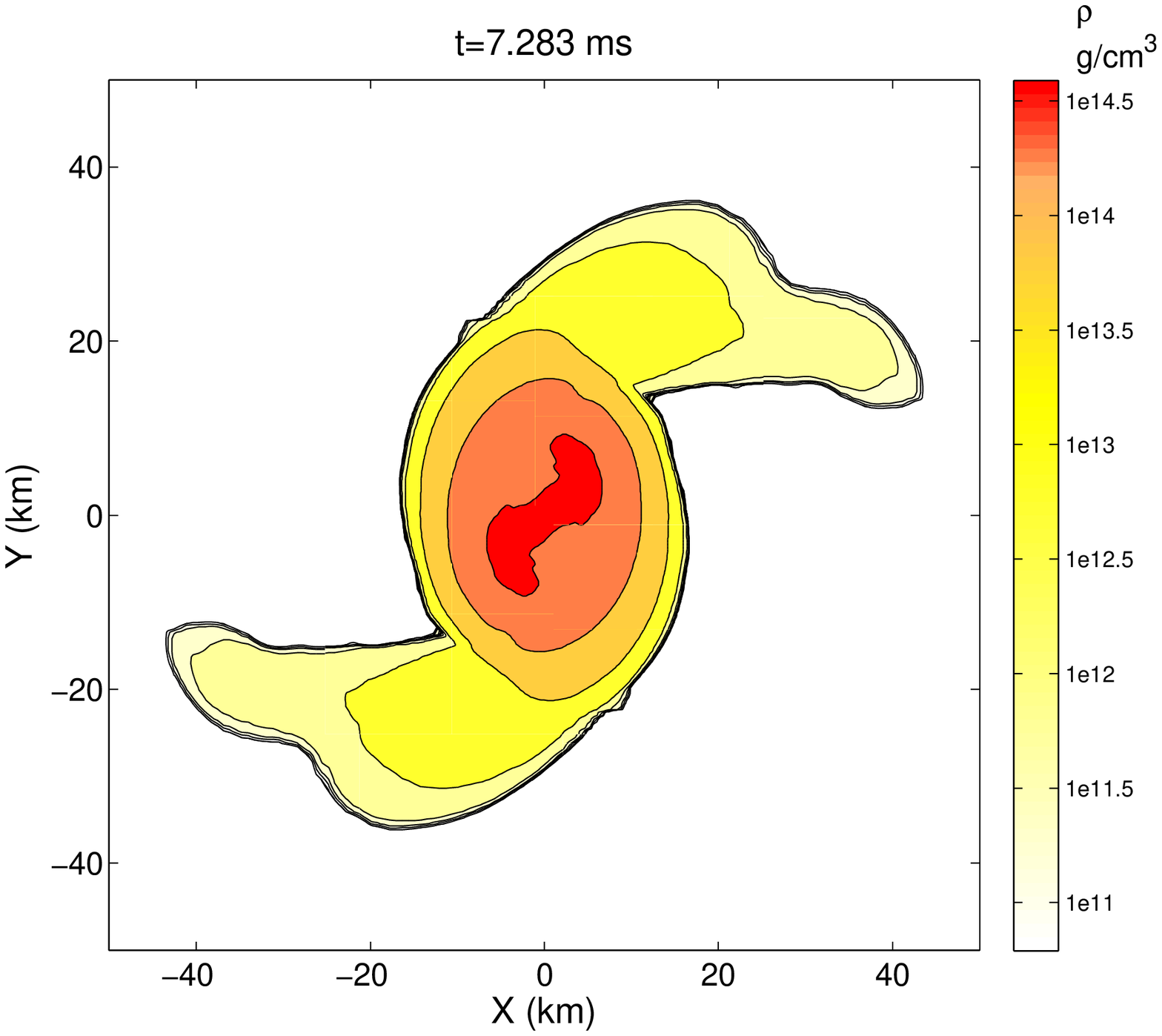} 
   \includegraphics[width=0.3\textwidth]{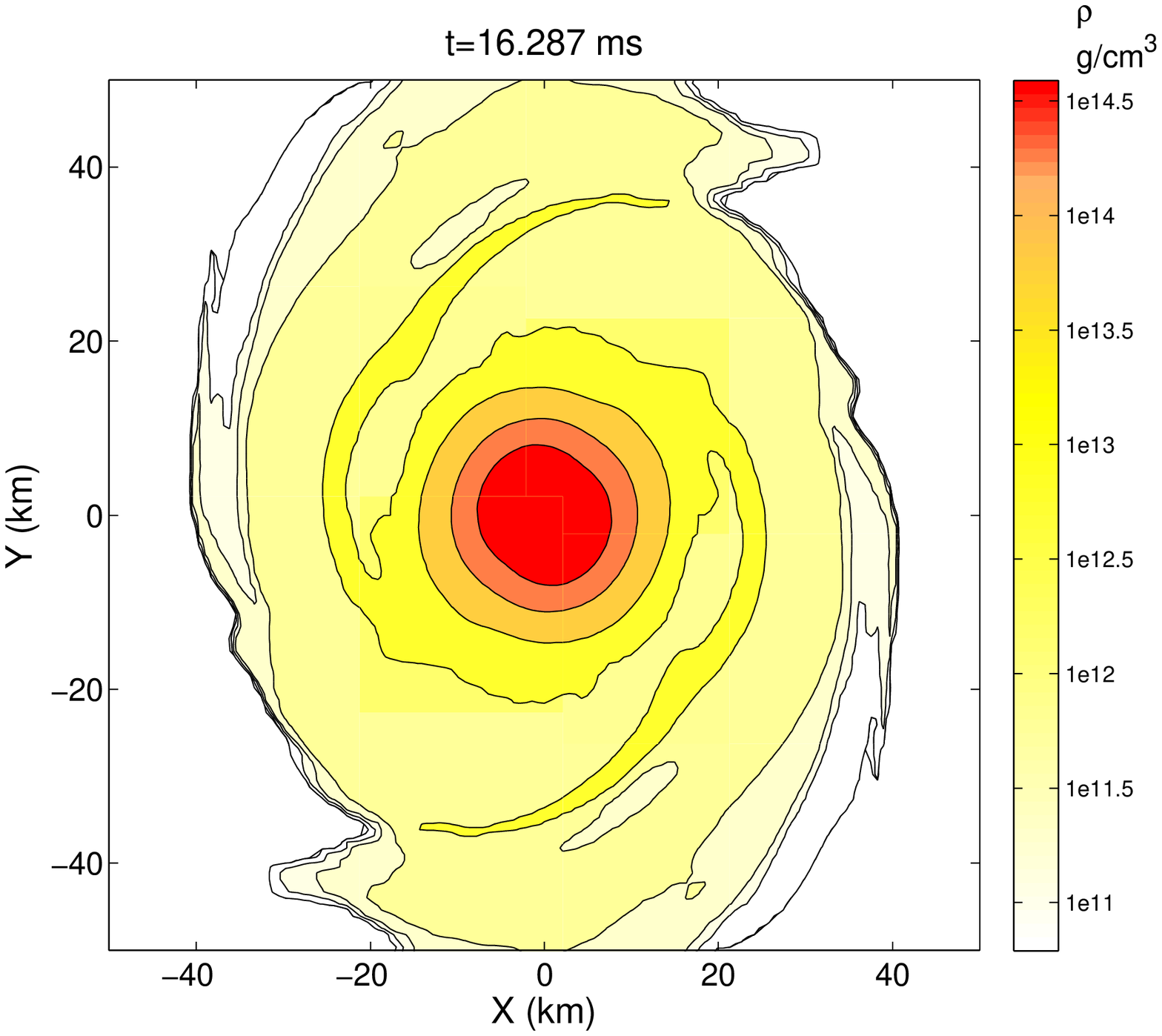} 
   \hskip 0.2cm             
   \includegraphics[width=0.3\textwidth]{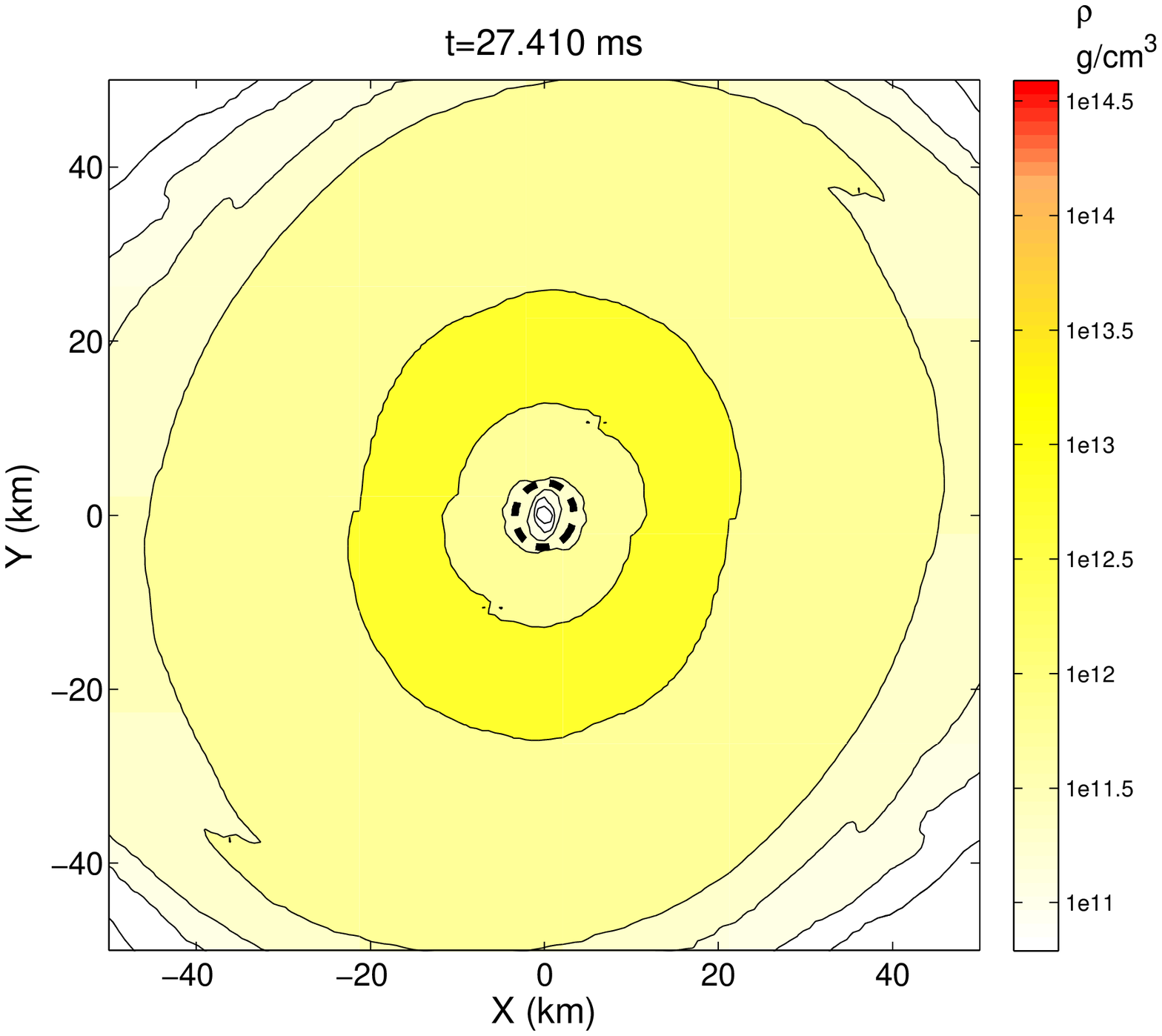} 
\end{center}
   \caption{Isodensity contours on the equatorial plane for the
     evolution of a BNS with the polytropic EoS.
     The thick dashed line in the last panel represents the apparent
     horizon. (From Ref. \cite{Baiotti08})\label{fig:bns-dynamics}}
\end{figure*}

\subsection{The dynamics of neutron-star binaries}
\label{binary dynamics}

For a more in-depth overview on the state of the art of numerical studies
on BNS systems, the reader is referred to recent review articles, like
Refs. \cite{Baiotti2016, Paschalidis2016, Duez2019}. Here, I will only
give a brief overview of their dynamics, also by showing images from a
representative simulation taken from
Ref. \cite{Baiotti08}. Fig. \ref{fig:bns-dynamics} collects some
representative isodensity contours (\ie~contours of equal rest-mass
density) on the equatorial plane.
The initial coordinate separation between the maxima in the rest-mass
density (defining the stellar centres) is $45\,{\rm km}$. As the stars
inspiral with increasing angular velocity, each one becomes more and more
tidally deformed by the gravity of its companion. This leads to an
increase in the inspiral rate \cite{Flanagan08}, which also depends, of
course, on the total angular momentum of the system, the orbital one and
the spins of the stars.

During the merger, when regions of the two stars with density around a
factor of a few less than their maximum density come into contact, a
noticeable vortex sheet (or shear interface) develops, where the
tangential components of the velocity exhibit a discontinuity. This
condition is known to be unstable to very small perturbations and it can
develop a Kelvin-Helmholtz instability, which will curl the interface
forming a series of vortices~\cite{Chandrasekhar81}. This is indeed what
is observed in all simulations of this kind, with features that are not
much dependent on the mass or on the EoS used. Even if this instability
is purely hydrodynamical, it can have strong consequences when studying
the dynamics of BNS systems in the presence of magnetic fields, because
it leads to an exponential growth of the toroidal component of the
magnetic field even if the initial magnetic field is a mostly or purely
poloidal one\footnote{Realistic descriptions of NSs, isolated or in
  binaries, probably require poloidal-toroidal mixed field configurations
  \cite{BraithwaiteNord2006, Ciolfi2009, Lander2012}.
} \cite{Price06, Anderson2008, Baiotti08}. Through this
mechanism, the instability can lead to an overall amplification of the
magnetic field of about three orders of magnitude \cite{Obergaulinger10,
  Giacomazzo2011b, Rezzolla:2011, Neilsen2014, Kiuchi2014, Kiuchi2015a,
  Kiuchi2017}. Such high magnetic fields are presumed to be behind the
phenomenology of magnetars \cite{Duncan1992, Kaspi2017} and short hard
gamma-ray bursts \cite{Eichler89, Narayan92, Nakar:2007yr, DAvanzo2015}.

\begin{figure}[bt]
\begin{center}
\begin{minipage}[t]{15 cm}
  \epsfig{file=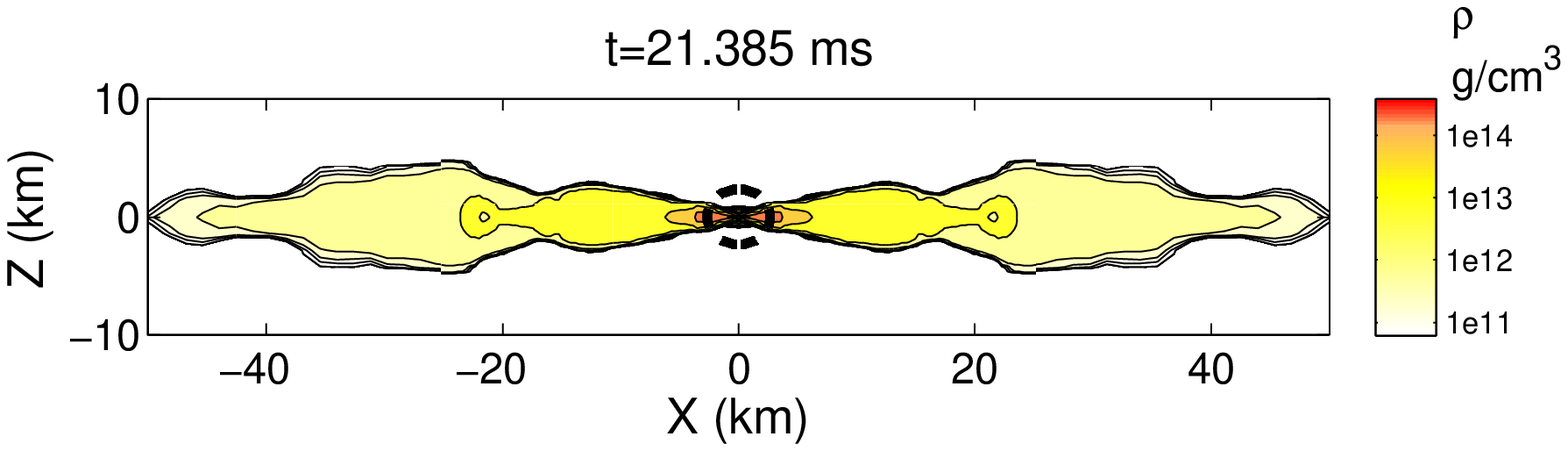,scale=0.4}
  \epsfig{file=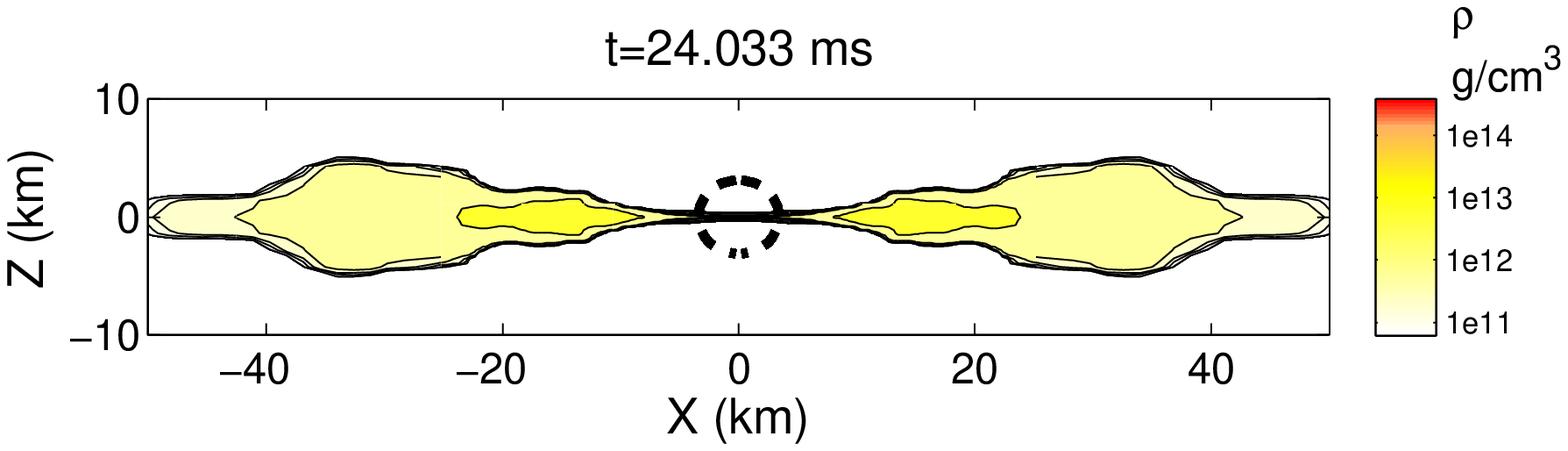,scale=0.4}
  \epsfig{file=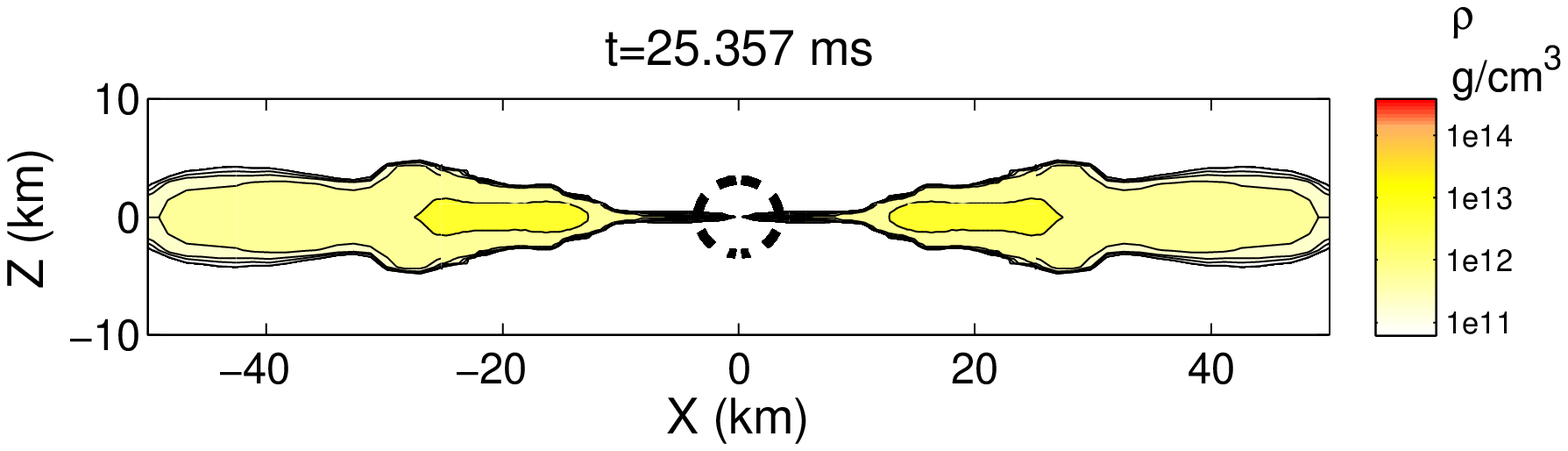,scale=0.4}
  \epsfig{file=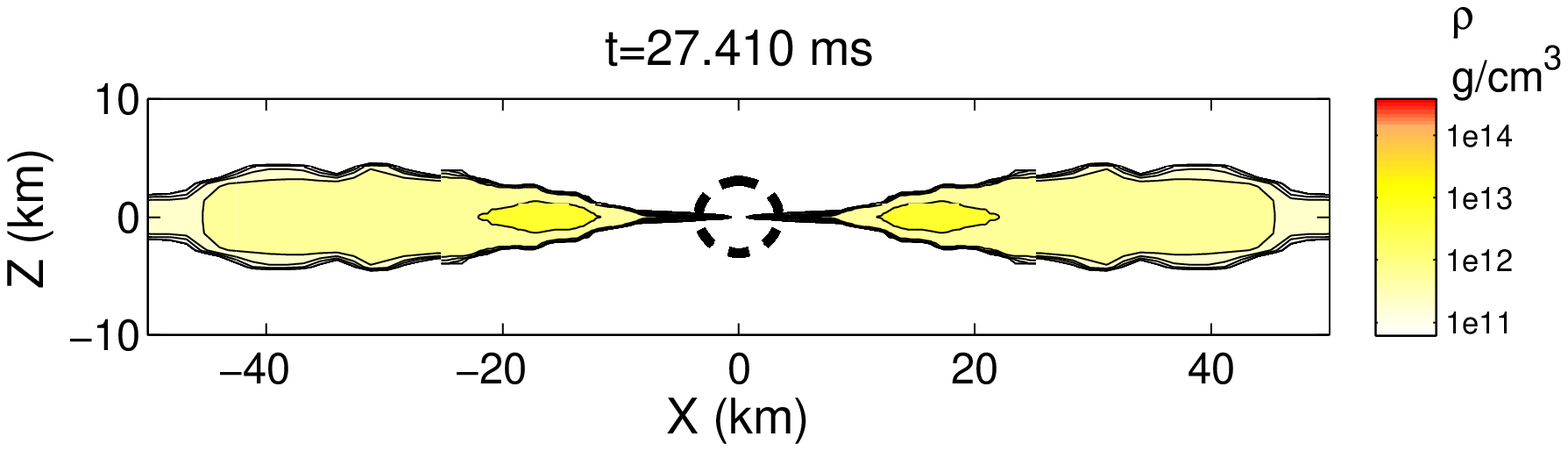,scale=0.4}
\end{minipage}
\begin{minipage}[t]{16.5 cm}
  \caption{Isodensity contours on a meridional plane, highlighting the
    formation of a torus surrounding the central black hole, whose
    apparent horizon is indicated with a thick dashed line. The data
    refer to a low-mass binary evolved with a polytropic EoS
    (\cf~Fig.~\ref{fig:bns-dynamics}). (From Ref.
    \cite{Baiotti08}) \label{fig:bns-torus}}
\end{minipage}
\end{center}
\end{figure}

After the merger, the cores of the two NSs coalesce. During their rapid
infall they experience a considerable decompression of $\approx 15\%$,
for about $1$ ms. Given the choice of mass of the initial configuration
and of the EoS, in the simulation shown in Fig. \ref{fig:bns-dynamics}
the merged object is initially a {\it hypermassive neutron star} (HMNS),
\ie~a NS with mass above the upper limit for uniformly rotating NSs and
that is temporarily supported against collapse by differential rotation
and thermal gradients \cite{Baumgarte00b}.
The HMNS undergoes a number of violent non-axisymmetric oscillations,
with a dominant overall $m=2$
deformation\label{stellar modes}\footnote{As widely known, stellar deformations can be
  described decomposing the linear perturbations of the energy or
  rest-mass density as a sum of quasi-normal modes that are characterized
  by the indices ($l$,$m$) of the spherical harmonic functions. Then the
  $m$ mentioned in the text is the dominant term of such expansion. $m=0$
  is a spherical perturbation, $m=1$ is a one-lobed (or one-armed)
  perturbation, $m=2$ is a bar-shaped perturbation. See
  Ref. \cite{Stergioulas03} for a review.}, \ie~a bar, as the system
moves towards an energetically favourable configuration through the
rearrangement of the angular-momentum distribution \cite{Shibata:2000jt,
  Shibata:2003yj, Baiotti06b, DePietri06, Franci2013, Franci2013b,
  Siegel2013, Kiuchi2014, DePietri2014, Loeffler2015}. As the bar
rotates, it also loses large amounts of angular momentum through
gravitational radiation. $m=1$ deformations have also been reported
\cite{Ou06, Corvino:2010, Anderson2008, Bernuzzi2013, Kastaun2014,
  Paschalidis2015, East2016, Dietrich:2015b, Radice2016a, Lehner2016a,
  East2016a, Paschalidis2017b}. Together with these oscillations, a
secular increase of the central rest-mass density is also observed. Then,
about $15\, \ms$ after the merger, the maximum rest-mass density is seen
again to increase rapidly, exponentially, and the object collapses to a
rotating black hole (Kerr black hole). This is expected from the fact
that the HMNS born after the merger, while initially not beyond the
stability limit for gravitational collapse, has lost its original
differential rotation\footnote{Note that the physical non-axisymmetry of
  the HMNS and coordinate effects due to the choice of gauges (gauges are
  usually not fixed in numerical relativistic simulations, but are
  evolved together with the other spacetime quantities
  \cite{Alcubierre:2008, Baumgarte2010, Gourgoulhon2012,
    Rezzolla_book:2013, Shibata_book:2016}) make it difficult to provide
  a unique measure of the degree of differential rotation (but see Refs.
  \cite{Kastaun2014, Kastaun2016}). On average, however, the angular
  velocity decreases of about one order of magnitude between the rotation
  axis and the surface.}. The collapse is marked by the appearance of an
apparent horizon\footnote{Black holes are defined through their {\it
    event horizons}, which are surfaces bounding a region in spacetime
  inside which events cannot affect an outside observer. In numerical
  computations it is possible \cite{Diener03a} but not easy to find event
  horizons, because information from all times are necessary for this and
  so it can be computed only at the end of the simulation. Instead, {\it
    apparent horizons} are usually searched for in numerical
  simulations. These are surfaces that are the boundary between light
  rays that are directed outwards and move outwards, and those that are
  directed outwards but move inwards. Apparent horizons are a local
  concept and thus they can be computed immediately for each time step of
  a simulation, with efficient techniques \cite{Thornburg95,
    Shibata-Uryu-2000b, Thornburg2003:AH-finding_nourl, Thornburg2007}.}.

When an apparent horizon appears, a large amount of
high--angular-momentum matter remains outside of it in the form of an
accretion torus. In the example presented here, it has an average density
between $10^{12}$ and $10^{13}\ {\rm g/cm^3}$, a vertical size of a few
${\rm km}$ and a horizontal extension of a few tens of ${\rm km}$. The
initial rest mass of the torus is an important quantity to be determined
in simulations, since it is related to the ejecta and to electromagnetic
emission in general. In particular, the existence of a massive torus
around the newly formed rotating black hole is a key ingredient in the modelling
of short gamma-ray bursts.

The dynamics of the torus are summarized for the representative case
adopted here in Fig.~\ref{fig:bns-torus}, which shows the isodensity
contours on a meridional plane.
Note that the panels refer to times between the times of the last two
panels of Fig.~\ref{fig:bns-dynamics}. Overall, the torus has a dominant
$m=0$ (axisymmetric) structure but, because of its violent birth, it is
very far from an equilibrium. As a result, it is subject to large
oscillations, mostly in the radial direction.

In the example simulation presented above, the merged object is a HMNS
that collapses to a rotating black hole in a few tens of
milliseconds. More in general, a HMNS can exist for up to $\sim 1$ s,
during which time cooling through neutrino emission, angular-momentum
transport associated with magnetic-field effects (such as the
magneto-rotational instability and magnetic braking), and the
gravitational torque resulting from its non-axisymmetric structure lead
the remnant to collapse \cite{Shapiro00, Rezzolla:2011,
  Hotokezaka2013c}. As it can be easily imagined, depending on the
initial mass of the system and on the EoS, other outcomes are possible
for the merger. The merger of BNS with higher masses and/or softer EoSs
ends in a prompt collapse to a black hole, while binaries with lower rest
masses and/or stiffer EoSs produce a merged object that does not collapse
for a longer time. This would likely be a {\it supramassive} NS
\cite{Cook92b}, namely an axisymmetric, uniformly rotating NS with mass
exceeding the upper limit for nonrotating NSs. After losing angular
momentum through secular mechanisms likely related to electromagnetic
emission, it would then collapse to a black hole on timescales of
$\sim10-10^4$ s \cite{Ravi2014}. Finally, in possibly marginal cases, the
merged object may not collapse at all, namely it may become a stable
NS. See Fig. 1 of Ref. \cite{Baiotti2016} for a schematic view of what
said in this paragraph about the possible outcomes of the merger.

\subsection{Gravitational-wave emission from binary--neutron-star mergers}
\label{sec:gwe}

\begin{figure}[htb]
\begin{center}
\begin{minipage}[t]{15 cm}
  \includegraphics[width=1.0\textwidth]{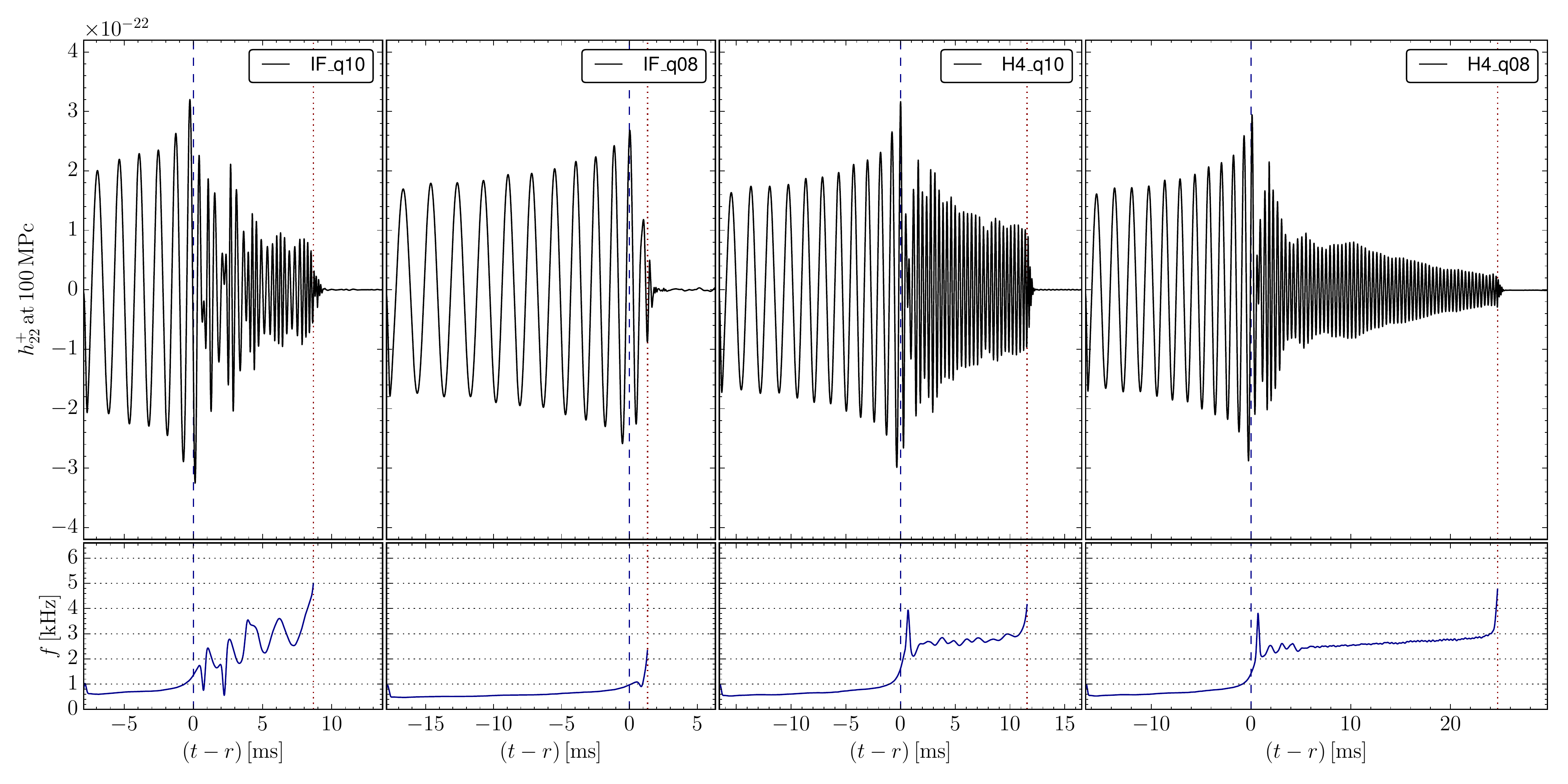}
\end{minipage}
\begin{minipage}[t]{16.5 cm}
  \caption{Some representative plots of the GW strain ($l=2$; $m=2$ mode
    only) taken from Ref. \cite{Kawamura2016}. Signals are for (from left
    to right): an equal-mass binary evolved with an ideal-fluid EoS, a
    binary with a mass ratio of the components of $q=0.8$ evolved with an
    ideal-fluid EoS, an equal-mass binary evolved with the H4 EoS
    \cite{GlendenningMoszkowski91}, a binary with a mass ratio of the
    components of $q=0.8$ evolved with the H4 EoS.  The top panels show
    the strain at nominal distance of $100\, \Mpc$. The lower panels show
    the instantaneous frequency. (From
    Ref. \cite{Kawamura2016}) \label{fig:GW-strain} }
\end{minipage}
\end{center}
\end{figure}

Representative waveforms from the merger of different BNS systems are
shown in Fig. \ref{fig:GW-strain}, taken from
Ref. \cite{Kawamura2016}. Reflecting the dynamics of matter as described
above, gravitational waveforms increase in amplitude and in frequency
during the inspiral (the so-called {\it chirp} signal\footnote{From the
  chirp signal, the chirp mass $M_{\rm chirp} =
  \frac{(m_Am_B)^{3/5}}{(m_A+m_B)^{1/5}}$ (the dominant parameter of the
  inspiral) of the binary system can be measured accurately
  \cite{Flanagan1994, Blanchet95, Blanchet:1996pi}, while the component
  masses $m_A$ and $m_B$ have much larger errors. For GW170817, the
  LIGO-Virgo Collaborations determined $M_{\rm chirp}
  =1.186\pm 0.001 \Msun$ \cite{Abbott2018a}.}), while the
waveforms after the merger, reflecting the oscillations of the merged
object, are much more varied and in many cases terminate with the
ringdown\footnote{At its birth from collapse (or merger), the black hole
  {\it rings}, namely oscillates in shape. Since this ringing is rapidly
  damped through the emission of GWs, it is called {\it ringdown}. Its
  frequencies are given by its quasinormal modes of oscillation,
  calculable through perturbation theory (see Ref. \cite{Kokkotas99a} for
  a review).} signal of the black hole, during which the distortions with
respect to a Kerr black hole are damped in a characteristic GW
signal. The ringdown signal for black holes formed in BNS mergers is at
frequencies of several $\kHz$ and so not easily measurable by current
detectors. The post-merger signal from before the collapse is at lower
frequencies than the ringdown, but still so high that detection is
probably limited to close sources \cite{Messenger2013, Clark2014,
  Clark2016, Yang2017, Chatziioannou2017, Abbott2018,
  Torres-Rivas2019}. The inspiral signal, instead, can be better measured
in current detectors, because of its duration (tens of seconds or even
more) and its frequency range of up to $\sim 1\, \kHz$. Detectors that
are currently active (Advanced LIGO \cite{Aasi:2014}, Advanced Virgo
\cite{Acernese:2014}), under construction (KAGRA \cite{Aso:2013},
Indigo/LIGO India \cite{Fairhurst2014}) and most of those that are
planned (LIGO Voyager \cite{McClelland2015}, Einstein Telescope
\cite{Punturo2010b, Sathyaprakash:2009xs}, Cosmic Explorer
\cite{McClelland2015}), in fact, are projected to have maximum
sensitivity around $1\, \kHz$ (see also Ref. \cite{Martynov2019} for a
recent proposal on further detector developments).

\subsection{Describing equations of state with generic parameterizations}
\label{eos-parameterization}

Given the complexity of the interactions and the structure of NSs, all of
the (supra)nuclear EoSs proposed so far use certain approximations and
involve large numbers of fundamental parameters. Researchers have indeed
been checking some of the proposed EoSs against the data we could gather
from GW170817, but another more general informative approach is to
parameterize possible EoSs with a few phenomenological parameters and to
analyse which constraints are imposed by observations on these
parameters. These results are less model dependent, since they do not
make use of a specific EoS among those proposed. However, the choice of
parameters for the parameterization is itself a model and actually the
form of the parameterization chosen has consequences on the Bayesian
inference of EoS parameters made using astrophysical data
\cite{Riley2018, Raaijmakers2018, Greif2019}.

A few parameterizations have been proposed and used, with different
degrees of complexity and fidelity to realistic EoS models
\cite{Lindblom2018a}. One is the parameterization with piecewise
polytropes \cite{Read:2009a, Steiner2010, Steiner2013a, Hebeler2013,
  Lackey2015, Raithel2016, Carney2018}, which, as the name says, consists
in stitching together polytropic EoSs with different adiabatic indexes
$\Gamma(\rho)$ for different ranges of density. It has been found that
the optimal number of polytropic {\it pieces} is five, each describing a
density interval between one and several (eight) times the nuclear
saturation density \cite{Raithel2016, Raithel2017}. It is a rather
straightforward method, which is used in most numerical-relativity
simulations because of its ease and efficiency of implementation, and
which allows to parameterize a large portion of the space of possible
EoSs with a few parameters (usually only three or four {\it pieces} are
used and necessary to get satisfactory agreement with all EoSs computed
from fundamental principles), but it has drawbacks as well. One of them
is that when comparing parameterized EoSs with specific EoSs computed
from fundamental principles, the accuracy drops near the fixed joining
densities \cite{Lackey2015}. Another drawback is that the
piecewise-polytropic parameterizations of the EoS may be
thermodynamically non-convex\footnote{The convexity of any EoS is
  mathematically defined in terms of the value of the {\it fundamental
    derivative}, which measures the convexity of the isentropes in the
  pressure-density plane. If the fundamental derivative is positive,
  isentropes in the pressure-density plane are convex, and thus
  rarefaction waves are expansive and shocks are compressive
  \cite{Rezzolla_book:2013, Ibanez2018}. This is the usual regime in
  which many astrophysical scenarios develop. However, some EoSs may
  display regimes in which the fundamental derivative is negative,
  namely, the EoS is non-convex. In this case rarefaction waves are
  compressive and shocks are expansive. These non-classical or exotic
  phenomena have been observed experimentally \cite{Cinnella2007,
    Cinnella2011}. Also, it has been shown that any first-order
  transition (see Sect. \ref{hybrid stars}) would lead to a non-convex
  thermodynamics \cite{Aloy2019}.}  at some of the juncture points even
if the physical EoS is convex \cite{Ibanez2013, Ibanez2018, Aloy2019} and
this may also lead to spurious deformations of the emitted GWs
\cite{Aloy2019}.

Furthermore, it was found that, in
principle, in some cases a piecewise-polytropic parameterization does not
admit an invertible mapping between the phenomenological EoS parameters
and quantities like gravitational mass, equatorial radius, and moment of
inertia of the star \cite{Riley2018, Raaijmakers2018}. However, this may
not be a serious problem, as in practice Bayesian analysis of
observations needs to be done \cite{Miller2019}, and it has been shown
that it is feasible to use gravitational signals to solve the
relativistic inverse stellar problem for piecewise-polytropic
parameterizations, \ie, to reconstruct the parameters of the EoS from
measurements of the stellar masses and tidal deformability (or tidal Love
number; see Sect. \ref{premerger-basic}) with a few observations made by
advanced interferometers \cite{Abdelsalhin2018}.

Another parameterization is the spectral parameterization
\cite{Lindblom2010, Lindblom2012, Lindblom2014}, which expresses the
logarithm of the adiabatic index of the EoS, $\Gamma(p, \gamma_i)$, as a
polynomial of the pressure $p$; $\gamma_i$ are the free EoS
parameters. This can match a wide variety of EoSs, usually better than
piecewise-polytropic models with the same number of parameters (however,
in cases in which the candidate EoSs contain phase transitions the
residuals are similar) \cite{Lindblom2010} and without the
above-mentioned problems at the joining points. The spectral
parameterization has recently seen further developments, including the
self-imposition of the causality constraint \cite{Lindblom2018a,
  Lindblom2018}, which would lead to improved computational
efficiency. However, the improved version has not yet been used in
applications to real data. It was also shown that the spectral
parameterization is free from the invertibility problems that may affect
the piecewise-polytropic representation (as long as the number of
measured data points is greater than or equal to the number of EoS
parameters) \cite{Lindblom2018}.

Both the piecewise-polytropic and the spectral parameterizations can be
stitched to an EoS known to be realistic at lower densities (\eg, below
half the nuclear saturation density), like the SLy EoS
\cite{Douchin01}. When they are used with random parameters to survey a
large portion of the space of EoSs, rather than used to mimic specific
tabulated EoSs, this connection at low density is actually the only
detailed input that these representations get from known nuclear
physics. Another limitation of these parameterizations is related to the
fact that they automatically include the assumption of
$\beta$-equilibrium\footnote{$\beta$-equilibrium is the condition of
  equilibrium between electrons, protons and neutrons with respect to the
  $\beta$-decay and inverse $\beta$-decay processes:
  \bea
  n &\rightarrow& p + e +\bar{\nu} \, , \nonumber\\
  e^- + p &\rightarrow& n + \nu \, , \nonumber
  \eea
  respectively. Neutrinos escape from the system. Inverse $\beta$-decay
  can proceed whenever the electron has enough energy to balance the
  mass difference between the proton and the neutron. $\beta$-decay is
  blocked if the density is high enough that all electron energy levels
  in the Fermi sea are occupied up to the one that the emitted electron
  would fill. See, \eg, Ref. \cite{Shapiro83}.}. After they form, NSs
cool rapidly via neutrino emission and quickly reach $\beta$-equilibrium,
therefore thermal and dissipative corrections to the EoS are negligible,
and NS matter is typically modelled as a perfect fluid. However, during
tidal disruption and merger of a BNS system, matter heats up and is not
in such an equilibrium \cite{Alford2018a}. For some time interval, then,
the piecewise-polytropic and the spectral parameterizations may not offer
a good description of matter. The common workaround used for this in
numerical simulations is the addition by hand of a thermal component of
the EoS modelled as an ideal fluid, in which the pressure $p$ is related
to the internal energy density $\epsilon$ and the rest-mass density
$\rho$ through the adiabatic index $\Gamma$ as $p=(\Gamma
-1)\rho\epsilon$.

The two representations introduced above do not allow for a simple
connection to knowledge on nuclear physics and they cannot bring
information about matter composition, such as the proton fraction. A
third and complementary way to obtain a parameterization of the EoS is to
Taylor expand the energy per nucleon $E$ of isospin asymmetric nuclear
matter near the saturation density $\rho_0$ (see, \eg,
Refs. \cite{Fattoyev2014, Margueron2018a, Margueron2018b, Zhu2018,
  Malik2018, Carson2019, Zhang2019a}). The expansion is carried out in
the isospin symmetry parameter $\delta \equiv (\rho_n-\rho_p)/\rho$ (with
$\rho_p$ and $\rho_n$ representing the proton and neutron number
densities respectively and $\rho \equiv \rho_p + \rho_n$) around the
symmetric nuclear matter case $\delta=0$ \cite{Vidana2009}:
\begin{equation}
  E(\rho,\delta)=E(\rho,0)+E_{\rm sym}(\rho)\delta^2+ \mathcal{O}(\delta^4),
\end{equation}
where $E(\rho,0)$ corresponds to the energy of symmetric nuclear
matter. $E(\rho,0)$ and $E_{\rm sym}(\rho)$ are then expanded themselves
around the saturation density
\bea
  E(\rho,0)&=&E_0+\frac{K_0}{2} y^2 + \frac{Q_0}{6}y^3 + \mathcal{O}(y^4),\\
  E_{\rm sym}(\rho)&=&E_{{\rm sym},0}+L y + \frac{K_{{\rm sym}}}{2} y^2 + \frac{Q_{{\rm sym}}}{6} y^3+\mathcal{O}(y^4),
\eea
where $y \equiv (\rho - \rho_0)/3 \rho_0$. These lowest-order parameters
are known as the energy per particle $E_0$, the incompressibility
coefficient $K_0$, the third derivative of the energy of symmetric matter
$Q_0$, the symmetry energy $E_{{\rm sym},0}$, its slope $L$, its
curvature $K_{{\rm sym}}$, and its skewness $Q_{\rm sym}$ at saturation
density. Some of these parameters ($E_0$, $K_0$, and $E_{{\rm sym},0}$)
are either strongly constrained by terrestrial experiments or have no
noticeable impact on the bulk properties of NSs (see, \eg,
Refs. \cite{Margueron2018b, Zhang2018, Carson2019, Li2019b} and
Ref. \cite{Li2019c} for a recent review.). The slope of the incompressibility
\bea
  M_0 &= Q_0 + 12 K_0
\eea
is also addressed in some works \cite{Alam2014, Alam2016, Malik2018,
  Carson2019}. For some results on constraints on the above parameters
from GW observations see Sect. \ref{sec:correl nuclear param}.

A different way of parameterizing the nuclear-matter EoS in terms of
nuclear parameters consists in expanding in the proton fraction $x_p =
\rho_p/(\rho_n + \rho_p)$ from pure neutron matter to symmetric neutron
matter \cite{McNeilForbes2019}.

The EoS representations discussed above are used only or mainly to
describe smooth, nucleonic EoSs and may not work very well if, at higher
densities, hadrons undergo deconfinement and NSs have a quark-matter core
\cite{Ivanenko1965, Itoh70, Baym1976}. In the literature, three main
effective models have been proposed and used to address this phase (see,
\eg, Refs. \cite{Buballa2014, Alford2016a} for reviews): (i) the
relatively simple MIT bag model \cite{Alcock86, Haensel1986} (and its
many variations), where quarks are treated as massless particles confined
in a {\it bag} of finite dimension through a {\it bag pressure} $B$
(called the {\it bag constant}), a phenomenological quantity introduced
to take into account the nonperturbative effects of QCD; (ii) the much
more complicated Nambu--Jona-Lasinio model \cite{Nambu1961a, Nambu1961b},
and (iii) a simpler, phenomenological description that can mimic the
above two sophisticated models by assuming a density-independent speed of
sound, known as constant-speed-of-sound parameterization
\cite{Chamel2013, Zdunik2013, Alford2013}, and that is much better suited
for numerical computations \cite{Montana2018, Han2018}. The speed of
sound may be a better parameter also because it is physically constrained
by both causality requirements and an asymptotic limit at ultrahigh
density, $c_s^2\rightarrow 1/3$, suggested by theoretical insights
\cite{Weinberg72} and perturbative QCD calculations \cite{Kurkela2010,
  Fraga2014}\footnote{Note that not all agree on this limit
  \cite{Olson2001, Bedaque2015}. Furthermore, such an asymptotic value
  would probably be reached only at densities orders of magnitude higher
  than the maximum densities attained in NSs, even after the merger
  (see, \eg, Refs. \cite{Tews2019, Fujimoto2019}).
  It is likely that the speed of sound reaches a maximum at some
  density before tending to the asymptotic limit
  \cite{Tews2019}. Ref. \cite{Annala2019} has recently discussed the
  constraints that astrophysical observations place on the speed of sound
  of quark matter in compact stars \label{sound speed limit 1/3}.}.
Alternative parameterizations of the speed of sound have been proposed
and used recently in other works \cite{Tews2018a, Tews2018,
  Christian2018, Greif2019, McNeilForbes2019, Annala2019}; see
Sect. \ref{hybrid stars}.

Other models to parameterize EoSs in general have been proposed, but are
currently rarely or not used in simulations and data analysis. One is a
model in which the EoS is divided into four density regimes: a fixed
crust below the nuclear saturation point, one pressure-energy relation
depending on nuclear-physics parameters (as symmetry energy and
proton/electron fraction) for densities around the saturation point, and
two polytropic relations at larger densities to fit the inner core
\cite{Steiner2010}\footnote{This representation turned out to be a
  generalization of that of Refs. \cite{Ozel2009, Ozel:2010}.}. This
parameterization was devised from considering data from
photospheric-radius-expansion bursts and quiescent thermal emission from
x-ray transients \cite{Steiner2010} and cannot include EoS that allow for
twin stars \cite{Glendenning2000, Schaffner-Bielich2002, Fraga2002} (see
Sect. \ref{hybrid stars} for more on twin stars).

Another method combines the better-known EoS of nuclear
matter at lower densities (as done in the representations described
above) with the requirement that the pressure approaches that of
deconfined quark matter at high densities \cite{Kurkela2014}. This places
significant further restrictions on the possible EoSs, if one accepts the
assumptions made.

Up to this point I have summarized parametric models that describe the
EoS. However, as mentioned above and as recently pointed out in
Ref. \cite{Landry2018}, parametric EoS inference has limitations, due to
unavoidable modelling errors \cite{Lindblom2010, Lackey2015, Carney2018},
especially for sharp features like first-order phase transitions (see
Sect. \ref{hybrid stars} for more on phase transitions and hybrid
stars). Given the extremely varied phenomenology of (hybrid) EoSs, it is
unlikely that any single parametric model will be able to faithfully
represent the full range of EoS variability with only a few
parameters. In order to remedy these drawbacks, non-parametric
representations of the EoS have been proposed \cite{Landry2018,
  Margueron2018a, Margueron2018b}. Non-parametric representations do
effectively involve some parameters (called hyperparameters) that control
allowed types of functional behaviour, but their coverage of the space of
EoSs is much larger than that provided by a parametric model. The key
difference between parametric and non-parametric models is that the
faithfulness of the representation increases with the number of
parameters in the former case, while in the latter case it scales not
with the number of hyperparameters but with how representative the input
knowledge (in the specific case, the set of EoSs) used to constrain the
functional behaviour is. It can then be expected that systematic errors
are much smaller in non-parametric representations, if the {\it a priori}
knowledge is realistic and accurately input.

One example of non-parametric representations is the metamodel for the
nucleonic EoS of Refs. \cite{Margueron2018a, Margueron2018b}. A
metamodel, or surrogate model, is a model of models, \ie~a simplified
model of actual models, EoS representations in this case. Conceptually, a
metamodel is like building a hypersurface from a limited amount of known
data from the underlying models and approximating the output over a much
wider parameter space. Metamodels have to be evaluated with respect to
their goodness (how well they fit a set of data) and there is no proof of
existence or of uniqueness in general. Some of the advantages of this
metamodel of Refs. \cite{Margueron2018a, Margueron2018b} are
that (i) it covers a wider space of EoSs, which may not
be covered by the underlying models or by specific elements like specific
EoSs (as opposed to parameterized EoSs) or experimental data; (ii) it
allows the incorporation of knowledge on nuclear physics acquired from
laboratory experiments or the results of complex {\it ab initio} models; (iii)
it can easily be used in the Bayesian framework, facilitating the
estimation of the experimental and theoretical error bars into confidence
levels for the astrophysical observables. Refs. \cite{Margueron2018a,
  Margueron2018b} even argue that predictions done with their metamodel
are without assumptions on the functional form of the EoS and only
require that the EoS is nucleonic and satisfies basic physical
constraints, like thermodynamical stability and causality, and therefore
can be qualified as model independent.

As a final addition to this Section, note that all recent works on NS
EoS, especially those reviewed in Sect. \ref{constraints}, impose
physical limits on the parameterizations used. In particular, the
following is required for viable EoSs.

(1) Causality: The speed of sound must be less than the speed of light
everywhere, in particular up to the central pressure of the heaviest NS
supported by the EoS\footnote{General limits imposed by causality have
  been known for a long time \cite{Rhoades1974} and applied to the tidal
  deformability \cite{Moustakidis2017, vanOeveren2017} also before the
  GW170817 event.}.

(2) Observational consistency: The EoS allows for a maximum stellar mass
equal to or larger than the largest observed mass of a NS. Among the
observations and inferences therefrom that are well-understood and
therefore reliable, it has been recently announced in April 2019 that a
NS of estimated mass $2.17\pm^{0.11}_{0.10} \Msun$ was observed in PSR
J0740+6620 \cite{Cromartie2019}. The other highest NS masses reliably
observed up to now are $1.908\pm 0.016 \Msun$ in PSR J1614−2230
\cite{Demorest2010, Arzoumanian2018} and $2.01\pm 0.04 \Msun$ in PSR
J0348+0432 \cite{Antoniadis_fulllist:2013}. Different works published
before April 2019 use different, more or less conservative, values for
such lower limit to the maximum mass to be reached by NSs, spanning from
$1.93$ to $2.01 \Msun$. Note that in Bayesian analyses also observations
of record masses should be treated with a likelihood-based procedure,
which includes multiple observations and the error on their measurements,
rather than strict bounds \cite{Miller2019}, as already done in
Ref. \cite{AlvarezCastillo2016}.

(3) All NSs are described by the same EoS\footnote{This last requirement
  was actually not imposed in the detection article for GW170817
  \cite{Abbott2017}, but only later, in analyses by the LIGO-Virgo
  Collaborations \cite{Abbott2018a, Abbott2018b} and by others (see
  Sect. \ref{nucleonic}).}.

\subsection{Other compact objects possibly similar to neutron stars}
\label{other-compact-objects}

Up to this point, I have focused only on compact stars made of ordinary
matter, usually referred to in general as NSs. In this subsection I will
briefly introduce other possible exotic objects that could mimic NSs (and
some of them also black holes, actually), to the point that current or
near-future observations may not be able to make a distinction. Such
exotic objects are of course interesting on their own and not only in
comparison to NSs. Research on these astrophysical objects may, for
example, give hints on the nature of dark matter, help identify quantum
effects that could halt collapse to black holes and thus prevent the
formation of spacetime singularities, or contribute to solve the
information problem in black-hole physics (loss of unitarity in Hawking
evaporation). See Refs. \cite{Cardoso2019, Barack2019} for reviews on
all these topics and more.

Perhaps the most popular alternative proposal is that of boson stars
\cite{Schunck2003, Liebling2012}, which are compact, stationary
configurations of a scalar field bound by gravity. There have been
several proposals about what such a scalar field could be, like the Higgs
boson \cite{Brihaye2010, Vincent:2016a} or axion particles
\cite{Barranco2011}. Boson stars could either be composed of stable
fundamental bosonic particles bound by gravity, or of unstable particles
for which a process inverse to their decay is enhanced by gravitational
binding and becomes efficient enough to reach equilibrium, similarly to
what happens with $\beta$-decay in NSs. Boson stars may have
formed through gravitational collapse during the primordial stages of the
big bang \cite{Hogan88, Madsen1990, Kolb1993}.

Boson stars made of axions are also called axion stars \cite{Iwazaki1999,
  Eby2017}. While yet unobserved, axions are supported and motivated by
theory, since they solve the strong CP problem of QCD \cite{Peccei1977},
arise naturally in string theory compactifications (see, \eg,
Ref. \cite{Arvanitaki2010}), and are candidates for dark matter
\cite{Marsh2016}. Another example of boson stars mentioned later in this
review is Proca stars \cite{Brito2016}, which are the vector analogues
of the scalar boson stars.

Other hypothesized compact objects are gravastars (short for {\it
  gravitational vacuum condensate star}) \cite{Mazur2004}, which are made
of a thin shell of matter with radius very close to its Schwarzschild
radius. Inside it, a phase transition originated by the collapse of the
original massive star would form a core described by a metric (de Sitter
metric) that has a repulsive effect that stops the collapse, thus
preventing the birth of both a singularity and an event horizon (see
again Ref. \cite{Cardoso2019} for a review and, \eg,
Ref. \cite{Chirenti2016} for a shorter summary of work on gravastars).
Very recently, another type of compact objects that may mimic black holes
and compact NSs has been proposed in Ref. \cite{Buoninfante2019} and
named {\it nonlocal stars}.

See Sect. \ref{exotic binaries} for current results on the possibility to
distinguish observationally these exotic binaries from binaries composed
of stars made of ordinary matter. Some comments on the possibility and the
effects of accumulation of dark matter in NSs will be made in
Sects. \ref{postmerger basic ideas}, \ref{exotic binaries} and \ref{dark
  matter}.

\subsection{{\it Universal relations} for neutron stars}
\label{univ-rel}

In works about the connection between GW observations
from BNS systems and the EoS at supranuclear density, several empirical
relations (found through numerical-relativity simulations) between
physical quantities of NSs and observed quantities are used. Often the
term employed to address them is {\it universal relations}. This subsection
contains an introduction to such relations.

{\it Universal} here means {\it approximately independent of the
  EoS}. There are two types of such relations: relations that connect
different physical quantities of a NS isolated or in a binary system
(like radius, mass, moment of inertia, tidal deformability; see
Sect. \ref{premerger-basic} for definitions) among themselves and
relations that connect physical quantities of BNS systems (like the
compactness $C\equiv M/R$ of the component stars) with quantities that
can be measured directly (like the main frequency of the post-merger GW
spectrum) in GW detectors.

Relations of the former type often have interesting names, like the {\it
  I-Love-Q} relations \cite{Yagi2013a, Yagi2013b, Yagi2017, Pappas2014,
  Chakrabarti2014}, which connect the moment of inertia $I$, the Love
number [which is related to the tidal deformability; see
  Eq. \eqref{eq:Love number}] and the quadrupole moment $Q$ of a NS. Such
universality may originate from the fact that these relations depend most
sensitively on the internal structure far from the core, where all
realistic EoSs are rather similar. The universality is found to hold at
the $1\%$ level, except in exotic cases, like bare solid quark stars
\cite{Anglani2014, Lau2017, Lau2019} (but even in this example they hold
at the $20\%$ level), but it may be affected by strong magnetic fields
\cite{Haskell2014, Yagi2017} and large differential rotation
\cite{Yagi2017}. Also, theories of gravity different from general
relativity predict different relations\footnote{This fact can actually be
  taken advantage of for testing general relativity versus other theories
  of gravity \cite{Yagi2013a, Yagi2013b, Sham2014a, Pani2014, Gupta2018,
    Doneva2018}, if one can obtain for a given NS independent
  measurements of two of the quantities involved in the relations, \eg,
  the Love number through GW observations and the moment of inertia or
  the quadrupole moment through observations of binary pulsars
  \cite{Lattimer2005b, Ozel2016a} or x-ray binaries \cite{Watts2016,
    Gendreau2016, Arzoumanian2018, Degenaar2018, Watts2019}.}.

Other relations of this type are the {\it C-Love} relations
\cite{Yagi2013a, Yagi2013b, Maselli2013, Yagi2017, Biswas2019} between
the compactness and the Love number of a NS. They hold at the few
percent level for all EoSs on which they were tried, including those with
phase transitions \cite{Carson2019a}. Still another is the $I-C$
(moment-of-inertia--compactness) relation \cite{Ravenhall94, Bejger02,
  Lattimer2005b, Baubock2013}, which was actually the first to be found,
but possesses a lesser degree of EoS insensitiveness (accurate at the
$10\%$ level) \cite{Chan2016, Yagi2017}, unless improved (to the several
percent level) with a different normalization \cite{Breu2016}. More
general relations between the lowest few multipole moments of NSs have
also been found \cite{Pappas2014, Stein2014, Yagi2014}.
A relation, holding at the level of a few percent, between the total
mass of the binary and the angular momentum in the remnant has been
presented in Ref. \cite{Bauswein2017}.

Then, the {\it binary Love} relations, often used in the analysis of the
data of GW170817, connect the tidal deformabilities of the two stars in a
BNS system\footnote{Similar relations have been proposed that are valid
  for any two NSs (that have the same EoS), independently of whether they
  are in a binary system or not \cite{Landry2018a, Kumar2019}.}
$\Lambda_A$ and $\Lambda_B$ \cite{Yagi2016, Yagi2017a, Rezzolla2017,
  De2018, Zhao2018, Landry2018a, Kumar2019, Carson2019a}. They hold at
about the $20\%$ level for nucleonic EoSs, but it has been shown that, as
intuition may suggest, they do not hold for EoSs with phase transitions
when one star in the binary is a NS and the other a hybrid star
\cite{Han2018, Bhat2019, Sieniawska2019, Carson2019a} (see also
Sect. \ref{hybrid stars}). Making use of combinations of the
above-mentioned relations, {\it R-Love} (NS-radius--Love-number)
relations have been proposed \cite{Carson2019a}.

Belonging to the other type of universal relations is, for example, the
relation between the tidal deformability and the frequency of the merger
of a BNS system, defined as the instantaneous GW frequency at the time
when the amplitude reaches its first peak. This relation was first found
in Ref. \cite{Read2013} and later confirmed by more advanced and
comprehensive works \cite{Bernuzzi2014, Bernuzzi2015a, Takami2015,
  Rezzolla2016, Tsang2019, Kiuchi2019a}. It had been shown to hold at
around the $1\%$ level, but a work posted to arXiv.org just before this
review was accepted for publication revised the relative error between
data and (an {\it improved}) fitting formula to $3\%$, through resolution
studies that included also a more varied set of binaries, in particular
with more varied mass ratios \cite{Kiuchi2019a}. Furthermore, the
relation is found to hold only for equal-mass or very nearly equal-mass
binaries \cite{Rezzolla2016, Kiuchi2019a} and it has not been tested yet
for magnetized and/or highly spinning binaries.

A similar relation \cite{Read2013, Kiuchi2017a, Kiuchi2019a} was found
between the tidal deformability and the GW amplitude at its first peak,
which defines the merger of a BNS system, as said above. It has been
shown to hold at the $4\%$ level \cite{Kiuchi2019a}.

Another set of universal relations, with different degrees of
reliability, has been found to connect the frequencies of the main peaks
of the power spectral density of the post-merger GW signal with
properties (radius at a fiducial mass, compactness, etc.) of a spherical
star in equilibrium \cite{Bauswein2011, Bauswein2012, Bauswein2012a,
  Bauswein2014, Takami:2014, Takami2015, Bernuzzi2015a, Dietrich2015,
  Foucart2015, Lehner2016, DePietri2016, Rezzolla2016, Dietrich2017,
  Maione2017, Kiuchi2019a} (see Sect. \ref{postmerger}). The spin of the
NSs in the inspiral has been found to affect these relations
\cite{Bauswein2015b, Bernuzzi2015a, Dietrich2018a, East2019}, that
otherwise hold at the $\approx 10\%$ level for EoSs without phase transitions.

Other relations have been found between the threshold mass for prompt
collapse, the maximum mass for a nonrotating NS and its radius
\cite{Bauswein2013, Bauswein2017, Bauswein2017b, Koeppel2019} (see
Sect. \ref{postmerger}), between the quantity $\kappa_2^T$ [defined in
Eq. \eqref{eq:kappa 2 T}], which parameterizes the late-inspiral of tidally
interacting binaries, and the main peak of the post-merger GW spectrum
\cite{Bernuzzi2015a} (again, see Sect. \ref{postmerger}), and between the
total gravitational radiation emitted in a merger event and the angular
momentum of the remnant \cite{Zappa2018, Kiuchi2019a}.

Still other universal relations have been found between the frequencies
of the fundamental modes ({\it f-modes}) of oscillation of stars and
certain combinations of the stellar mass and radius \cite{Andersson1996a,
  Andersson1998b, Benhar1999, Benhar:2004xg, Lattimer04, Lau2010,
  Chan2014, Chirenti2015, Doneva2015a}. The best of these relations hold
at the $1\%$ level \cite{Lau2010, Chan2014, Chirenti2015}. Combining
these with the I-Love relation, it is possible to find a universal
relation between the f-mode oscillation frequency and the tidal
deformability \cite{Chan2014, Yagi2017, Wen2019}.

As mentioned above, these universal relations are in fact not literally
{\it universal}, since they have some (small) levels of dependence on the
EoS. Hence, if, on one side, using these relations in data analysis
allows to perform estimates impossible otherwise (see the rest of this
review), on the other side they do include further uncertainties in the
analyses. In some cases, the issue has been addressed by marginalizing
over the EoS variability, for example when inferring the NS radii of
GW170817 employing binary Love and C-Love relations
\cite{Chatziioannou2018}. Since in current detectors statistical
uncertainties in parameter estimation are much larger than systematic
errors added by using these universal relations, such a marginalization
procedure is not a noticeable handicap, but, in order to be useful with
the higher sensitivity of (near) future detectors and the large number of
expected observations, more accurate universal relations will be necessary
\cite{Carson2019a}. On the other hand, the observations themselves will
impose more and more stringent constraints on the EoS, and the reduced
allowed space of EoSs will allow to decrease the number of EoSs used for
computing the uncertainties in the universal relations and thus decrease
their variability \cite{Carson2019a}. For example, by imposing the
$90\%$ credible region constructed from the posterior probability
distributions on the pressure-density plane \cite{Abbott2018b,
  Carney2018} obtained from GW170817, it has been found that the EoS
insensitivity increases by a factor of $\sim 60\%$ in the binary Love
relations, by a factor of $\sim 70\%$ in the C-Love relations, and by
factors of $\sim 50\%$ in the I-Love-Q relations \cite{Carson2019a}.

\section{Extracting information on the equation of state from gravitational waves emitted before
the merger}
\label{premerger}

The dynamics of BNS systems around merger depend on the EoS of
ultrahigh-density matter. Essentially two methods to link the observed
GWs to the NS EoS have been studied. One method uses tidal deformations
during the last orbits before merger \cite{Flanagan08, Hinderer08,
  Read:2009b, Bernuzzi2012, DelPozzo2013, Bernuzzi2013, Read2013,
  Yagi2013b, Favata2014, Wade2014, Lackey2015, Agathos2015,
  Chatziioannou2015, Hotokezaka2016, Dietrich2017b, Kawaguchi2018,
  Dudi2018, Samajdar2018, Harry2018}; tidal deformations have been measured for
GW170817\footnote{Note that one analysis suggested that the noise in the
  high-frequency region of the data from the Livingston LIGO
  interferometer may have corrupted information about the tidal
  deformability estimated for GW170817 \cite{Narikawa2018}.}
\cite{Abbott2017, Abbott2018a, Abbott2018b}.

The other method uses the spectra of the gravitational radiation of the
post-merger object (if it does not collapse to a black hole too soon)
\cite{Bauswein2011, Bauswein2012, Bauswein2012a, Bauswein2014,
  Takami:2014, Takami2015, Bernuzzi2015a, Dietrich2015, Rezzolla2016,
  Dietrich2017}. More energy may be emitted in GWs in this phase than in
the inspiral, but, because of their higher frequency, their
signal-to-noise ratio in current and projected detectors is smaller than
in the inspiral.

In this Section I will focus on the first method.

\subsection{The basic idea}
\label{premerger-basic}

Stars in a binary system undergo tidal deformations that become larger as
they get closer. These deformations affect the orbital trajectory of the
binary and thus the emitted GWs \cite{Kochanek92, Lai1996a}, encoding in
the latter the NS EoS. Tidal deformations are described through the tidal
deformability coefficient defined as the proportionality constant
$\lambda$ between the external tidal field $\mathcal{E}_{ij}$ (the field
generated by the companion star) and the quadrupole moment of the star
$Q_{ij}$ \cite{Flanagan08, Hinderer08, Read:2009b, Read2013, Yagi2013b}:
\begin{equation}
Q_{ij} = - \lambda \mathcal{E}_{ij}\, .
\end{equation}
However, tidal deformations are more usefully described through the
dimensionless tidal deformability\footnote{Here and in the
  other equations below I use geometric units, in which $G=c=1$, unless
  otherwise stated.}
\begin{equation}
  \Lambda \equiv \frac{\lambda}{M^5}\, ,
\end{equation}
where $M$ is the stellar mass. Equivalently, this can be written as
\begin{equation}
  \label{eq:Love number}
  \Lambda \equiv \frac{2}{3}\kappa_2\bigg(\frac{R}{M}\bigg)^5\, ,
\end{equation}
where $\kappa_2$ is the quadrupole Love number and $R$ the stellar
radius. This gives the quadrupole component of $\Lambda$, which can be
calculated via the following expression \cite{Hinderer08, Damour:2009,
  Yagi2013b}
\bea 
\Lambda &=& \frac{16}{15} (1-2C)^2[2+2C(y(R)-1)-y(R)]\cdot \nonumber\\
&&\Big\{2C[6-3y(R)+3C(5y(R)-8)]+4C^3[13-11y(R)+C(3y(R)-2)+2C^2(1+y(R))]+ \nonumber\\
&& 3(1-2C)^2[2-y(R)+2C(y(R)-1)]\ln{(1-2C)}\Big\}^{-1}\, ,
\eea
where $C \equiv M / R$ is the stellar compactness and $y(r)$ satisfies the
differential equation
\bea
\label{eq-zzzz} 
\frac{dy}{dr}= \frac{4(m+4\pi r^3 p)^2}{r(r-2m)^2}+\frac{6}{r-2m} -\frac{y^2}{r} - \frac{r+4\pi r^3 (p-\rho)}{r(r-2m)}y
-\frac{4\pi r^2}{r-2m}\left[5\rho+9p
+\frac{\rho+p}{(dp/d\rho)}\right]\, ,
\eea
where $p$ and $\rho$ represent pressure and mass density,
respectively, and
\bea
m(r) \equiv \frac{\left[ 1- e^{-\gamma (r)} \right] r}{2}\, ,
\eea
where $\gamma (r)$ is in the definition of the metric coefficients \cite{Hartle67}
\bea
ds_0^2 = -e^{\nu(r)} dt^2 + e^{\gamma (r)} dr^2 + r^2 (d\theta^2 + \sin^2
\theta d\varphi^2)\, .
\eea
Eq.~\eqref{eq-zzzz} can be solved for a given EoS together with the
Tolman-Oppenheimer-Volkoff (TOV) equations \cite{Tolman39,
  Oppenheimer39b}, which describe spherically symmetric stars in static
equilibrium in general relativity. Appropriate boundary conditions need
to be imposed \cite{Hinderer08}.

In general, the extraction of higher-order GW parameters (like tidal
deformabilities) from the GW signal is difficult because these parameters
can be efficiently extracted only in the late part of the inspiral, which
may not be very long, and because there exist degeneracies between
different higher-order parameters, like the individual NS spins and the
tidal polarizability\footnote{This is why the tidal deformabilities
  estimated by the LIGO-Virgo Collaborations were separated in a low-spin
  and a high-spin scenario \cite{Abbott2017, Abbott2018a, Abbott2018b}.}.
Moreover, because of the strong correlation between the two tidal
deformabilities, $\Lambda_A$ and $\Lambda_B$, associated with each star
in a binary, it is challenging to extract them separately from the
gravitational waveform, unless one assumes some empirical {\it universal
  relations} connecting $\Lambda_A$ and $\Lambda_B$ \cite{De2018,
  Yagi2016, Yagi2017a, Zhao2018, Carson2019a}, which may add
uncertainties to the estimates, as said in Sect. \ref{univ-rel}. Also, as
mentioned earlier, it has been shown that such {\it universal relations}
do not necessarily hold for EoSs with phase transitions \cite{Han2018,
  Sieniawska2019, Carson2019a}.

What can be more easily directly measured from BNS waveforms is the
dominant\footnote{More in detail, the lowest-order post-Newtonian
  correction (see Sect. \ref{note:postNewtonian}) of $\Lambda_A$ and
  $\Lambda_B$ can be written as $\tilde{\Lambda}$ and another parameter
  $\delta\tilde{\Lambda}$, whose contribution, however, is very small
  ($\delta\tilde{\Lambda}/\tilde{\Lambda} \lesssim 0.01$) and can be
  ignored, also because it cannot be extracted from data of current
  detectors \cite{Favata2014, Wade2014}.} tidal parameter in the
waveform, corresponding to the mass-weighted average tidal deformability
(also called effective tidal deformability) given by \cite{Flanagan2008}
\begin{equation}
  \label{Lambda-tilde}
  \tilde{\Lambda} = \frac{16}{13} \frac{(m_A+12m_B) m_A^4\Lambda_A+(12m_A+m_B)m_B^4\Lambda_B}{(m_A+m_B)^5}
\end{equation}
or, equivalently,
\begin{equation}
  \tilde{\Lambda} = \frac{16}{13} \frac{(1+12q) \Lambda_A+(12+q)q^4\Lambda_B}{(1+q)^5},
\end{equation}
where $q \equiv m_B/m_A (\leq 1)$ is the ratio of the masses of the two stars in the binary.

Another useful related quantity that is part of some universal relations
(see Sects. \ref{univ-rel} and \ref{postmerger basic ideas}) is
%
\begin{equation}
  \label{eq:kappa 2 T}
  \kappa_2^T \equiv 2\Big(\frac{1}{q^4(1+1/q)^5}\frac{\kappa^A_2}{C^5_A} + \frac{1}{q(1+1/q)^5}\frac{\kappa^B_2}{C^5_B}\Big)\, ,
\end{equation}
where $\kappa^{A,B}_2$ are the quadrupole Love numbers of the two stars
of the binary, respectively, and $C_{A,B}$ are their compactnesses
\cite{Bernuzzi2015a}. It has been shown that the dimensionless
GW frequency depends on the stellar EoS, binary mass and
mass ratio, only through this {\it tidal coupling constant} $\kappa_2^T$
\cite{Bernuzzi2015a}.

An alternative way to gain information on the tidal deformability from
detector data would be to extract the (mass-independent) coefficients of
a Taylor expansion of the tidal deformabilities about some fiducial mass
\cite{Messenger:2011, DelPozzo2013, Yagi2016, Yagi2017a}. This
parameterization, however, can be efficiently applied only to systems
whose NS masses are close to the fiducial mass, otherwise, the systematic
error on the leading tidal coefficient due to mismodelling the tidal
deformability can dominate the statistical one. This drawback, on the
other hand, may be compensated and overcome by combining the information
from multiple events (also with different masses), which can be done
easily with this method. However, a sensitivity higher than that of
current detectors would be necessary to accurately measure any of these
coefficients.

Since the Taylor expansion of the tidal deformabilities just mentioned
has not been used yet in practice, I will focus on the use of the
mass-weighted average tidal deformability. Reducing the tidal parameters
to one ($\tilde{\Lambda}$) allows for better statistical estimations, but
of course some physical information about the two stars is lost. As
mentioned in Sect. \ref{univ-rel}, a way around this problem was proposed
in Ref. \cite{Yagi2016}, that showed the existence of an EoS-insensitive
relation (with variations of at most $20\%$) between symmetric and
antisymmetric combinations of the tidal deformabilities
\bea
\Lambda_s\equiv \frac{\Lambda_A+\Lambda_B}{2}\, , \qquad
\Lambda_a\equiv \frac{\Lambda_A-\Lambda_B}{2} \, .
\eea
These {\it binary Love} relations [\eg~$\Lambda_a(\Lambda_s)$] allow to
compute the individual tidal deformabilities from the mass-weighted
average tidal deformability. A simple Fisher analysis has shown that the
binary Love relations improve parameter estimation of the individual
tidal deformabilities by up to an order of magnitude with respect to
estimations done by extracting $\Lambda_A$ and $\Lambda_B$ from the data
directly \cite{Yagi2016, Yagi2017a}. Modified binary Love relations for
specific purposes have also been proposed \cite{Rezzolla2017,
  Landry2018a, Kumar2019}.

As an important note on this introductory material, it may be useful to
notice that the relationship between the tidal deformability $\Lambda$
and the radius $R$ cannot be simply derived from Eq. \eqref{eq:Love
  number}, namely $\Lambda$ is not necessarily proportional to $R^5$,
because the quadrupole Love number $\kappa_2$ also depends (and
differently for each EoS) on the radius in a complicated manner,
determined by differential equation \eqref{eq-zzzz} coupled to the
Tolman-Oppenheimer-Volkov equations \cite{Tolman39, Oppenheimer39b}. A
few empirical model-dependent estimations for such a relation
\cite{Annala2017, Fattoyev2017, Zhou2017, Lim2018, Malik2018, De2018,
  Tews2019} found exponents between $5.28$ and $7.5$.  Even if the
correlation between radii and tidal polarizabilities for different EoSs
is tight, these two quantities provide complementary information, since
for a given tidal polarizability, different EoSs may lead to somewhat
different radii \cite{Tews2019}.  Also, it has been shown that these
relations have errors of $5\%$ or more for hybrid EoSs
\cite{Sieniawska2019}.

\begin{figure}[htb]
\begin{center}
\begin{minipage}[htb]{12 cm}
   \includegraphics[width=1.0\textwidth]{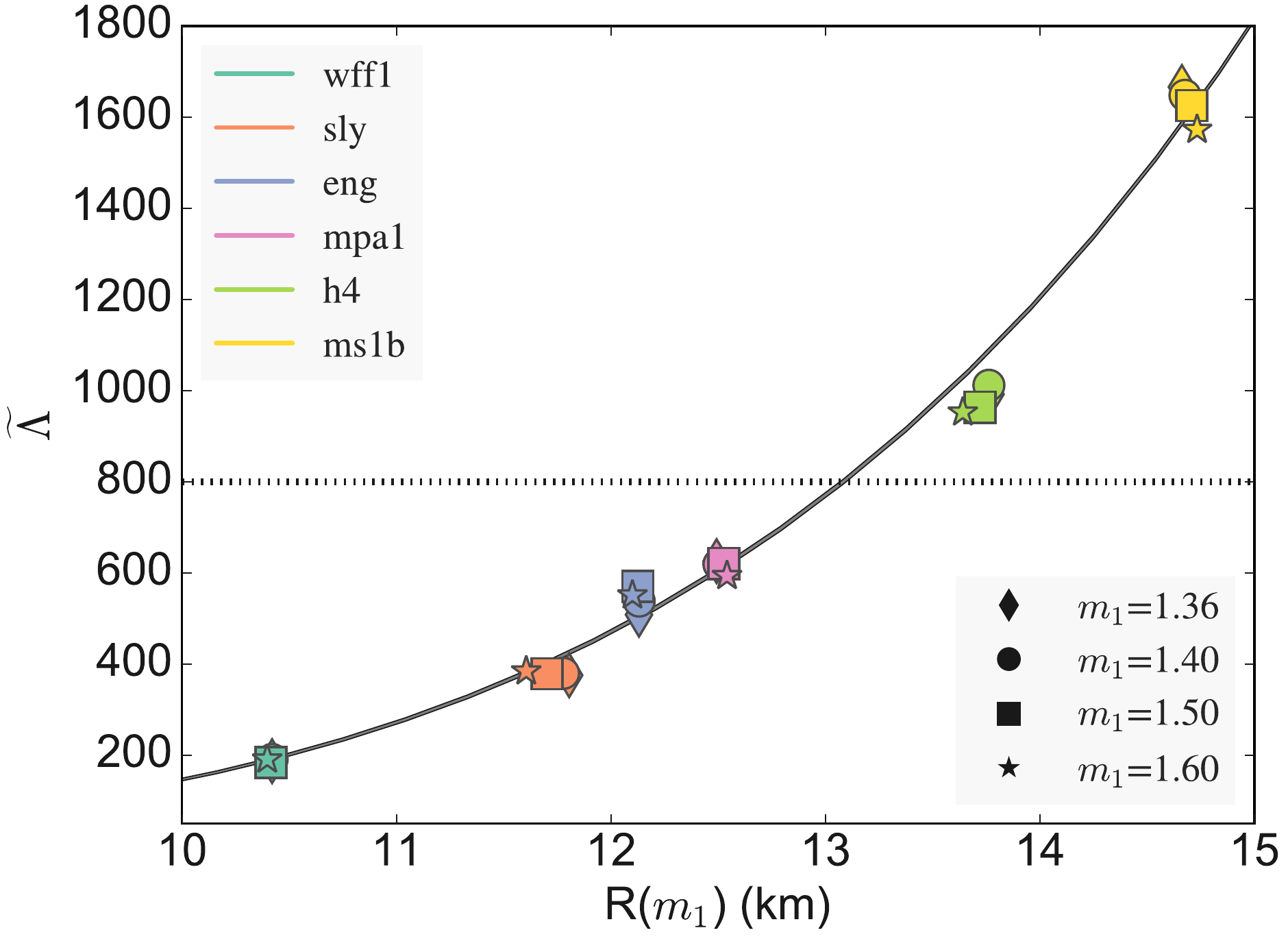}
\end{minipage}
\begin{minipage}[t]{16.5 cm}
\caption{Mass-weighted average tidal deformability
  [Eq. \eqref{Lambda-tilde}] of a binary system as a function of the
  radius of the primary NS. The tidal deformability is calculated for
  various primary masses (corresponding to the different symbols) using
  several proposed EoSs (corresponding to the different clusters of
  radii). The mass of the secondary NS is computed assuming the chirp
  mass. The horizontal dotted line indicates the observed $90\%$
  confidence upper limit on the effective tidal deformability (from the
  original LIGO-Virgo analysis \cite{Abbott2017}). The narrow solid band
  (which is indistinguishable from a single curve) is some
  “quasi-Newtonian” expression for the binary tidal deformability
  \cite{Raithel2018} for $0. 7 < q <1.0$. (From
  Ref. \cite{Raithel2018})\label{fig:Raithel2018}}
\end{minipage}
\end{center}
\end{figure}

If, on one hand, the mass-weighted tidal deformability depends strongly on
stellar radius, on the other hand it was recently found that, if the
chirp mass\footnote{As mentioned earlier, the chirp mass can be extracted
  very accurately.} is specified, the mass-weighted tidal deformability
is approximately independent of the component masses for a BNS merger
\cite{Raithel2018}. This empirical result was obtained by calculating the
tidal deformabilities for several EoSs after choosing various values for
one of the component masses in such a way that they lie within the mass
range inferred for GW170817. The corresponding values for the mass of the
other component star were calculated from the chirp mass measured for
GW170817.
The results of Ref. \cite{Raithel2018} are summarized in
Fig.~\ref{fig:Raithel2018}, where
one sees, for example, that the upper limit\footnote{Note that the value
  $\tilde\Lambda < 800$ was incorrectly reported in the detection article
  of GW170817 \cite{Abbott2017}: The corrected value in the case of the
  low-spin prior was $\tilde{\Lambda}\leq 900$, but the analysis of
  Ref. \cite{Raithel2018} and others made before the publication of the
  correction \cite{Abbott2018a} could not but use the mistaken value. See
  also Sect. \ref{nucleonic}. Ref. \cite{Raithel2018} and others also used
  the original value given for the chirp mass $M_{\rm chirp}
  =1.188^{+0.004}_{-0.002} \Msun$ in Ref. \cite{Abbott2017},
  which was later revised in Ref. \cite{Abbott2018a} to $M_{\rm chirp}
  =1.186\pm 0.001 \Msun$.} of $\tilde\Lambda < 800$ \cite{Abbott2017}
immediately excludes radii above $\approx 13 \, \km$ at the $90\%$
confidence level, without requiring detailed knowledge of the component
masses ($m_1$ in Fig.~\ref{fig:Raithel2018}).

This idea of Ref. \cite{Raithel2018} came from using the I-Love-Q
relations between stellar compactness and tidal deformability
\cite{Yagi2013a, Yagi2013b, Yagi2017} (see Sect. \ref{univ-rel})
to express the mass-weighted tidal deformability of the binary as a
function of component masses and stellar radii. 

Let me conclude this introductory Subsection by mentioning that, in
addition to measurements based on the tidal deformability, other ways to
gain information on the interiors of NSs from GW observations have been
proposed, even if they require higher sensitivities and thus may be
applicable only when third-generation detectors \cite{Punturo2010b,
  Sathyaprakash:2009xs, McClelland2015}) become operational. Some of
these studies involve tidal excitations of resonant modes
\cite{Shibata1994, Reisenegger1994, Lai1994, Kokkotas1995, Ho1999,
  Hinderer2016, Steinhoff2016, Yu2017a, Yu2017b, Xu2017, Schmidt2019},
gravitomagnetic excitations of resonant modes \cite{Flanagan2007},
resonant shattering of the NS crust by tides \cite{Tsang2012, Tsang2013}
and non-linear tidal effects \cite{Essick2016}.

\subsection{Applications to data analysis}

One needs to treat carefully many aspects of the basic idea delineated in
the previous Subsection when applying it concretely to GW
data. Estimating the parameters of BNS systems during the inspiral phase
is based on matched filtering: the GW data stream is cross-correlated
with theoretically predicted template waveforms (approximants) for
different possible physical parameters. These trial waveforms need to be
accurate to allow for correct estimates of the stellar masses and spins,
and of the internal structure of the stars. This is especially true for
the very last orbits before the merger, where instead approximants become
increasingly inaccurate (see, \eg, Refs. \cite{Bernuzzi2012, Favata2014,
  Wade2014, Hotokezaka2016, Dudi2018, Samajdar2018, Nagar2018,
  Abdelsalhin2018a, JimenezForteza2018}). As it can be easily imagined,
approximants that do not consider tidal effects are not sufficient,
especially for spinning BNS systems or stiff EoSs. For example, it was
estimated that for searches in Advanced LIGO at design sensitivity
neglecting tidal effects would cause roughly a $5\%$ additional loss of
signals \cite{Cullen2017}. Furthermore, for spinning systems (spin
parameter $\chi\equiv J/M^2 \gtrsim 0.1$, where $J$ is the NS angular
momentum and $M$ its mass), in order to reduce mismatches to an
acceptable level, it is crucial to include spin-induced and EoS-dependent
higher order terms \cite{Poisson:1997ha, Bohe2015} in the waveform
approximants \cite{Harry2018, Dietrich2019, Nagar2018, Abdelsalhin2018a,
  JimenezForteza2018, Samajdar2019, Tsokaros2019}. Approximants including
nonprecessing (namely aligned to the orbital angular momentum) stellar
spins and self-spin effects have become available recently
\cite{Nagar2018, Abdelsalhin2018a, JimenezForteza2018}.

\begin{figure}[htb]
\begin{center}
\begin{minipage}[tb]{12 cm}
\includegraphics[width=1.0\textwidth]{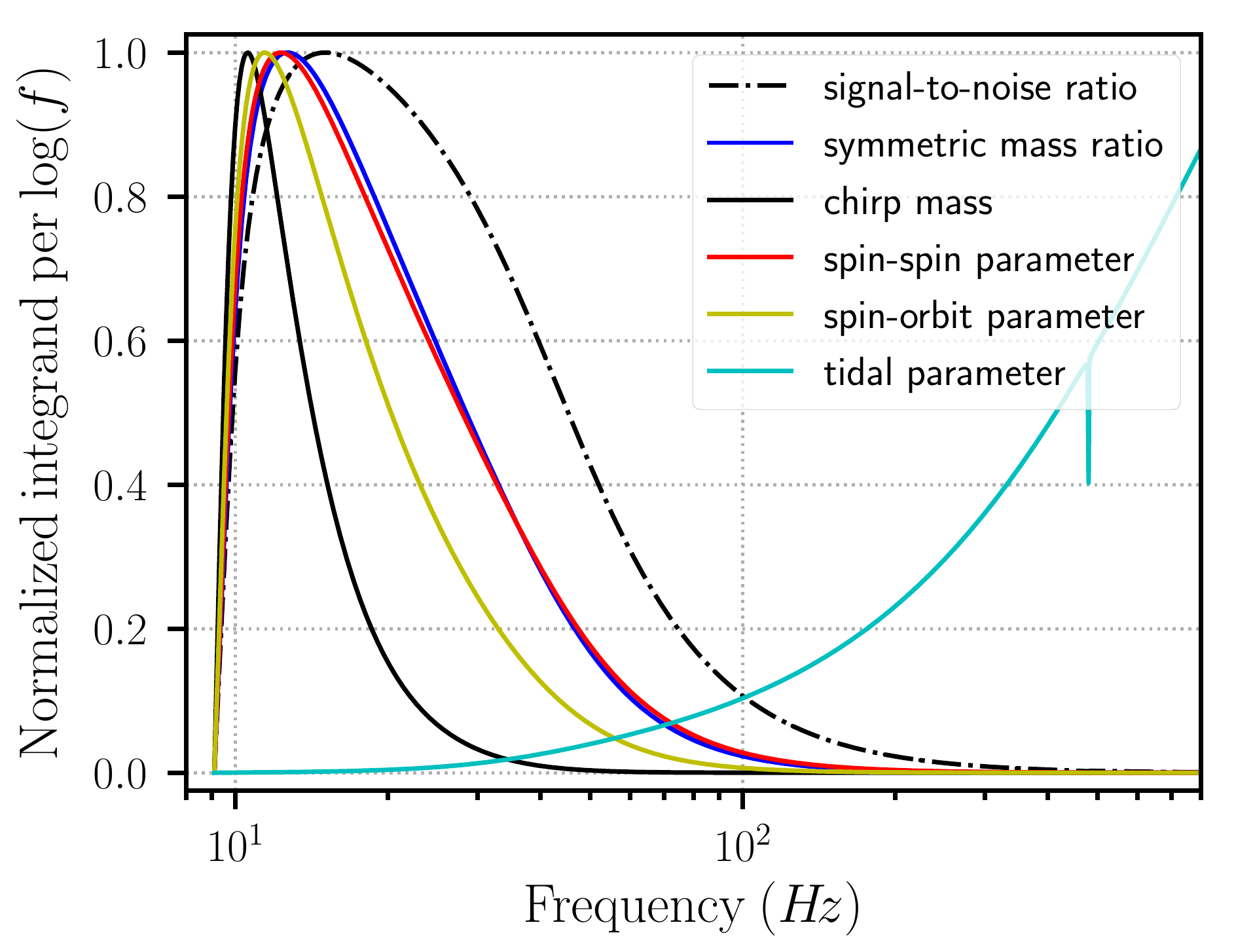}
\end{minipage}
\begin{minipage}[t]{16.5 cm}
  \caption{Illustration of where in frequency the information about
    intrinsic binary parameters predominantly comes from. The quantity
    shown on the y-axis is a normalized quantity characterizing the
    accumulation of information about the binary parameters (see
    Ref. \cite{Harry2018} for details) per logarithmic frequency
    interval. The zero-detuned high-power configuration of Advanced LIGO
    is used and each curve is normalized to its maximum value. (From Ref.
    \cite{Harry2018}) \label{fig:parameter-info}}
\end{minipage}
\end{center}
\end{figure}

Figure \ref{fig:parameter-info} offers an illustration, based on
computations, of which frequency ranges are the most important for
extracting information about intrinsic binary parameters. Considering
that the maximum sensitivities of interferometric detectors are for
frequencies $\lesssim 1\, \kHz$, one immediately sees that information
about tidal parameters accumulates where the detectors are not at
their best sensitivity.

Another issue is the fact that the computation of trial waveforms needs
to be efficient and fast, because source properties are generally
inferred via a coherent Bayesian analysis that involves repeated
cross-correlation of the measured GW strain with predicted waveforms
\cite{Veitch2015}. Computational efficiency is crucial also because BNS
systems are visible by GW detectors for a long time, several seconds or
even minutes before the merger. Another development that would help data
analysis would be producing approximants that include also the
post-merger regime consistently, but no such model exists yet (see the
end of Sect. \ref{post-merger basics application} and Sect. \ref{pre+post
  merger} for related work).

A few types of waveform models have become the standard in GW analysis:
post-Newtonian, effective-one-body (EOB) \cite{Buonanno:1998gg,
  Buonanno00a} and the {\it Phenom} models \cite{Ajith:2007qp:longal,
  Ajith:2007kx:longal}. Post-Newtonian expansions are approximate
solutions, valid for weaker fields, of the Einstein field equations. The
expansion is made in parameters that are small when the approximation is
valid, like the velocity $v$ of the objects (for the binary systems
considered in this review it would be the relative velocity of binary
constituents) with respect to the speed of light or deviations from a
background metric. A post-Newtonian term of order $n$ is proportional to
$v^{2n}$ relative to the leading-order term in the expression (see
Ref. \cite{Blanchet2014} for an introduction). \label{note:postNewtonian}
In post-Newtonian expansions, the GW signal can be approximated by
imposing that the power radiated by a binary system in GWs is equal to
the change in the energy of the binary. Even if physically tidal effects
are present even in a Newtonian description, in the post-Newtonian
framework, the lowest-order tidal effects appear as terms of the fifth
post-Newtonian order. However, point-mass terms for binaries are
currently only known to fourth order in the dynamics \cite{Damour2016a,
  Marchand2018} and only to order 3.5 in the GW phasing
\cite{Blanchet2014}, and this lack of complete information has raised
concerns about systematic errors in GW measurements of tidal effects
\cite{Wade2014, Favata2014, Yagi2014}.

In order to solve this problem, two classes of effective or
phenomenological models for point-mass binaries that to some extent
include all post-Newtonian orders in an approximate way have been
developed: the EOB model \cite{Buonanno:1998gg, Buonanno00a} and the
so-called {\it Phenom} models \cite{Ajith:2007qp:longal,
  Ajith:2007kx:longal, Schmidt2015, Khan2016}. Both approaches combine
analytical results with information from numerical relativity simulations
and, while addressing the problems of the systematic errors of the
post-Newtonian expansion, they may introduce other, smaller, systematic
errors (which can be assessed by comparison with numerical
relativity simulations).

The Phenom models are phenomenological waveforms that approximate a set
of hybrid waveforms constructed by matching numerical-relativity
waveforms with analytical post-Newtonian waveforms\footnote{In general,
  hybrid waveforms are constructed by matching numerical waveforms with some
approximant of the general-relativistic equations.
}
\cite{Ajith:2007qp:longal, Ajith:2007kx:longal, Schmidt2015, Khan2016,
  Dietrich2017b}.

The fundamental idea of the effective-one-body method consists in
representing the two-body dynamics by those of a single effective
particle in an effective potential. In practice, the dynamics and GWs
from a binary are computed by solving the coupled system of ordinary
differential equations for the orbital motion, GW generation, and
radiation backreaction in the time-domain. Tidal effects have been fully
incorporated in the effective-one-body model \cite{Damour:2009wj,
  Vines:2010ca, Damour:2012, Bini2012, Bini2014a, Bernuzzi2015,
  Steinhoff2016, Hinderer2016}, including for (nonprecessing) spinning
binaries \cite{Nagar2018, Abdelsalhin2018a,
  JimenezForteza2018}. Refinements and calibrations of the models are
performed by comparison with numerical-relativity simulations
\cite{Bohe2017, Babak2017, Nagar2017, Cotesta2018}, which suggest that
further improvements of the tidal effective-one-body models are still
necessary for a satisfactory description of the signal
\cite{Hotokezaka2015, Dietrich2017, Samajdar2018, Dietrich2019}, at least
in some regions of the parameter space of BNS systems. For example, a
full Bayesian analysis determined that different tidal and
point-particle/binary--black-hole descriptions for the waveform
approximant yield estimated tidal parameters that can differ by more than
a factor of two \cite{Samajdar2018} (see also
Sect. \ref{sec:inspiral}). Furthermore, the effective-one-body model
still has a rather high computational cost per waveform \cite{Dietrich2019}.

Solutions to the latter problem may come from reduced-order-modelling
techniques \cite{Lackey2017}, which, however, also add further 
complexity, from additional inclusion of numerical-relativity results, or
from other modelling techniques complementary to the effective-one-body model, like those
in Refs. \cite{Pannarale2012, Lackey2012, Lackey2013, Barkett2016}. In
the future numerical-relativity--based waveform models will likely be the ones that allow
more precise and stringent extraction of the source properties
\cite{Dietrich2019}.

\subsection{Estimates about inspiral waveforms}
\label{sec:inspiral}

Starting from the work\footnote{Previous works \cite{Flanagan08,
    Hinderer09, Damour:2012} had already suggested that information on
  EoSs from the inspiral waveforms could be obtained with future
  detectors, but these were based solely on theoretical considerations
  and did not employ numerical-relativity simulations and a detailed
  treatment of detector noise.} of Ref. \cite{Read2013}, which quantified
data-analysis estimates of the measurability of matter effects in
gravitational waveforms from NS binaries with different EoSs
(approximated as piecewise polytropes) by analysing numerical waveforms
produced with codes of different groups, with different numerical setups,
and combined with detector noise curves, it was made clear that it is
actually possible to measure the tidal deformability (and the radius) of
compact stars from BNS inspirals with current detectors.

By using hybrid waveforms constructed as a match between the numerical
waveforms at higher frequencies and some approximant waveform (see
previous Section) at lower frequency, Ref. \cite{Read2013} found that for
Advanced LIGO the radius of a NS can be estimated with an error of
$\delta R \simeq 0.5 \, \km \times (D_{\rm eff}/100\, \Mpc)$, or $\delta
R/R \simeq 5\% \times (D_{\rm eff}/100\, \Mpc)$, where $D_{\rm eff}$ is
the effective distance to the source. Hybrid waveforms have been later
noticeably improved by using better numerical-relativity simulations
(higher resolutions, smaller initial orbital eccentricity\footnote{In BNS
  systems born bound, GW emission reduces the eccentricity
  \cite{Peters:1963ux, Peters:1964, Kowalska2011} to smaller values, but
  compact binaries formed through dynamical scattering and dynamical
  capture in dense stellar environments could have non-negligible
  eccentricity \cite{Fabian1975, Pooley2003, Oleary2009, Lee2010,
    Kocsis2012, Samsing2014, Fragione2019} in the LIGO band (see, \eg,
  Ref. \cite{Kowalska2011, Seto2013}).}), better approximants (resummed post-Newtonian
expressions \cite{Dietrich2017b}, tidal effective one body
\cite{Kawaguchi2018}), either in the time domain \cite{Dietrich2017b} or
directly in the frequency domain \cite{Kawaguchi2018}. Through comparison
with numerical-relativity simulations, these works obtained tidal
corrections to the GW phase and amplitude that can be efficiently used in
data analysis.

The studies of Ref. \cite{Read2013} relied on the Fisher matrix
approximation, which holds for loud signals. However, making strong
statements about estimating source parameters requires a full Bayesian
analysis. With more sophisticated Bayesian statistical analyses and/or
increasing amounts of (physical and detector) details taken into account,
other works later confirmed that for NS binaries with
individual masses around $1.4 \Msun$, the dimensionless tidal deformability
$\Lambda$ could be realistically determined with about $10\%$ accuracy by
combining information from about $20-100$ sources, depending on
assumptions about the BNS population parameters (for example, if one
considers also nonzero spins for the initial NSs the necessary
number of sources is higher) \cite{DelPozzo2013, Bernuzzi2013, Wade2014,
  Favata2014, Lackey2015, Agathos2015, Hotokezaka2016}.

In particular, the first Bayesian study for the estimation of tidal
deformability parameters was carried out in Ref. \cite{DelPozzo2013}.
Using post-Newtonian approximants that included tidal corrections up to
the sixth post-Newtonian order, they concluded that tens of detections
are required to constrain the tidal deformability parameter to an
accuracy of $\approx 10\%$. If one expresses the tidal deformability as a
linear expansion in mass, this would allow to distinguish between
different types of EoSs at that accuracy level; however a caveat is
necessary here, since EoSs with phase transitions do not necessarily allow
for such a linear expansion (see Sect. \ref{hybrid stars} for further
discussion).

Following Ref. \cite{DelPozzo2013}, an analytical approach to study the
systematic and statistical uncertainties arising from neglecting physical
effects in the estimation of the Love number of NSs (and so of their
tidal deformability) was made by Refs. \cite{Favata2014, Wade2014}. It
was found that relevant estimation biases that exceed statistical ones
are introduced (i) if post-Newtonian terms of order 4 and higher are
neglected, (ii) if even relatively small spins ($\chi\geq 0.003$) are not
taken into account, and (iii) if eccentricities larger than about
$10^{-3}$ are neglected. Other biases may originate from inaccurate
numerical simulations if they are used to tune the approximants, even if
in the most advanced numerical-relativity codes these are relatively
under control \cite{Radice2013b, Kiuchi2017a, Kiuchi2019a}.

In further developments, Ref. \cite{Lackey2015}, provided a method to
estimate the EoS parameters for piecewise polytropes by stacking
tidal-deformability measurements from multiple detections and concluded,
in accordance with previous estimates, that a few bright sources would
allow to constrain the NS EoS. Ref. \cite{Chatziioannou2015}, however,
showed that stacking multiple detections with moderately low
signal-to-noise ratio should be carried out with caution as the procedure
may fail when the prior information dominates over new information from
the data.

Ref. \cite{Agathos2015} then found similar results after extending the
previous analysis \cite{Lackey2015}
by including a larger number of simulated BNS signals and taking into
account more physical ingredients, such as spins, the quadrupole-monopole
interaction (this did not affect parameter estimation for the considered
configurations), and tidal effects to the highest known order. In
addition, the tidal deformability parameter was expanded to include up to
a quadratic function of mass.

Using Bayesian model selection analysis, Ref. \cite{Chatziioannou2015}
found that detectors like Advanced LIGO and Virgo can heavily
constrain EoSs containing only quark matter (with a signal-to-noise ratio
of $\approx 20$),
but hybrid stars (see Section \ref{hybrid stars}) would be more
difficult to distinguish from hadronic stars, because they differ only at
higher densities, that contribute less to the tidal deformability. They
considered kaon, hyperon, and hybrid EoSs with exotic matter parameters
within the range allowed from experiments and theoretical
calculations. They concluded that the presence of kaon and hyperons
cannot be easily confirmed (but easily excluded).

Many of the works mentioned above (which date to before the detection
event GW170817) were carried out by using piecewise-polytropic
parameterizations of the EoS \cite{Read:2009a}. As already mentioned in
Sect. \ref{eos-parameterization}, however, it had been shown in
Ref. \cite{Lackey2015} that near the fixed joining densities of the
piecewise-polytropic representation systematic error arises that may
exceed statistical error. Refs. \cite{Abbott2018a, Carney2018}
extended the work of Ref. \cite{Lackey2015} by using also a spectral EoS
parameterization\footnote{However, Refs. \cite{Abbott2018a, Carney2018}
  have not adopted the latest developments in such a spectral
  representation, namely the self-imposition of the causality constraint
  \cite{Lindblom2018a, Lindblom2018}.} \cite{Lindblom2010} (see Section
\ref{eos-parameterization}), which does not have such problems, for
analysing the data of GW170817 (for more on estimates from GW170817 see
Sect. \ref{constraints}). In particular, in Ref. \cite{Carney2018} it
was found that both the piecewise-polytropic and spectral EoS
parameterizations allow to recover consistent tidal information from the
simulated signals, but, as expected, the spectral model allows for
smaller errors.

In Ref. \cite{Abbott2018a}, in addition to the adoption of the
spectral representation of the EoSs, by probing different mass-ratios for
non-spinning signals with a tidal effective-one-body--based model \cite{Hinderer2016}, it
was concluded in an exhaustive Bayesian analysis that for GW170817 the
systematic uncertainties due to the modelling of matter effects are
smaller than the statistical errors in the measurement.
Also through Bayesian inference, Ref. \cite{Dudi2018} investigated
numerical-relativity--based tidal waveform models and showed the
importance of the inclusion of tidal effects for the extraction of the NS
masses and spins from the GW signal for high signal-to-noise ratios. They
found that inaccurate modelling of tidal effects in the analysis of the
inspiral GW signal for stiff EoSs leads to a large bias in the measurement of
the masses and spins. They also studied whether omitting the post-merger
waveforms from the global analysis of the signal leads to
significant loss of information, or possibly to biases in the estimation
of the source properties, and concluded that, as expected but not
rigorously shown before, the post-merger part has no impact on
GW measurements with current detectors for the
signal-to-noise ratios considered.

Ref. \cite{Dudi2018} stated that systematic errors in the analysis are
under control, but Ref. \cite{Samajdar2018} pointed out that in some cases there
may be potential systematic biases during the extraction of parameters,
from non-spinning sources as well. By experimenting with many different
approximants, they found that for unequal-mass binaries\footnote{For
  equal-mass binaries sufficiently similar predictions with any modelling
  were found.}, while the mass and spin recovery shows almost no
systematic bias with respect to the chosen waveform model, the extracted
tidal effects can be significantly biased, up to a point where the
injected value is not contained within the $90\%$ credible interval. In
particular, they concluded that, generally, post-Newtonian approximants
predict NSs with larger deformability and radii than models tuned by
using numerical-relativity simulations. Furthermore, it was highlighted
that the use of higher post-Newtonian orders in the tidal phasing does
not lead to a monotonic change in the estimated properties
\cite{Samajdar2018}. Their remarkable note of caution states that for a
signal with strength similar to GW170817, but observed with design
sensitivity by the Advanced LIGO and Virgo detectors, different tidal
descriptions of the waveform approximant yield estimated tidal parameters
that can differ by more than a factor of two (see also
Ref. \cite{Harry2018}). This could lead to misclassification of the
observed system (NS binaries, black-holes binaries, or NS--black-hole
binaries). Thus, further improved waveform models with improved tidal
descriptions are imperative for characterizing unequal-mass mergers.

Other notes of caution have been rung, among others, by
Ref. \cite{Dai2018}, which found that the posterior distribution of the
tidal deformability is strongly affected by the choice of the maximum
frequency considered in the analysis, and by Ref. \cite{Greif2019}, which
found that the choice of parameterization of the EoS (see
Sect. \ref{eos-parameterization}) can have a significant effect on
posterior EoS constraints inferred from GW data. Also, the effect of
precession in the late inspiral has been found, through numerical
simulations, to be visible in the GW signal \cite{Dietrich2018a}, even if
it may be of secondary importance in practical cases, since it is
appreciable only for edge-on systems, which are harder to detect because
of the smaller observable GW strain for such inclinations.

More comprehensive Bayesian analyses have been performed that include not
only gravitational emission from GW170817, but also the related
electromagnetic emission \cite{Radice2017b, Bauswein2017b, Radice2018c,
  Coughlin2018, Coughlin2018a, Wang2018, Margalit2019}, but these are not
treated in detail in this review (see also Sect. \ref{constraints}).


In the previous paragraphs, I reported the main advances in estimations
from the inspiral waveforms, especially as far as statistical treatments
(Bayesian techniques) are concerned. In addition to these, the work of
Ref. \cite{Read2013}, described at the beginning of this Subsection, has
been improved in other ways as well. Ref. \cite{Hotokezaka2016} has
quantitatively improved its computations and estimations in two principal
directions. First, they employed new numerical-relativity simulations of
irrotational binaries with longer inspirals (\ie~$14-16$ orbits) and
higher accuracy both in the initial-data setup (\ie~residual eccentricity
of the order of $10^{-3}$). Second, they included in the analysis lower
frequencies, down to $30\, \Hz$ , to which ground-based detectors like
Advanced LIGO and Virgo are reasonably sensitive. They also adopted
additional EoSs. Results were very similar to those of
Ref. \cite{Read2013}, namely that deformability $\Lambda$ and radius can
be determined to about $10\%$ accuracy for sources at $200\, \Mpc$.

More recent works \cite{Dietrich2017b, Kawaguchi2018} noticeably improved
procedures for efficient data analysis of the pre-merger signal in
relation to tidal deformations. In particular, Ref. \cite{Kawaguchi2018}
found that the statistical error for the measurement of the mass-weighted
tidal deformability is more than six times larger than the systematic
error for a signal-to-noise ratio of $50$. They also showed that the
statistical error for the measurement of the mass-weighted tidal
deformability is larger than the variation of the mass-weighted tidal
deformability with respect to the mass ratio even for signal-to-noise
ratio $100$ (see also the last paragraphs of Sect. \ref{premerger-basic}
and Fig. \ref{fig:Raithel2018}). This suggests that even for events with
signal-to-noise ratio $100$, the systematic error in current waveform
models is unlikely to cause serious problems in the parameter estimation
\cite{Kawaguchi2018}.

\subsection{Eccentric mergers}
\label{eccentric}

As briefly mentioned in Sect. \ref{sec:inspiral}, BNS systems that have
evolved without close interaction with other stars are thought to have
lost, by the time their GW signals enter the detectability range, any
significant initial eccentricity they may have had through the emission
of gravitational radiation \cite{Peters:1963ux, Peters:1964} and
therefore to have very small orbital eccentricities, \ie~$\lesssim
10^{-3}$ \cite{Kowalska2011}. Instead, compact binaries formed in dense
stellar environments (globular clusters and galactic nuclei) through
dynamical scattering, dynamical capture and, in general, multibody
interactions could still have non-negligible eccentricity by the time
they merge \cite{Fabian1975, Pooley2003, Oleary2009, Lee2010, Kocsis2012,
  Samsing2014, Fragione2019}. These events are probably rarer, even
though the estimates of their event rates are very uncertain (see, \eg,
Refs. \cite{Lee2010, Tsang2013, Fragione2019}), and more difficult to
detect with detectors like Advanced LIGO and Virgo, mostly because the
signal power at frequencies around $\sim 100 \Hz$, where the detectors
perform best, is smaller (see, \eg, Refs. \cite{Seto2013, Gondan2017,
  Gondan2019}).

However, if detected in GWs, they would provide, possibly more easily,
measurements of several stellar physical quantities, from which then
information on the EoS may be gained. This comes from the fact that tidal
perturbations during pericentre passage can excite stellar fundamental
modes (the previously mentioned f-modes) of oscillation that have a
time-varying quadrupole moment and that therefore act as sources of
gravitational radiation, on top of the inspiral waveform\footnote{The
  excitation of f-modes in low-eccentricity inspirals can also be
  measured, but only with a network of third-generation detectors
  \cite{Pratten2019}.}. This f-mode signal depends on the EoS (in
general, stiff EoSs store more energy in the oscillations compared to
soft EoSs \cite{Chaurasia2018}) and can significantly affect the phase of
the gravitational radiation by enhancing the loss of orbital energy by up
to tens of percent over that radiated away by GWs during an orbit. Part
of the orbital angular momentum may also be transferred to the
stars. Measurements of the frequency, damping time, and amplitude of the
tidally excited f-modes could yield simultaneous measurements of their
masses, moments of inertia, and tidal Love numbers and thus present a
prime opportunity to test the I-Love-Q relations (see
Sect. \ref{univ-rel}) observationally. Mergers of eccentric BNS systems
may also produce brighter electromagnetic emission than quasicircular
mergers (see, \eg, Refs. \cite{East2012c, Gold2012, Rosswog2013,
  Dietrich:2015b, Radice2016, Papenfort2018}) and contribute to the
overall $r$-process \cite{Lattimer77, Eichler89, Li1998,
  Kulkarni2005_macronova-term, Arnould2007, Metzger:2010} element
abundances \cite{East2012c, East2016, Radice2016, Papenfort2018}.

Actually, if accurate enough templates for eccentric BNS systems are
used, these binaries can be detected from farther away, and parameters
such as the chirp mass and sky localization can be estimated more
accurately \cite{Gondan2017, Gondan2019}, mostly because of the their
richer structure that breaks parameter degeneracies. Detection of a few
highly eccentric BNS mergers per year might be possible with
third-generation detectors \cite{Chaurasia2018, Papenfort2018} or even
with the LIGO-Virgo-KAGRA detector network at design sensitivity
\cite{Gondan2019}.

Several numerical studies have been performed on eccentric binaries
\cite{Turner1977a, Turner1977b, East2012c, Gold2012, Rosswog2013,
  Dietrich:2015b, Radice2016, East2016a, Chirenti:2016b,
  Chaurasia2018, Yang2018, Papenfort2018}. Ref. \cite{Yang2018} proposed
a basic model that, using a post-Newtonian description, can predict the
timing of different pericentre passages within a factor of two with
respect to results of numerical simulations. The error is probably
dominated by systematic effects in measuring orbital properties from the
numerical-relativity results \cite{Yang2018}. A refined version of this
model (that takes into account post-Newtonian corrections to the tidal
coupling and the oscillations of the stars) may serve as a template for
detection and analysis of gravitational radiation from eccentric systems
\cite{Yang2018}.

Extensive studies on the subject have found that, on one side, the same
qualitative relation between the merger frequency and the stiffness of
the EoS that is known for quasicircular binaries is valid for eccentric
binaries \cite{East2016a, Chaurasia2018}, while, on the other side, there
is no direct correlation between initial eccentricity and post-merger
frequencies \cite{Chaurasia2018} (see Sect. \ref{postmerger}), and
post-merger peak frequencies do not follow the approximate universal
relations with stellar properties like compactness
\cite{Papenfort2018}. Actually for eccentric orbits the position of the
main peak in the power spectrum has been found to vary with the
eccentricity of the orbit, as a result of the different angular momentum
of the merger remnant \cite{Papenfort2018}.

\section{Extracting information on the equation of state from gravitational waves emitted after
the merger}
\label{postmerger}

Also GW signals from the merger and post-merger phases of GW170817 have
of course been searched in the data from Advanced LIGO and Advanced
Virgo, but no signal was found \cite{Abbott2018a, Abbott2018c}. In fact,
the strain upper limits set by the detector were found to be about one
order of magnitude above the numerical-relativity expectations for
post-merger emission from a hypermassive NS at the distance of GW170817.

The first detection of a BNS post-merger signal is still to come, but
simulations and investigation on their possible connections with
observations of GWs from the post-merger phase of BNS mergers has been
very active. The strong interest springs also from the fact that such
observations would probe densities higher (several times the nuclear
density) than typical densities in inspiralling stars and would also
probe effects of temperature, which becomes much higher (up to $\sim 50
\MeV$) after the merger. However, even though the energy emitted in GWs
after the merger may be higher than that emitted during the inspiral (if
there is not a prompt collapse), since post-merger GW frequencies are
higher (from $1$ to several $\kHz$ \cite{Bauswein2011, Takami:2014,
  Maione2017, Paschalidis2017b}), their signal-to-noise ratio in current
and projected detectors is smaller than that of the inspiral and they are
probably only marginally measurable by detectors like Advanced LIGO or
Advanced Virgo. Third-generation detectors, such as Einstein Telescope
\cite{Punturo2010b, Sathyaprakash:2009xs} and Cosmic Explorer
\cite{McClelland2015}, are needed.

Also on the theoretical side, the post-merger phase presents more
difficulties. Numerical simulations of the merger and post-merger
dynamics are more difficult than for the inspiral part, because of
strong shocks, turbulence, large magnetic fields, various physical
instabilities, neutrino cooling, viscosity and other microphysical
effects\footnote{It has been hinted at that viscosity and neutrino
  cooling are probably subdominant with respect to the emission of
  gravitational radiation in the dynamics of the post-merger phase, at
  least for the first $20\, \ms$ after merger \cite{Kiuchi2017,
    Alford2018, Zappa2018, Radice2018, Alford2019}, but this is not true in all cases. In
  particular, it was found that thermal conduction and shear viscosity
  are not subdominant if both neutrino trapping and short-distance
  gradients (for example due to turbulence on a scale of $\sim 10$ m) are
  present \cite{Alford2018}. Moreover, a more recent analysis, assuming
  that nuclear matter remains neutrino transparent up to temperatures of
  a few MeV, found that bulk viscosity can damp oscillations on
  timescales of $\sim 10\, \ms$, comparable to those of a BNS merger, for
  nuclear matter at temperatures of $2-4 \MeV$ and for densities between
  $0.5 - 2$ times the saturation density \cite{Alford2019a}. This means
  that bulk viscous damping should probably be considered seriously for
  inclusion in future simulations.}. Therefore, their accuracy is not as
good as for the inspiral. For example, currently there exist no reliable
determinations of the phase of post-merger gravitational radiation, but
only of its spectrum. Furthermore, the complicated morphology of
post-merger signals makes constructing accurate templates challenging.

The basic idea for obtaining information on the supranuclear EoS from GWs
emitted after the merger is that the main peak frequencies of the
post-merger power spectrum (see Fig. \ref{fig:Clark2016-Fig1}) strongly
correlate with properties (radius at a fiducial mass, compactness, etc.)
of a zero-temperature spherical equilibrium star in a rather
EoS-insensitive way. These are also called {\it universal relations}, as
illustrated in Sect \ref{univ-rel}, even though they depend on the spin
of the stars in the inspiral and hold only approximately even for
irrotational binaries. These relations have been employed to estimate
constraints on the NS radius from the post-merger signal of future
observations \cite{Bose2017, Chatziioannou2017, Torres-Rivas2019}.

Some details of the post-merger spectrum are still debated, but there is
widespread consensus that (i) the post-merger GW signal possesses
spectral features that are robust (present in all simulations,
irrespective of the numerical methods, codes used and numerical settings)
and that emerge irrespective of the EoS or the mass ratio of the original
binary; (ii) for the frequencies of the most salient peaks, analytic
fitting functions can be obtained in terms of the stellar tidal
deformability or compactness.

\subsection{Basic ideas}
\label{postmerger basic ideas}

A first attempt to study systematically the post-merger waveforms was put
forward by Ref. \cite{Hotokezaka2013c}, that tried to codify the whole
spectrum, instead of singling out individual peaks. They decomposed the
merger and post-merger GW amplitude into three parts: (i)
a peak in frequency and amplitude that appears soon after the merger starts; (ii) a
decrease in amplitude during the merger and a new increase when the
compact star forms; (iii) a final decrease in the amplitude during the
hypermassive NS phase, which is either monotonical or with
modulations. They also identified a damping of the oscillations of the
frequency during the phase (spanning several oscillation periods) in
which the merged object is a compact star; the frequency eventually
settles to an approximately constant value (although a long-term secular
change associated with the change of the state of the hypermassive NS is
always present). Based on this, they found an optimal 13-parameter
fitting function, using which it may be possible to constrain the
NS radius with errors of about 1 km.

The first to propose relations between single peak frequencies and
stellar properties (mass, radius, compactness) were
Refs. \cite{Bauswein2011, Bauswein2012, Bauswein2012a, Bauswein2014},
using simulations with approximate treatment of general
relativity. Subsequent analyses were performed by a few groups with
general-relativistic codes \cite{Takami:2014, Takami2015, Bernuzzi2015a,
  Dietrich2015, Foucart2015, DePietri2016, Rezzolla2016, Maione2016,
  Dietrich2017, Feo2017, Maione2017, Kiuchi2019a},
which confirmed that the conformally flat approximation employed by other
authors provided a rather accurate estimate of the frequencies of the
largest peak in the power spectral densities.

In more detail, Refs. \cite{Bauswein2011, Bauswein2012}
pointed out that the largest peak in the power spectral
density (whose frequency is dubbed there as $f_{\rm peak}$ and as $f_2$
in other works, while in this review I will call it $f_{\rm main\;
  peak}$ for the sake of the reader) correlates with the radius $R_{\rm
  max}$ of the maximum-mass nonrotating star for a given EoS and with
other similar quantities, like the radius of a $1.6 \Msun$ NS
with a given EoS.
It was found later that the scatter in the relations of $f_{\rm main\;
  peak}$ with these stellar properties was such that it does not allow to
determine the radius with an accuracy below $1 \km$, even if $f_{\rm
  main\; peak}$ were measured exactly \cite{Hotokezaka2013c,
  Kiuchi2019a}.
It was also found later that the relation between $f_{\rm main\; peak}$
and the dimensionless tidal deformability is more universal and holds at
the $\approx 10\%$ level, but only for equal-mass or very nearly
equal-mass binaries \cite{Rezzolla2016, Kiuchi2019a}.

\begin{figure}[tb]
\begin{center}
\begin{minipage}[t]{15 cm}
\epsfig{file=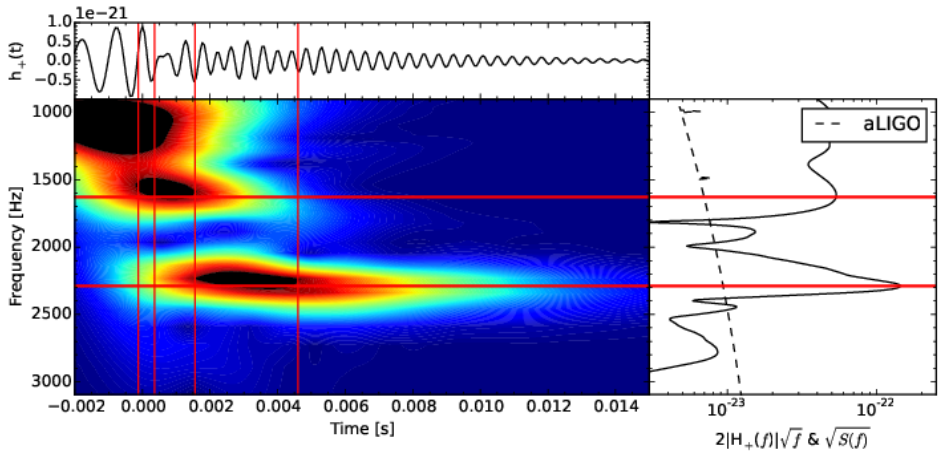,scale=0.5}
\end{minipage}
\begin{minipage}[t]{16.5 cm}
  \caption{Time-frequency analysis for the waveform of an equal-mass BNS
    merger for an optimally-oriented source at $50 \Mpc$. The top and
    right panels show the time-domain waveform-component $h_+$ and its
    Fourier magnitude spectrum, respectively.
    The time-frequency map is constructed from the magnitudes of the
    coefficients of a continuous wavelet transform (see
    Ref. \cite{Clark2016} for details). Horizontal straight lines
    emphasize the locations of the main peak frequency $f_{\rm main\;
      peak}$ and a secondary peak which the authors of
    Ref. \cite{Clark2016} call $f_{\mathrm{spiral}}$ (a peak with similar
    frequency and amplitude has been called $f_1$ in other works; see
    main text for discussion). The vertical lines have no meaning out of
    the context of the original article. (From
    Ref. \cite{Clark2016}) \label{fig:Clark2016-Fig1} }
\end{minipage}
\end{center}
\end{figure}

Refs. \cite{Takami:2014, Takami2015} then presented another method based
on results form a large number of numerical-relativity simulations of
equal- and unequal-mass irrotational binaries with different EoSs and
different masses. They identified two distinct and robust main spectral
features in the post-merger power spectral density: the largest peak
$f_{\rm main\; peak}$ mentioned above and a smaller peak at lower
frequencies (see Fig. \ref{fig:Clark2016-Fig1}), dubbed $f_1$.

Other works then pointed out that what was called $f_1$ in
Refs. \cite{Takami:2014, Takami2015} may be composed of two peaks, either
or both present depending on the total mass and EoS \cite{Bauswein2015,
  Bauswein2015b, Rezzolla2016, Maione2017, Bauswein2019b, Kiuchi2019a}. One of these
subdominant lower-frequency peaks is thought to originate from the
quasi-linear interaction between the dominant quadrupolar oscillation and
the quasi-radial mode of the remnant \cite{Stergioulas2011b} (even if the
quasi-radial mode by itself may not be visible in the power
spectrum). Its frequency, named $f_{2-0}$, is equal to the difference of
the frequencies of these two modes. The $f_{2-0}$ feature is particularly
pronounced for relatively high total masses of the binary system and soft
EoSs, while it is often not visible for less compact stars in the binary,
either low-mass stars or stars described with a stiff EoS
\cite{Bauswein2015, Maione2017, Bauswein2019b, Kiuchi2019a}. The other peak, called
$f_{\rm spiral}$, is thought to be produced by the orbital motion of {\it
  antipodal bulges} that form during the merger and persist for a few
milliseconds \cite{Bauswein2015} (see Fig. \ref{fig:Clark2016-Fig1}).

Different theoretical toy models were proposed to explain the origin of
the $f_1$ or $f_{\rm spiral}$ peak \cite{Takami2015, Bauswein2015}, but
one thing on which everyone agrees is that it originates (or they
originate) only from the dynamics immediately following the merger
(within $3$ or $4\, \ms$ after the merger). Also, in practical uses of the
relations involving $f_1$ or $f_{\rm spiral}$, the different origin of these
subdominant features or whether the lower frequency post-merger peak has
composite substructure have not been taken into
account. Ref. \cite{Torres-Rivas2019}, for example, defines what they
call $f_{\rm sub}$ as the frequency of the second highest peak with a
frequency at least $400\, \Hz$ below the main peak.

Coming back to the findings of Refs. \cite{Takami:2014, Takami2015}, it
was shown that, at least for the sample of EoSs tested, a single function
relates the $f_1$ frequency to the average compactness $\bar{M}/\bar{R}$
(where $\bar{M} \equiv (M_A+M_B)/2$ and $\bar{R} \equiv (R_A+R_B)/2$,
where $M_{A,B}$ and $R_{A,B}$ are the masses and the radii of the
nonrotating stars associated to those of the binary). Knowing the masses
from measurements of the inspiral waveform and the $f_1$ frequency from
measurements of the post-merger waveform would then allow to compute the
radius of the NSs. Also, since the $f_1$ or $f_{\rm spiral}$ peak is
produced soon after the merger, it should not be affected significantly
by magnetic fields and radiative effects, whose modifications emerge on
much larger timescales. However, a work (already mentioned in
Sect. \ref{univ-rel}) posted to arXiv.org just before this review was
accepted for publication pointed out that the existence and accurate
location of this peak depends significantly on the numerical resolution
of the simulations \cite{Kiuchi2019a}.

Also, it has to be noted that, especially in the first milliseconds after
the merger, when also the subdominant modes are active and when the merger
remnant is rapidly evolving towards a more stable configuration,
post-merger frequencies evolve in time (see, \eg, spectrograms in
Refs. \cite{Rezzolla2016, Dietrich2017, Maione2017} and
Fig. \ref{fig:Clark2016-Fig1}), albeit only slightly. In particular, the
frequency of the main peak increases up to the formation of the black
hole, while its amplitude decreases, as physically expected from the
increase of rotational velocity and compactness of the merged object as
it loses angular momentum \cite{Dietrich2017}. Hence, the spectral
properties of the GW signal can only be asserted reliably when the
signal-to-noise ratio is sufficiently strong so that even these changes
in time can be measured in the evolution of the power spectral
densities. In light of these considerations, as mentioned earlier, the
prospects for high-frequency searches for the post-merger signal are
limited to rare nearby events.

An interesting extension of the works described above has been done in
Ref. \cite{Bernuzzi2015a}.
They found that the coupling constant $\kappa_2^T$ [defined in
  Eq. \eqref{eq:kappa 2 T}] efficiently parametrizes the late-inspiral of
tidally interacting binaries and observed that it can also be used to
determine the main features of the post-merger GW spectrum, instead of
the tidal deformability parameter $\Lambda$. The relation $f_{\rm main\;
  peak}(\kappa_2^T)$ depends very weakly on the total mass of the binary,
mass-ratio, and EoS. However, there is dependence on stellar spins. The
proposed physical explanation is that at fixed separation, the tidal
interaction is more attractive for larger values of
$\kappa_2^T$. Binaries with larger $\kappa_2^T$ merge at lower
frequencies (larger separations). As a consequence, the remnants of
binaries with larger $\kappa_2^T$ are less bound and have larger
angular-momentum support at formation. The $f_{\rm main\; peak}$
frequency seems mainly determined by these initial conditions, other
physical effects having negligible influence on the value of the
frequency.

Dependence of the main peak frequency $f_{\rm main\; peak}$
itself on the initial state of rotation, especially for very rapidly
rotating NSs, has been pointed out in several works \cite{Bernuzzi2013,
  Bauswein2015b, Bernuzzi2015a, Dietrich2017, Dietrich2018a, East2019}. In particular, 
Ref. \cite{Dietrich2018a}
emphasized that their spectra are qualitatively similar, but
quantitatively different, from the non-spinning cases. More in detail,
the frequencies of all peaks are about $200\, \Hz$ higher and
the $f_1$ peak is significantly more pronounced for systems in which
the stellar spins are aligned with the orbital angular momentum and less
pronounced for anti-aligned system. This means that actually the
quasi-universal relations found for irrotational binaries
\cite{Bauswein2011, Bauswein2012, Bauswein2012a, Bauswein2014,
  Takami:2014, Takami2015, Bernuzzi2015a} might contain systematic biases
for spinning systems. Furthermore, Ref. \cite{East2019} found that the
secondary peaks are less significant in mergers of NSs with higher spin
magnitude.

Additional checks on the validity of the relations discussed above and
estimations of their error were done independently through simulations
with codes different from the ones used in the first place to obtain the
fitting parameters in those relations \cite{Maione2017, Kiuchi2019a}.
Furthermore, it was found in the same works that the post-merger
frequencies of unequal-mass binaries differ at most of $\approx 10\%$
from those of equal-mass binaries, as long as the differences in the
masses of the stars in the binary are within $10\%$ \cite{Lehner2016,
  Rezzolla2016, Dietrich2017, Maione2017, Kiuchi2019a}. For lower mass
ratios the spectra become more complicated: for fixed EoS and total mass,
spectra for smaller mass ratios show less power in the main peak and more
peaks at frequencies below it.

Incidentally, there is actually a third, higher-frequency peak, named
$f_3$, often identifiable in computed spectra. It may be explained,
together with $f_{2-0}$ and perhaps $f_1$ (as mentioned above), as the
result of the combinations of the quasiradial oscillation mode and the
fundamental mode $f_{\rm main\; peak}$ \cite{Stergioulas2011b}. In fact,
in most cases, the lower- and higher-frequency subdominant peaks are
almost equidistant from the dominant one and the frequency predicted for
the mode combination is a very good approximation for these subdominant
peaks. As said above, however, this is not observed in binaries composed
of less compact stars \cite{Bauswein2015, Maione2017, Bauswein2019b}. Also, the
dependency of $f_1$ and $f_3$ on the binary mass ratio seems to be
similar (and much stronger than that of the $f_{\rm main\; peak}$ mode):
The more unequal a system is, the lower both frequencies seem to be. In
any case, the $f_3$ peak is too weak and at too high frequencies to
foreseeably help in parameter estimation from GW observations.

Another peak in the spectrum may be visible in some cases because of an
$m=1$ (see note on page \pageref{stellar modes}) deformation of the merged
object \cite{Ou06, Corvino:2010, Anderson2008, Bernuzzi2013, Kastaun2014,
  Paschalidis2015, East2016, Dietrich:2015b, Radice2016a, Lehner2016a,
  East2016a, Paschalidis2017b}, especially in mergers of eccentric
binaries \cite{East2016a}. This deformation, due to the so-called
one-armed spiral instability, is found to be present generically in BNS
simulations and to carry information about the EoS\footnote{See
  Sect. \ref{exotic binaries} for a possible relation of the $m=1$ mode
  to dark matter \cite{Bezares2019}.}, but its peak in the spectrum,
located at about half the frequency of $f_{\rm main\; peak}$, has a much
smaller amplitude and the prospects of observations in GWs appear
unlikely in current detectors \cite{Radice2016a}. Third-generation
detectors may be able to target this signals.

Talking about additional peaks, it has been proposed that the presence of
dark-matter cores inside NSs may produce one or two supplementary peaks
in the post-merger GW spectrum of NS mergers, with frequencies between
$f_{\rm main\; peak}$ and $f_3$ \cite{Ellis2018a}. This result was
obtained by adding the effects of dark matter (in amounts up to $10\%$ of
the stellar mass\footnote{Note that dark matter cannot condensate inside stars in
quantities as large as this, therefore non-standard properties or
mechanism involving dark matter have to be thought \cite{Foot2004,
  Fan2013, Pollack2015}.}) into the mechanical model proposed in
Ref. \cite{Takami2015}, and mentioned above, that captures the essential
features of the dynamics of the hyper-massive NS formed in the first
instants after the merger.
Again, because of their higher frequency and lower power, it will be
difficult to distinguish such peaks observationally in the near future.
See Sects.  \ref{exotic binaries} and \ref{dark matter} for other results
involving the hypothetical presence of dark matter in NSs.

Still further additional peaks may appear if the lifetime of the merged
stellar object before collapse is long enough that convective excitation
of inertial modes occurs, as pointed out in
Ref. \cite{DePietri2018}. These are related to the thermal properties of
the EoS but a systematic study of these modes and of how they are related
to the properties of the stellar EoS is still missing.

A different way to gain knowledge on the mass-radius relation of NSs was
proposed in Refs. \cite{Bauswein2013, Bauswein2017, Bauswein2017b}. It
consists in using some empirical relations obtained from numerical
simulations of BNS mergers between threshold mass for prompt collapse,
maximum mass for a nonrotating NS, and its radius
\cite{Bauswein2013, Bauswein2017}. Considering that observations of the
electromagnetic emission associated with GW170817 may imply (but see
Ref. \cite{Kiuchi2019} for alternative interpretations) that GW170817
was not a prompt collapse\footnote{In general, of course, GW observations
  as well can inform us about the fate of the merged object (prompt or
  delayed collapse, or maybe no collapse).} \cite{Shibata2017c,
  Bauswein2017b} and that therefore the total mass of GW170817 is a lower
bound on the threshold mass described above, an estimate was found for
the maximum mass for a nonrotating NS and its radius
\cite{Bauswein2017b} (see Table \ref{table:radii} for results on radii). The authors
remark that the constraints they set are particularly robust because they
only require a measurement of the chirp mass and a distinction between
prompt and delayed collapse of the merger remnant.
Ref. \cite{Koeppel2019} later used the same arguments to update the
obtained results in a fully general-relativistic framework.

\subsection{Applications to data analysis}
\label{post-merger basics application}

For data-analysis standards and in comparison to the inspiral, numerical
simulations of the post-merger phase are still sparse and not accurate
enough (as explained above) and this contributes to the lack of
analytical, physically parameterized waveform templates.
This reduces the feasibility of matched-filtering. Generic analyses that
target signals of unknown morphology might be less efficient than
matched-filtering, but they have been shown to be able to extract the
main features of post-merger signals, such as its main frequency
components \cite{Clark2014, Clark2016, Klimenko2016, Vinciguerra2017,
  Chatziioannou2017, Easter2018, Tsang2019}. Below, more details on these
works are given.

Ref. \cite{Clark2014} was the first to carry out a systematic study of
the detectability of the high-frequency content of the merger and post-merger
parts of BNS post-merger GW signals with methodologies used to search for unmodelled
GW transients. Using a morphology-independent algorithm based on
constrained likelihood statistics that identifies whether a signal is
significant with respect to the noise, they focused on the problem of
discriminating among different post-merger scenarios (prompt collapse or
not) and on measuring the dominant oscillation frequency in the
post-merger signal, in case of non-prompt collapse (see
Sect. \ref{post-merger-estimates} for a summary of their results too).

With the goal of reducing the complexity of the problem from a
high-dimensional physical parameter space, where the waveforms are
modelled directly through numerical simulations, to a lower-dimensional
problem that includes only the dominant features of the waveform, the
same group then proposed a low-dimensionality frequency-domain template
based on principal component analysis \cite{Clark2016}. They constructed
a catalogue of numerical waveforms and decomposed the magnitude and phase
spectra into orthogonal bases, as per the principal component analysis.
By excluding each element of the catalogue one at a time and checking how
well just the first principal component from the principal component
analysis could reproduce that element, a match of $0.93$ was found. The
usually accepted desired threshold is $0.97$, but Ref. \cite{Clark2016}
suggested that the first principal component is robust enough to model
the high-frequency GW spectrum for BNS mergers for data-analysis
purposes.

Ref. \cite{Chatziioannou2017},
then, also proposed a method that relies less on numerical-relativity
simulations: reconstructing the post-merger GW signal in observed data
through a sum of wavelets without assumptions on the morphology of the
system. The algorithm they use is BAYESWAVE \cite{Cornish2015,
  Littenberg2015} and it was shown to be capable of reconstructing the
dominant features of the injected signal with an overlap of above $90\%$
for post-merger signal-to-noise ratios above $\approx 5$. This would
provide a measurement of the dominant post-merger frequency $f_{\rm
  main\; peak}$ to within a few tens of $\Hz$. Additional information
about the signal, such as its broadband structure or the finite extent of
the post-merger peak, could increase the sensitivity of the analysis,
making it easier to detect and characterize the post-merger signal.
After the reconstruction an empirical relation based on
numerical-relativity calculations to relate radius and $f_{\rm main\;
  peak}$ was used \cite{Chatziioannou2017}. The final results (namely
that the radius can be measured to a few hundred metres for signals with
a signal-to-noise ratio $\approx 5$) are comparable to the methods that
are based on a description of the post-merger power spectral density
obtained from numerical-relativity simulations \cite{Bauswein2011,
  Bauswein2012, Bauswein2012a, Bauswein2014, Takami:2014, Takami2015,
  Bernuzzi2015a, Kiuchi2019a}. It is to be noted that their error on the
radius is dominated by the scatter in the {\it universal relation} used,
rather than the statistical error of the reconstruction itself
\cite{Chatziioannou2017}.

Very recently, Ref. \cite{Tsang2019} has done a similar work based on a
Bayesian analysis that employs simplified Lorentzian model functions and
found that the main emission frequency of the post-merger remnant, for
signal-to-noise ratios of $8$ and above, can be extracted within a
$1\sigma$ uncertainty of about $100 \Hz$ for Advanced LIGO and Advanced
Virgo at design sensitivities.

The methodology described in Ref. \cite{Chatziioannou2017} was applied to
the data of GW170817 and results were reported by the LIGO-Virgo
Collaborations \cite{Abbott2018a, Abbott2018c}. That is how it was found
that upper limits on energy are at least an order of magnitude larger
than expectations based on simulations performed with the EoSs they chose
\cite{Abbott2018c}.

Recently, the first draft suggestion of a way of generating reliable
post-merger spectra rapidly enough to be useful for templated data
analysis was proposed in Ref. \cite{Easter2018}. This is done through a
hierarchical model built to represent the amplitude spectra and trained
on the numerical-relativity simulations of Ref. \cite{Rezzolla2016}. They
report a mean of $0.95$ for the noise-weighted fitting factors across all
tested spectra, for sources simulated at a distance of $50\,
\Mpc$. However, they do not test their results in a Bayesian framework to
find its actual efficiency in parameter estimation of detected
events. Actually, in order to be useful in search and
parameter-estimation studies, the model of Ref. \cite{Easter2018} should
be extended to include more numerical simulations for training.

\subsection{Estimates about future observations}
\label{post-merger-estimates}

Realistic estimates of what information can be obtained with the methods
described in the previous subsections require Bayesian analyses, like
those performed in Refs. \cite{Messenger2013, Clark2014, Clark2016,
  Yang2017, Chatziioannou2017, Torres-Rivas2019}. As mentioned at the
beginning, GW measurements at the expected frequencies and amplitudes for
$f_1$ and  $f_{\rm main\; peak}$ are very difficult; in practice they are limited to
sources within $\sim 30\, \Mpc$ for second generation detectors, as shown
by Refs. \cite{Clark2014, Clark2016, Yang2017}
via large-scale Monte Carlo studies. As mentioned above,
Refs. \cite{Clark2014, Clark2016} provided the first systematic studies
of the detectability of the high-frequency content of the merger and
post-merger parts of GW signals from BNS systems. These investigations
did not rely on waveform models and optimal filtering, but exploited
methodologies used to search for unmodelled GW transients. They concluded
that second generation detectors could measure post-merger signals and
constrain the NS EoS for sources up to a distance of $10-25\, \Mpc$
(assuming optimal orientation). Ref. \cite{Clark2016} also showed that
the error in the estimate of the NS radius would be of the order of
$400\, {\rm m}$ in Advanced LIGO. Ref. \cite{Yang2017} proposed methods
that stack the post-merger signal from multiple BNS observations to boost
the post-merger detection probability. They found that, after one year of
operation of Cosmic Explorer \cite{McClelland2015}, the dominant-peak
frequency can be measured to a statistical error of $\approx 4-20\, \Hz$
for certain EoSs, corresponding to a radius measurement to within
$\approx 15-56\, {\rm m}$, a fractional error of about
$4\%$. They showed that errors in the universal relations between
post-merger oscillation frequency and total mass of the binary, and in
the template construction dominate over the statistical
error. Detectability of individual events could potentially improve if
one considers all components/peaks that arise in the post-merger
waveform, and not only the dominant peak.

As said in Sect. \ref{post-merger basics application}, waveform templates
that span the pre- and post-merger signals are currently not available,
but observations of the pre-merger signal from GW170817 has been used to
inform expectations about the properties of the undetected post-merger
signal. Ref. \cite{Torres-Rivas2019} used the posterior samples for the
masses, radii and tidal parameters of the inspiral of GW170817 to
estimate its expected post-merger signal to be approximately $2.5 \, \kHz
\leq f_{\rm main\; peak} \leq 4 \, \kHz$.

It was also pointed out that, the full LIGO-Virgo network operating at
design sensitivity may provide reasonable estimates of the dominant
post-merger oscillation frequency and corresponding constraints on the NS
EoS \cite{Abbott2018a} and this will be certainly measured if the
sensitivity in the $\kHz$ regime of the detectors are improved by a
factor 2 or 3 over their current design sensitivity
\cite{Torres-Rivas2019}. With further improvements and next-generation
detectors, it is reasonable to believe that we will also be able to
extract subdominant frequencies \cite{Torres-Rivas2019}, as long as the
numerical waveforms currently available approximately reflect the true
strength of the emitted signal (which could be weaker if physical
viscosity is very high; see also Ref. \cite{Kiuchi2019a}) and the
expected calibration of the detectors at high frequencies are realized.

\subsection{Investigating phase transitions with post-merger waveforms}
\label{postmerger phase transitions}

Hadron-quark (and other strong) phase transitions or phase transition to
hyperonic matter may occur at some high-density threshold. Since the
densities reached after the merger are larger than those in the original
stars in the binary, it is possible that phase transitions occur only
after the merger. In this case, which is the focus of this Section, a
measurement of the tidal deformabilities, of course, cannot contain
information on phase transitions. The case in which such phase
transitions occur in both or one of the stars in the binary already
before the merger will be treated in Sect. \ref{hybrid stars}.

Several simulations of BNS mergers with EoSs containing a phase transition
to hyperonic matter (see the end of Sect \ref{nucleonic} for more
comments on stars containing hyperonic matter)
have been carried out, starting with the work of
Refs. \cite{Sekiguchi2011, Kiuchi2012},
where it was found that the post-merger GW frequency significantly
changes during the time between merger and collapse. If it were so, the
above-mentioned relations between post-merger spectral properties and
stellar physical quantities may have to be taken with care. However, more
recent works \cite{Radice2017a, Bauswein2019}
found that also in case hyperons appear the post-merger main GW frequency
remains rather constant over time and that its frequency is very similar
to (and, in the foreseeable future, indistinguishable from) that given
from BNS mergers with the corresponding purely nucleonic EoS. What
differentiates hyperonic EoSs in the post-merger GW signal is the
amplitude and phase modulation and the total luminosity
\cite{Radice2017a}, but these quantities are both more difficult to
measure\footnote{Ref.  \cite{Radice2017a} estimated that Advanced LIGO
  could distinguish between the two EoSs they employed with a single
  merger at a distance of up to $\sim 20 \Mpc$, depending on the total
  mass of the binary.}
and more difficult to obtain accurate estimates of from numerical
simulations.

In a strong phase transition, like one from nucleonic to quark matter,
the main feature of the power spectral density of the post-merger phase,
$f_{\rm main\; peak}$, may instead change rapidly, because of the abrupt
speed up of the rotation of the differentially rotating core of the
remnant \cite{Bauswein2019}. The lifetime of the merged object before
collapse and the black-hole ringdown waveforms may also be rather
different from the ones expected from purely nucleonic EoSs
\cite{Most2018b}. In addition to GWs, the rearrangement of the angular
momentum in the remnant as a result of the formation of a quark core
could be accompanied by a prompt burst of neutrinos followed by a
gamma-ray burst \cite{Cheng1996, Bombaci2000, Mishustin2003}. However, a
preliminary study found that there would be no significant qualitative
differences in the electromagnetic counterpart of NS mergers between a
system undergoing a phase transition to quark matter and purely hadronic
mergers \cite{Bauswein2019a}.

These indications come from simulations of merging NSs described by EoSs
that include a phase transition \cite{Most2018b, Bauswein2019,
  Bauswein2019a}. As mentioned in Sect. \ref{simulations}, such
simulations have been performed for the first time very recently and
published at the same time by two groups, one that used a
smoothed-particle-hydrodynamics code with approximate treatment of
general relativity \cite{Bauswein2019, Bauswein2019a} and the other that
carried out fully general-relativistic simulations
\cite{Most2018b}. Neither work includes magnetic fields.

\begin{figure}[tb]
\begin{center}
\begin{minipage}[t]{14 cm}
   \includegraphics[width=1.0\textwidth]{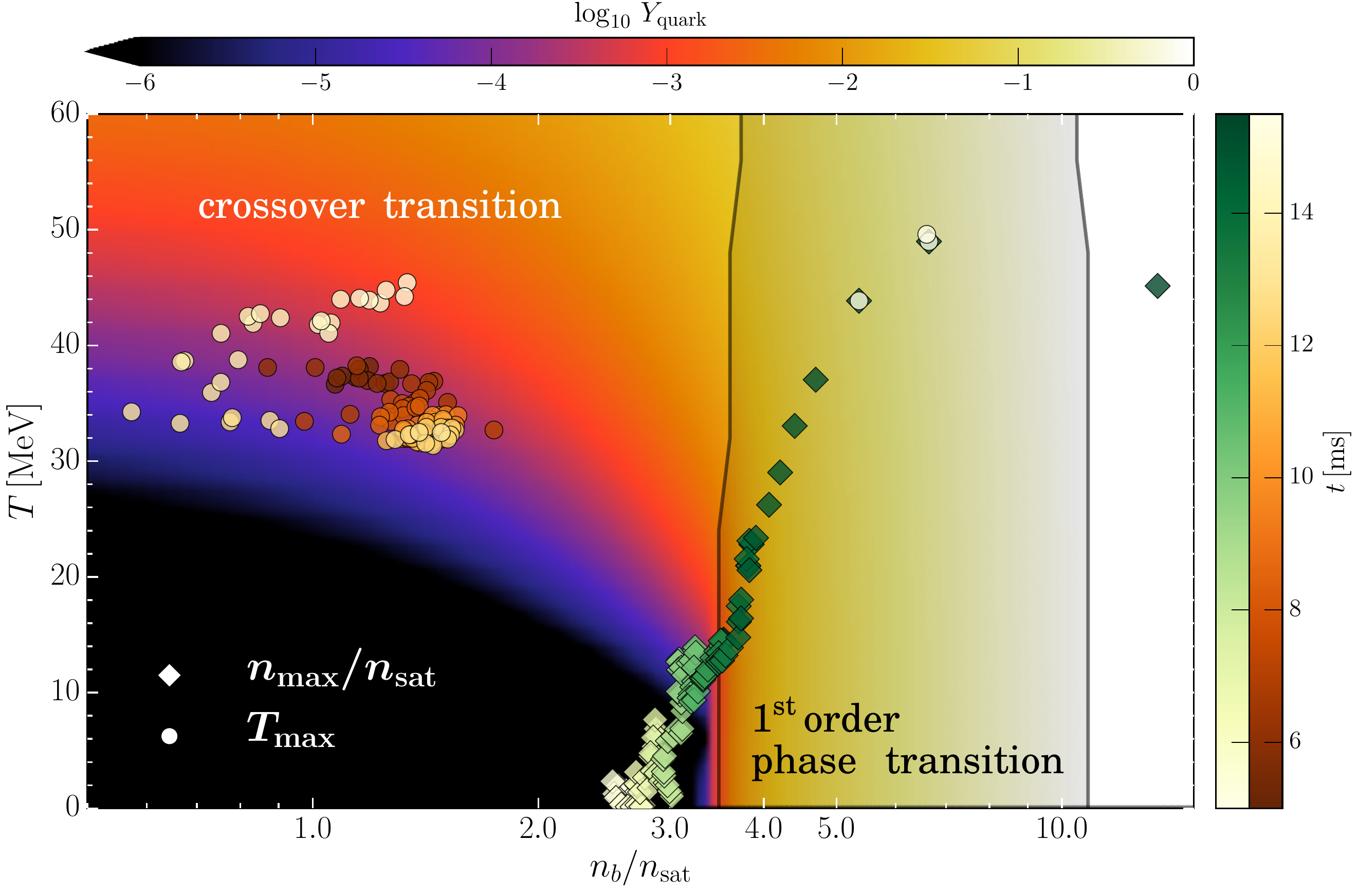}
\end{minipage}
\begin{minipage}[t]{16.5 cm}
  \caption{Phase diagram illustrating the properties of the phase
    transition by showing the evolution of the temperature and density in
    the merger remnant. Diamonds refer to the maximum normalized
    baryon-number density and circles to the temperature after the merger
    of a BNS system described with an EoS that includes a strong
    first-order phase transition to quark matter. Different times of the
    evolution are represented with a colour code, together with the quark
    fraction $Y_{\rm quark}$ . The grey shaded area shows the first-order
    phase-transition region. See Ref.\cite{Most2018} for details. (From
    the the version of Ref. \cite{Most2018} on
    arXiv.org) \label{fig:Most2018-Fig3}}
\end{minipage}
\end{center}
\end{figure}

The latter simulations included quarks at finite temperatures (within a
temperature-dependent chiral mean field model\footnote{The EoS employed
  does not produce gravitationally stable hybrid stars, therefore the
  effects of such a hadron-quark phase transition can only be observed in
  a BNS merger.} \cite{Dexheimer:2009}) and were used to present for the
first time a phase diagram (see Fig. \ref{fig:Most2018-Fig3})
illustrating the properties of the phase transition, as often done in
studies about heavy-ion collisions. This work then showed that a
quark-hadron phase transition (occurring after the merger, in this case)
would indeed leave its strong imprint in the GW signal (collapse time,
ringdown), but has only a small influence on the post-merger GW
frequencies. This is because the EoS used in Ref. \cite{Most2018b}
produces a small (see below for a comparison with another EoS with a
strong phase transition, the DD2F-SF EoS \cite{Fischer2017}) quark-matter
fraction during the early phases of the post-merger. When, at later
times, the quark matter fraction increases, it immediately induces the
gravitational collapse of the remnant.

\begin{figure}[tb]
\begin{center}
\begin{minipage}[t]{12 cm}
   \includegraphics[width=1.0\textwidth]{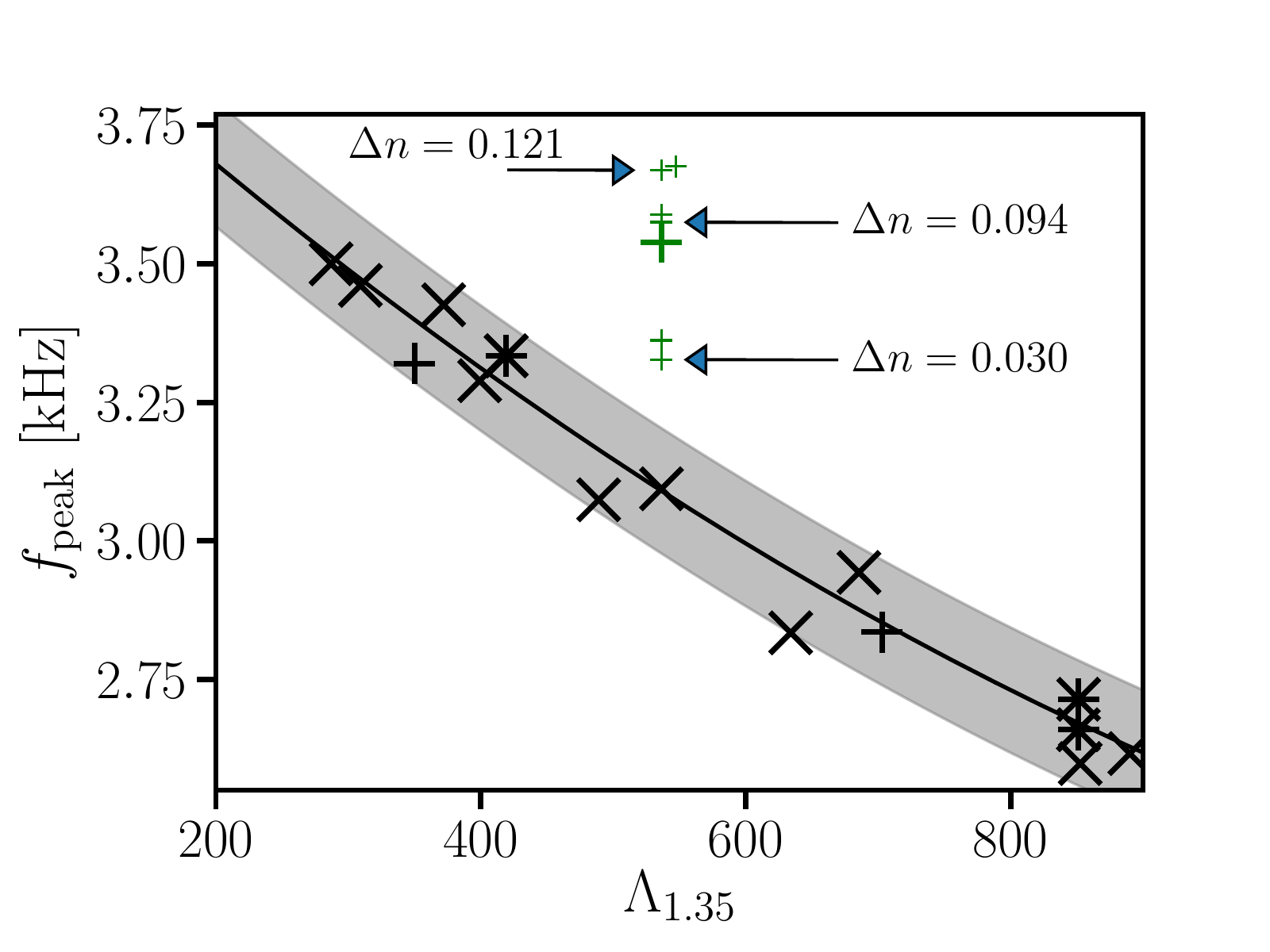}
\end{minipage}
\begin{minipage}[t]{16.5 cm}
  \caption{Dominant post-merger GW frequency $f_{\rm main\; peak}$ as a
    function of tidal deformability $\Lambda$ for mergers of two $1.35
    \Msun$ stars. The DD2F-SF \cite{Fischer2017} models with a phase
    transition to deconfined quark matter appear as clear outliers. The
    solid curve displays the least square fit for all purely hadronic
    EoSs (including three models with hyperons, marked by
    asterisks). Arrows mark DD2F-SF \cite{Fischer2017} models with
    roughly the same onset density and stiffness of quark matter, but
    with different strength of the density jumps. (From
    Ref. \cite{Bauswein2019}) \label{fig:Bauswein2019-Fig3}}
\end{minipage}
\end{center}
\end{figure}

Ref. \cite{Bauswein2019} more systematically studied the effects of phase
transitions on the post-merger phase by adopting purely hadronic EoSs,
EoSs with a second-order phase transition to hyperonic matter
\cite{Banik2014, Fortin2018, MARQUES2017} and EoSs with a first-order
phase transition to quark matter (the DD2F-SF EoS \cite{Fischer2017}),
which produce a non-negligible amount of quark matter soon after the
merger. It was shown that the phase transition increases the dominant
post-merger GW frequency $f_{\rm main\; peak}$ if and only if a strong
first-order-like phase transition leads to the formation of a
gravitationally stable extended quark-matter core in the post-merger
remnant \cite{Bauswein2019}. This happens because the formation of a
quark-matter core makes the remnant more compact. Also, comparing the
outcomes of all their simulations on the $\Lambda$-$f_{\rm main\; peak}$
plane, they found that only the DD2F-SF models with a phase transition to
deconfined quark matter lie clearly outside the relation visible for
other EoSs\footnote{Note, however, that error bars on the points in the
  graph have not been computed, but similar computations by the same
  authors had given uncertainties a few $10 \Hz$ \cite{Bauswein2011}. See
  also Ref. \cite{Kiuchi2019a} for more on computations of error bars in
  these kinds of relations.}, including those with a second-order phase
transition (see Fig. \ref{fig:Bauswein2019-Fig3}). The maximum deviation
from this relation is about $100 \Hz$ for EoS models without strong
first-order phase transitions and about $500 \Hz$ for those with a strong
phase transition to deconfined quark matter. Namely, it was found that
only a sufficiently strong first-order phase transition (or, actually,
any transition, even not formally first-order, which causes a strong
softening of the EoS like the one seen for DD2F-SF models) has a
noticeable impact on the stellar structure and thus alters the
post-merger GW signal in a measurable way. What would be determinant for
the discovery of such a phase transition is a shift of the $f_{\rm main\;
  peak}$ frequency as compared to that expected from the tidal
deformability of the inspiralling NSs. Such a shift of the dominant
post-merger GW frequency might be revealed by future GW observations
using second- and third-generation GW detectors \cite{Bauswein2019} and
would also allow to constrain the density at which the phase transition
occurs.

\section{Combining analyses of waveforms emitted before and after the
  merger}
\label{pre+post merger}

Even if (as mentioned in Sect. \ref{sec:inspiral}), with current
detectors, omitting the post-merger waveforms from the global analysis of
the signal does not lead to significant loss of information or biased
estimation of the source properties \cite{Dudi2018}, it would be ideal to
have waveform templates that consistently and exhaustively cover the
inspiral, merger, and post-merger phases, so that one could perform
matched-filter searches, as mentioned in Sect. \ref{post-merger basics
  application}. This is true especially for the shorter-duration signals
of the post-merger phase. However, complete numerical-relativity
waveforms are too sparse and semianalytical models for the whole signal
are in their infancy.

One work has carried out a global analysis that combines the two types of
estimates from the pre-merger and post-merger waveforms
\cite{Bose2017}. Through Monte-Carlo simulations that combine
measurements of the total mass from the inspiral phase with those of the
compactness from the post-merger oscillation frequencies, improved
estimates were found for the mean population radius of NSs: With about
$50$ observations, the error on the average radius was computed to be
$2-5\%$ for stiff EoSs and $7-12\%$ for soft EoSs.

\section{The merger of binaries of stars not made of ordinary matter}
\label{exotic binaries}

As mentioned in Sect. \ref{other-compact-objects}, there may exist
compact objects similar in mass and size to NSs, but not made of ordinary
matter, like boson stars or gravastars. A few recent works have proposed
preliminary studies about whether current and future observations can
distinguish between inspirals and mergers of NSs and boson stars
\cite{Cardoso2017, Palenzuela2017, Sennett2017, Clough2018, Bezares2018,
  Dietrich2019a} or gravastars \cite{chirenti_2007_htg, Chirenti2016,
  Cardoso2019}. GW detectors may also be able to probe the structure of
these different compact objects (if they exist) through their tidal
interactions in the inspiral of binary systems and through the
phenomenology of their merger, post-merger and ringdown phases.

The first step of such studies is the computation of the tidal
deformability of these objects \cite{Cardoso2017, Sennett2017,
  Mendes2017,
  Pani2015, Uchikata2016}. Boson stars, actually, have no surface, so
defining their radius (and hence compactness) may be ambiguous. One
common convention is to define the radius as that of a shell containing a
fixed fraction of the total mass of the star. Another option consists in
using only quantities that can be extracted from the asymptotic geometry
of the boson stars, like the total mass and the dimensionless tidal
deformability \cite{Sennett2017}. The boson-star tidal deformability can
be obtained similarly to how it is done with NSs (see
Sect. \ref{premerger-basic}).
Different varieties of boson stars have been considered in the literature
(see Refs. \cite{Schunck2003, Liebling2012, Cardoso2019} for full reviews
and, \eg, Ref. \cite{Sennett2017} for a short summary), but in general,
the compactness and tidal deformability of boson stars is comparable to
that of NSs, with most types of boson stars having tidal deformability as
much as $\approx 25\%$ smaller than that of a NS of similar
compactness. Solitonic boson stars \cite{Friedberg1987} and Proca stars
\cite{Brito2016} can have larger compactness and smaller tidal
deformability than any realistic NS.

Numerical simulations of the merger of boson stars have been performed
for some years, with the first works focussing on head-on collisions
\cite{Lai2005, Palenzuela06, Palenzuela2008}, which are computationally
less difficult. More recent simulations have investigated collisions of
very compact boson stars and compared them to collisions of black holes
\cite{Cardoso2016a, Helfer2019, Sanchis2019}.
In one work, numerical simulations of the head-on collision of a NS with
an axion star \cite{Dietrich2019a} were performed. In this particular
example, it was found that the GW emission after the merger extends for a
much longer duration with respect to a NS head-on collision, thus
releasing more energy in GWs. Assuming that a binary merger of a NS and
an axion star would show similar dynamics, it has been estimated
\cite{Dietrich2019a} that it would be observable up to about $100\, \Mpc$
with current GW detectors and up to $1\, \Gpc$ with the Einstein
Telescope \cite{Punturo2010b, Sathyaprakash:2009xs}.

Simulations of binary mergers of a NS and an axion star have not been
performed yet, but those of the merger of two inspiralling boson stars
have. The first such simulations were done with the main goal to study
the properties of the merged object \cite{Bezares2017}. One interesting
finding was that, if collapse to black hole does not occur, a nonrotating
boson star forms from the merger, having lost all the original angular
momentum by emitting scalar-field and gravitational waves. This is
obviously a very different outcome from that of BNS mergers. Other works
then studied the frequency of the GW emission that carries away the
angular momentum \cite{Palenzuela2017}, finding that it depends on the
compactness of the initial boson stars and that it is in most cases
higher than the main post-merger frequency in BNS mergers (that do not
end in a prompt collapse). As for BNS mergers (see Sect. \ref{univ-rel}),
Ref. \cite{Palenzuela2017} found that the main post-merger frequency can
be related to the frequency of the merger, defined as the instantaneous
GW frequency at the time when the amplitude reaches its first peak after
the inspiral phase \cite{Read2013}. It was suggested that
third-generation detectors like the Einstein Telescope
\cite{Punturo2010b, Sathyaprakash:2009xs} or Cosmic Explorer
\cite{McClelland2015} may be able to distinguish post-merger
gravitational signals of boson-star mergers from those of NS mergers
\cite{Palenzuela2017}.

Ref. \cite{Bezares2018} considered simulations of binaries of {\it dark}
boson stars, namely binaries composed of two boson stars each described
by a different complex scalar field and thus not interacting with each
other except through gravity. It was found that their merger produces a
gravitational signature in principle distinguishable from that of
binaries composed of black holes, NSs and also interacting boson
stars. Similar results were found for mergers of BNS with dark matter
particles trapped on their interior \cite{Bezares2019}. The distinctive
feature of this scenario is the generic development of a strong $m=1$
one-armed instability, in principle distinguishable from that of mergers
of NSs (see Sect. \ref{postmerger basic ideas}). I have already reviewed
in Sect. \ref{postmerger basic ideas} some works that found other
distinctive features in the post-merger spectrum due to the possible
presence of dark matter \cite{Ellis2018a}.

Finally, Refs. \cite{Cardoso2017, Sennett2017} investigated more
systematically for different types of boson stars the extent to which
tidal effects in the inspiral GW signal can be used to discriminate
between standard sources and boson stars. Only some types of boson
stars and only nonrotating boson stars were studied. It was found that
Advanced LIGO could differentiate between massive boson stars and NSs or
black holes, but only in some cases, namely only for some types of boson
stars and for systems with large mass asymmetry. For example, it was
found that the lower limit for the tidal deformability for boson stars
with a quartic self-interaction is $\approx 280$, while that for boson
stars with a solitonic interaction is $\approx 1.3$ \cite{Sennett2017},
implying that Advanced LIGO is not sensitive enough to discriminate
between solitonic boson stars and black holes. Third-generation detectors
like the Einstein Telescope \cite{Punturo2010b, Sathyaprakash:2009xs} or
Cosmic Explorer \cite{McClelland2015} should be able to distinguish
between them. A Fisher-matrix analysis confirmed these first estimates
\cite{Cardoso2017}.

\section{Constraints on stellar radius and equation-of-state parameters deduced from GW170817}
\label{constraints}

In this Section, I discuss the numerous works and results that appeared
after the GW observation of GW170817 \cite{Abbott2017} and tried to
estimate physical quantities from it and, ultimately, the EoS. Before
starting, however, let me note that some articles \cite{Zhang2018,
  Lim2018, Tews2019} claimed that actually GW170817 has not added new
insights about the EoS, because the constraints it imposes are less
stringent than those obtained from current knowledge in nuclear physics,
with the possible exceptions of the estimate of the lower limit for the
stellar radius \cite{Radice2017b, Bauswein2017b, Radice2018c, Koeppel2019}, which
however is said to be possibly affected by systematic errors
\cite{Tews2019}. In particular, Ref. \cite{Zhang2018} pointed out that
limits on the high-density EoS parameters from the constraints set by
GW170817 on the tidal deformability are weaker than the constraints
extracted from analysing other astrophysical observations unrelated to
GWs. Ref. \cite{Tews2019} then stated that the EoSs that are said to be
ruled out by the upper limit on the tidal polarizability derived from
GW170817 were actually already ruled out on the basis of state-of-the-art
nuclear theory describing microscopic EoSs at densities where error
estimates are still credible \cite{Tews2019}. More details on this view
are presented below.

In addition to the GW measurement, additional information has been
obtained through the electomagnetic emission (in the gamma-ray, x-ray,
ultraviolet, optical, infrared, and radio bands) related to GW170817,
named GRB 170817A (for the gamma-ray burst) and AT2017gfo for the rest of
the astronomical transient \cite{Abbott2017b, Abbott2017d, Alexander2017,
  Arcavi2017, Arcavi2017a, Arcavi2018, Chornock2017, Coulter2017,
  Covino2017, Cowperthwaite2017, Drout2017, Evans2017, Goldstein2017,
  Hallinan2017, Kasliwal2017, Margutti2017, Murguia-Berthier2017,
  Nicholl2017, Pian2017, Savchenko2017, Smartt2017, Soares-Santos2017,
  Tanaka2017, Tanvir2017, Troja2017}, alone or in combination with GW
constraints on the tidal deformability. However, electromagnetic
radiation from binary mergers is out of the scope of this review, so I
limit myself to refer the reader to some of the relevant references
\cite{Radice2017b, Bauswein2017b, Radice2018c, Coughlin2018,
  Coughlin2018a, Wang2018, Margalit2019} and to reporting estimates from
some of these works for the radius of a $1.4 \Msun$ star in Table
\ref{table:radii}. Note also that extracting accurate information from
electromagnetic radiation from GW170817 is difficult because the
underlying models are still based on assumptions that may significantly
introduce systematic errors \cite{Kiuchi2019}.

Refs. \cite{Kumar2019, Fasano2019, McNeilForbes2019}
combined the GW measurement with electromagnetic observations of other
NSs in accreting low-mass x-ray binaries (from
Ref. \cite{Ozel2015}) to give a more precise estimate of the tidal
deformability and Ref. \cite{Miller2019} proposed an affordable Bayesian
approach that can include diverse evidence, such as nuclear data and the
inferred masses, radii, tidal deformabilities, moments of inertia, and
gravitational binding energies of NSs obtained through different types of
observations \cite{Miller2019}. The method allows any parameterization of
the EoS.

Estimates and constraints on EoSs from measurements of NSs have been
derived on the basis of (i) bounds on the tidal deformability in GW170817
\cite{Abbott2018a, Abbott2018b, Annala2017, Most2018, Lim2018, De2018,
  Zhao2018, Dai2018, Tews2018, Zhang2018, Chatziioannou2018, Landry2018,
  Vuorinen2019, Margueron2018a, Margueron2018b, Zhu2018, Malik2018,
  Carson2019, Zhang2019a, Zhang2019b, Zhou2017, Lai2017b, Lau2019,
  Most2018, Paschalidis2018a, Nandi2018, Nandi2018a, Burgio2018,
  Tews2018, Gomes2018, AlvarezCastillo2018, Li2018b, Drago2018, Han2018,
  Montana2018, Wei2019, Sieniawska2019, Christian2019, Gomes2019,
  Tews2019}, (ii) the upper bound for the maximum mass of a cold
non-rotating compact star deduced in various ways from GW170817
\cite{Shibata2017c, Ruiz2017, Rezzolla2017, Margalit2017, Landry2018,
  Montana2018, Zhang2019b, Shibata2019}, (iii) the lower bound for the
maximum mass for a non-rotating compact star from pulsar
observations\footnote{As mentioned earlier, works published before the
  discovery of a $2.17 \Msun$ NS \cite{Cromartie2019} use for this limit
  different numerical values in the range $1.97 \pm 0.04 \Msun$.}, (iv)
the lower bound on the radius of a non-rotating compact star deduced from
GW170817 \cite{Radice2017b, Bauswein2017b, Radice2018c, Koeppel2019}. Condition (iii) requires stiff enough
EoS, while conditions (i) and (ii) require soft enough EoS.  According to
Refs. \cite{Tews2018, Tews2019}, condition (ii) is powerful for smooth
EoS models, because stars without phase transitions exhibit a strong
correlation between their maximum mass and the radii of a NS of typical
mass\footnote{For example, considering smooth EoSs,
  Ref. \cite{Zhang2019c} showed that the discovery of a $2.17 \Msun$ NS
  \cite{Cromartie2019} significantly reduced the allowed range for the
  skewness $Q_0$ (see Sect. \ref{eos-parameterization}).}, but it is not
very constraining for general EoS models that may contain strong
first-order phase transitions. Also note that there is an open debate on
what may be the (systematic) error bar of the upper bound for the maximum
mass of a cold non-rotating compact star, since some assumptions need to
be made in its calculation. Some of the the articles cited at point (ii)
above do report an error bar on the values they give, but a recently
posted work explained that all previous results were obtained with
oversimplified assumptions and that presently all one can reliably say is
that the upper bound for the maximum mass of a cold non-rotating compact
star is only weakly constrained as $\lesssim 2.3 \Msun$
\cite{Shibata2019}.

Since studies performed for
hadronic EoSs, quark EoSs, and EoSs with phase transitions lead to rather
different results, below I will separate them in different subsections.

\subsection{Hadronic equations of state}
\label{nucleonic}

As mentioned earlier, the first published analysis of GW170817 by the
LIGO-Virgo Collaborations \cite{Abbott2017} was very conservative: the
EoSs of the two stars were assumed to be completely independent, which
resulted in a not-much-constraining upper limit for the tidal
deformability and no estimate of radius.
Later, the LIGO-Virgo Collaborations published a revised analysis with
improved estimates \cite{Abbott2018a, Abbott2018b} (see below) and the
correction of an error in their reporting of the $90\%$ confidence upper
limit of
\label{bookkeeping error note}
the tidal deformability\footnote{A mentioned previously, the corrected
  value in the case of the low-spin prior is $\tilde{\Lambda}\leq 900$
  instead of $\leq 800$; see the caption of Table IV in
  Ref. \cite{Abbott2018a}.}, but in the meanwhile some notable works
were published that used the original very conservative (and affected by
the bookkeeping error) observational results \cite{Annala2017, Most2018,
  Lim2018, Wei2019}. Some of these works studied how parameterized smooth
EoSs fit with the constraints from GW170817.

The two most prominent articles to carry out the first systematic studies
of the statistical properties of the tidal deformability on this line
\cite{Most2018, Lim2018} generalized efforts of previous works that had
used a more limited set of EoSs to check whether the observation of
GW170817 allowed them or not \cite{Annala2017, Krastev2018,
  Wei2019}\footnote{Ref. \cite{Vuorinen2019} later updated the results of
  \cite{Annala2017} with the new LIGO-Virgo Collaborations estimates for
  GW170817.}.  Both Refs. \cite{Most2018, Lim2018} parameterized a very
large range of physically plausible EoSs for compact stars with a
piecewise-polytropic representation and computed equilibrium stellar
solutions for up to one million different EoS by numerically solving the
Tolman-Oppenheimer-Volkoff \cite{Tolman39, Oppenheimer39b} equations. By
imposing only one condition, that the sound speed is subluminal,
Ref. \cite{Lim2018} focussed on smooth EoSs and concluded that the $95\%$
credibility range of predicted tidal deformabilities for a $1.4 \Msun$ NS
is consistent with the upper bound deduced from GW170817
\cite{Abbott2017}. Ref. \cite{Most2018} considered also EoSs with phase
transitions and imposed some additional constraints, namely that the
tidal deformability for a $1.4\Msun$ NS is $\Lambda_{1.4}< 800$ (from the
first official analysis of GW170817 \cite{Abbott2017}), that the lower
bound on the maximum mass for a non-rotating NS is $ M_{\rm max}> 2.01$
and that an upper bound on the maximum mass for a non-rotating NS is
$M_{\rm max}< 2.16$ \cite{Rezzolla2017}.  Ref. \cite{Most2018} proposed
for the first time a lower limit for the tidal deformability
$\Lambda_{1.4} > 400$ at a $2\sigma$ confidence level (see Table
\ref{table:radii} for their radius estimates). They pointed out that the
distributions are very robust upon changes in the upper limit of the
maximum mass and of the tidal deformability and that different
prescriptions on the treatment of the nuclear matter in the outer core
($0.08 < \rho/{\rm fm}^{-3} < 0.21$) \cite{Most2018, Annala2017} may have
a large impact on the statistical properties of NS radii. Progress in
theoretical knowledge of the outer core is then important for a good
description of the NS structure. Also the NS crust may have a large
impact on the statistical properties of NS radii \cite{Lim2018,
  Gamba2019}, even if it does not strongly impact tidal deformability
\cite{Gamba2019, Biswas2019a, Kalaitzis2019} and thus it is not very
important in relation to current GW measurements from BNS systems.

Another work that was published before the updated analysis by the
LIGO-Virgo Collaborations \cite{Abbott2018a, Abbott2018b} compared the
tidal deformability observation from GW170817 (the first announcement
\cite{Abbott2017}) with data from the PREX experiment
\cite{Abrahamyan2012, Horowitz2012} (elastic scattering of polarized
electrons) on the neutron-skin thickness (the difference between the
root-mean-square radii of the distribution of neutrons and protons) of
$^{208}$Pb \cite{Fattoyev2017} (see also Refs. \cite{Fattoyev2013,
  Fattoyev2014}). In order to connect the observed tidal polarizability
to nuclear observables, they assumed EoSs described with a relativistic
mean-field approach and found that the central value of the neutron-skin
thickness of $^{208}$Pb measured in PREX is rather far from the one
derived from GW170817, even if there is no tension because the error bar
of PREX is very large. If this experimental datum is confirmed by more
accurate experiments (like PREX-II) in the future, then this might be in
tension with GW170817 and be indicative of a phase transition in the
interior of NSs \cite{Fattoyev2017}. An alternative view on this possible
tension has been expressed in Ref. \cite{Tews2019}, where it was stated
that actually
smooth EoSs (without phase transitions) different from those employed in
Ref. \cite{Fattoyev2017} can be found that are compatible with both
GW170817 and future more accurate nuclear experimental data that may
confirm the value of the neutron-skin thickness of $^{208}$Pb found in
PREX \cite{Tews2019}. Applying the calculation of
Ref. \cite{Fattoyev2017} to the revised analysis of GW170817
\cite{Abbott2018a, Abbott2018b} would probably not change their
conclusions.

As mentioned above, several months after the announcement of the
observation of GW170817, the LIGO-Virgo Collaborations published a revised
analysis \cite{Abbott2018a, Abbott2018b} that provided improved estimates
on the tidal deformability and on the radius by incorporating new
theoretical insights, improved analysis tools, re-calibrated data (also
starting at a lower frequency threshold), and some likely assumptions
that had not been assumed in the first article \cite{Abbott2017} in order
to maintain the widest generality and conservativeness. These assumptions
are the source location (known mostly by electromagnetic observations
\cite{Abbott2017b, Soares-Santos2017, Cantiello2018}), the fact the
two stars have the same EoS, the fact that the more massive
star in the binary has smaller radius and lower tidal deformability, the
fact that, because of causality, the speed of sound in the NS must be
less than the speed of light (plus $\approx 10\%$ to allow for imperfect
parameterization) up to the central pressure of the heaviest star
supported by the EoS, and the fact that the EoS must allow for
non-rotating stars of mass $>1.97 \Msun$.

The improved analysis tools consisted in more accurate post-Newtonian
waveform models, the use of approximately EoS-insensitive relations to
relate observed quantities to radii, and sampling the EoS also as a
spectral parameterization \cite{Lindblom2010}, which has the advantages
mentioned in Sect. \ref{eos-parameterization}. Such an analysis is more
model dependent than the original one, but apparently the different
models used in Refs. \cite{Abbott2018a, Abbott2018b} give the same
results. However, Ref. \cite{Landry2018} warned that, according to their
own Bayesian analysis, constraints on the EoS from GW170817 alone may be
relatively prior-dominated and thus should be interpreted with care (see
also the end of Sect. \ref{hybrid stars}).

The refined estimate for the mass-weighted average dimensionless tidal
deformability is $\tilde{\Lambda} = 300^{+420}_{-230}$ and that for the
dimensionless tidal deformability of a NS of mass $1.4 \Msun$ is
$\Lambda_{1.4}=190^{+390}_{-120}$ \cite{Abbott2018b}.
For comparison, the widely used candidate EoS Sly \cite{Douchin01}, one
of the softest theoretical models, has $\Lambda_{1.4}\approx 290$, while
the ms1b EoS \cite{Mueller1996}, one of the stiffest, has $\Lambda_{1.4}
\approx 1220$. Overall, these constraints favour a relatively soft NS EoS,
and are believed to rule out several candidate EoSs at $90\%$ confidence
level \cite{Abbott2018a, Abbott2018b} (but see below for ideas on how to
{\it revive} the excluded EoSs and Refs. \cite{Zhang2018, Lim2018,
  Tews2019} for an alternative view). See Table \ref{table:radii} for the
radius estimates.

Various other works re-analysed the publicly available data of GW170817
in similar but distinct ways and drew similar and consistent conclusions
\cite{De2018, Chatziioannou2018, Zhao2018, Dai2018, Tews2018}.
Ref. \cite{Chatziioannou2018} revisited the early analysis of the
LIGO-Virgo Collaborations \cite{Abbott2017} by incorporating some
physical assumptions that had been left out for conservativeness. First
of all, they assumed that the two stars in GW170817 had the same EoS
(this has then been assumed in all subsequent studies). Then, they made
use of the binary Love universal relation \cite{Yagi2016, Yagi2017a} (see
Sect. \ref{univ-rel}) between the dimensionless tidal deformabilities of
the two stars and the ratio of their masses.
The use of such a universal relation implies that, for a given realistic
EoS, the dimensionless tidal deformability is a decreasing function of
the mass of the NS (for stable NS configurations) and so that the most
massive component in a BNS has the smallest tidal deformability. The use
of this relation to link the tidal deformabilities of the two stars leads
to a reduction in the extension of the credible region by factors of $2$
to $10$ and to an improvement in the measurement of the individual tidal
parameters by up to an order of magnitude, depending on the EoS and the
mass ratio.

Also Ref. \cite{De2018} was published before the updated analysis of the
LIGO-Virgo Collaborations \cite{Abbott2018a, Abbott2018b} and assumed
that the two stars in GW170817 had the same EoS. This is the first
published work to place also lower bounds on the deformability and radii
of NSs on the basis of a revised statistical analysis of GW170817. Their
statistical errors are comparable to the error reported later by the
LIGO-Virgo Collaborations, but total errors were estimated to be larger
than in Refs. \cite{Abbott2018a, Abbott2018b} because systematic errors
(from unknown physics related to the EoS) of $0.2 \, \km$ were added as
conservative bounds.

In order to carry out their estimations, Ref. \cite{De2018} justified and
used the approximate relation $\Lambda_A = q^6\Lambda_B$, where $q$ is
the mass ratio and $\Lambda_{A,B}$ the dimensionless tidal
deformabilities of the two stars of the binary. Later, the treatment was
improved by considering instead analytic limits on the dependence of the
dimensionless tidal deformability on the stellar mass, $\Lambda(M)$
\cite{Zhao2018}. This method is alternative to that used in
Refs. \cite{Chatziioannou2018, Abbott2018a, Abbott2018b}, where the
binary Love relations between the dimensionless tidal deformabilities of
the two stars and the ratio of their masses \cite{Yagi2016, Yagi2017a}
were used. It is claimed that the latter method is more suitable to
describe modifications to deformability correlations because it also
contemplates the existence of a strong first-order phase transition (see
also Sect. \ref{hybrid stars}). In fact, if one of the stars in the
binary has undergone a strong phase transition and the other has not, the
correlation found in Ref. \cite{De2018} would be weakened
\cite{Abbott2018a, Abbott2018b}.

In a different line of research, Refs. \cite{Li2019a, Li2019b, Sun2019}
studied whether the constraints imposed by GW170817, together with those
from other astrophysical observations (in particular the lower limit on
the maximum NS mass) and from nuclear theory and experiments, can allow
for hyperons to appear in NSs. It is well known that, while, on one side,
beyond nuclear density the conversion of nucleons into hyperons is
energetically favourable (see, \eg, Ref. \cite{Weber2005} for a review),
on the other side it is not easy to explain how NSs containing hyperons
can have masses around $2 \Msun$, since hyperonic EoSs are usually too
soft at the required densities. A few ideas to solve this puzzle have
been proposed (see, \eg, Ref. \cite{Baym2017} for a review). One of these
consists in tuning the interactions in the hyperonic sector to allow for
high-mass NSs \cite{Weissenborn2012, Weissenborn2014, vanDalen2014,
  Oertel2015, Tolos2017a, Fortin2017, Li2018d, Sahoo2019}. However, most
of the hyperonic stars proposed have tidal deformabilities and radii that
are larger than what indicated by GW170817 (and other observations). A
hyperonic EoS should be sufficiently soft below 2-3 times the saturation
density, in order to have a tidal deformability compatible with GW170817,
and sufficiently stiff at higher densities to sustain NSs with masses
larger than $2 \Msun$. This is exactly what was found to be the effect of
considering $\Delta$-isobars in the EoS, in addition to hyperons
\cite{Li2018c, Li2019a, Li2019b, Sun2019, Drago2014, Zhu2016}. Namely,
EoSs including both hyperons and $\Delta$-isobars are found to satisfy
also all current astrophysical constraints \cite{Li2019a, Li2019b,
  Sun2019}.

\subsubsection{Correlations of nuclear-matter parameters of the equation
  of state}
\label{sec:correl nuclear param}

A few works focused on estimating physical parameters of the expansion of
the EoS, as introduced in Sect. \ref{eos-parameterization}, for the data
of GW170817. The first estimates, based on the LIGO-Virgo first analysis
release \cite{Abbott2017}, concluded that the limits on the high-density
EoS parameters from the $\Lambda_{1.4} < 800$ constraint\footnote{The
  corrected upper limit $\Lambda_{1.4} < 900$ (see note on page
  \pageref{bookkeeping error note}) would have been even less
  constraining.} alone are weaker than those imposed by measurements of
the NS radius performed through x-ray observations \cite{Zhang2018}.

Other works, based on ideas of previous studies \cite{Sotani2014,
  Silva2016, Alam2016} proposed before the observation of GW170817,
pointed out that actually the (mass-averaged) tidal deformability is
found to be weakly or only moderately correlated with the individual
nuclear-matter parameters of the EoS \cite{Margueron2018a,
  Margueron2018b, Zhu2018, Malik2018, Carson2019, Zhang2019a}, while
stronger correlation is found for specific combinations of the EoS
parameters, in particular for a linear combination of the slopes of the
incompressibility and symmetry-energy coefficients \cite{Malik2018,
  Carson2019}. The correlation coefficients found are only around $50\%$,
but the authors set conservative bounds on some nuclear parameters (the
incompressibility $K_0$, its slope $M_0$, and the curvature of symmetry
energy $K_{{\rm sym},0}$ at nuclear saturation density) to be $81\, \MeV
\leq K_0 \leq 362\, \MeV$, $1556\, \MeV \leq M_0 \leq 4971\, \MeV$, and
$-259\, \MeV \leq K_{{\rm sym},0} \leq 32\, \MeV$ at $90\%$ confidence
level from GW170817 \cite{Carson2019}.

Also Ref. \cite{Zhang2019a} found that the individual measurements of the
tidal deformability and stellar radius can only set limits on combinations
of nuclear parameters; in their case, stringent constraints on the
correlation between the slope $L$ of the symmetry energy and its
curvature $K_{{\rm sym}}$ were found. The same authors, by combining
several physical and astrophysical constraints, also claimed that the
symmetry energy at twice the saturation density is constrained to $46.9\pm
10.1\, \MeV$ \cite{Zhang2019b}.
 
It was also pointed out in a Bayesian study \cite{Miller2019} (already
mentioned above), which included nuclear data, NS masses, radii, tidal
deformabilities, moments of inertia, and gravitational binding energies
obtained through different types of observations, that more precise
measurements of the slope $L$ of the symmetry energy would strongly
constrain the EOS only below nuclear saturation density and would have
little implications for densities above saturation density.

An important caveat needs to be highlighted here: Most correlations of
the kind mentioned above that are described in the literature originate or may originate from the
lack of generality of existing phenomenological functionals describing
EoSs. Instead, in these estimates it is important to account for the
widest possible range of valid EoSs and EoS variation
\cite{Margueron2018a, Margueron2018b, Malik2018, Landry2018,
  Carson2019}. Furthermore, bounds like those mentioned above have been
derived only for NSs and may not be valid for hybrid stars with a quark
core and a nuclear matter envelope, or for compact stars containing
heavy baryons (hyperons and/or $\Delta$-isobars) \cite{Li2019a, Li2019b}
(see Sect. \ref{hybrid stars}).

\subsection{Quark stars}
\label{quark stars}

Some works have checked whether the data from GW170817 (the first
announcement \cite{Abbott2017}) are compatible with stars made purely of
free quarks or of strange-quark clusters (which have been called {\it
  strangeons} \cite{lai2009}). It is important to note that, because of
the finite surface density of quark stars and of strangeon stars, the
standard relations involving the tidal deformability for nucleonic EoSs
need to be modified; therefore constraints on NSs according to the
analysis of GW170817 performed by the LIGO-Virgo Collaborations
\cite{Abbott2017, Abbott2018a, Abbott2018b} and similar studies cannot be
simply applied in the scenario of binary--quark-star mergers
\cite{Zhou2017, Lai2017b, Lau2019} (see also Sect. \ref{hybrid stars}).

With this caveat in mind, Ref. \cite{Zhou2017}, making use of the MIT bag
model \cite{Alcock86, Haensel1986},
and considering the limits on the tidal deformability from the GW170817
discovery article \cite{Abbott2017}, inferred that the constraints on the
tidal deformability from GW170817 are compatible with a
binary--quark-star merger. This remains true also when considering the
updated constraints from the LIGO-Virgo Collaborations \cite{Abbott2018a,
  Abbott2018b}, but very possibly only for stars made of superfluid
quarks (where quarks form Cooper pairs) \cite{Zhou2017}.
Also Ref. \cite{Lau2019} showed that the tidal deformability of quark
stars may be reduced to values compatible with current GW observations by
the effect of elasticity generated if a crystalline colour-superconducting
phase occurs, like in the proposed solid quark stars \cite{Lau2017},
solid quark stars covered with a thin layer of nuclear matter
\cite{Lau2017}, or solid quark-cluster stars \cite{Xu2003}. It must be
said, anyway, that even if their tidal deformabilities were compatible
with GW170817, it is difficult to explain how quark stars can provide the
ejecta necessary to produce the observed macronova emission
\cite{Wang2017}.

Even if no detailed studies exist, it has been also claimed that the
tidal deformability of a strangeon star is compatible with GW170817
\cite{Lai2017b, Lai2019}, also when considering the updated constraints
from the LIGO-Virgo Collaborations \cite{Abbott2018a, Abbott2018b}. Also
for strangeon stars there are issues related to the ejecta: As for free
quarks, the ejecta composed of strangeon {\it nuggets} would not lead to
$r$-process nucleosynthesis. However, it has been claimed that different
components of the observed macronova AT2017gfo may be explained in the
strangeon scenario in the following way: The blue component could be
powered by the decay of ejected strangeon nuggets, while the late red
component could be powered by the spin-down of the remnant strangeon star
after merging \cite{Lai2017b}.

Finally, the so-called {\it two family} scenario \cite{Drago2014a,
  Drago2018, Drago2018a, Burgio2018, DePietri2019} involves also quark
stars, but I will describe it in the next Sect. \ref{hybrid stars}, for
convenience.

\subsection{Hybrid stars}
\label{hybrid stars}
Several works have investigated whether GW170817 could have originated by
the coalescence of a NS with a hybrid star or that of two hybrid stars
\cite{Most2018, Paschalidis2018a, Nandi2018, Nandi2018a, Burgio2018,
  Tews2018, Gomes2018, AlvarezCastillo2018, Li2018b, Drago2018, Han2018,
  Montana2018, Csaki2018, Annala2018, Wei2019, Sieniawska2019, Sen2019,
  Christian2019, Gomes2019, Bozzola2019}. All found that GW170817 is also
consistent with the coalescence of a binary system containing at least
one hybrid star.

As mentioned earlier and widely known, the neutron matter EoS is expected
to be reliable up to densities as high as twice the nuclear saturation
density of $0.16\, {\rm fm}^{−3}$. Beyond that not much is known for sure
and matter could also undergo a phase transition to quark matter
\cite{Ivanenko1965, Itoh70, Baym1976}, possibly giving rise to a hybrid
hadron-quark star, in which the EoS is nucleonic for lower densities and
contains quark matter at higher densities. The possibility of more than
one phase transition is also envisaged \cite{Alford2008, Alford2017a,
  Han2018}. Such phase transitions would change drastically the
properties of the star, in particular because they would produce drastic
softening or stiffening of the EoS (according to the type of the
discontinuity, namely whether it is sharp or smoothed, the Love number of
strange quark stars can be smaller or larger than that for hadronic stars
\cite{Postnikov2010}). Multiple detections of GWs from different BNS
mergers with the same value for the tidal deformability of one star
$\Lambda_A$ but considerably different values of the tidal deformability
of the second star $\Lambda_B$ would probably indicate the existence of a
strong first-order phase transition in NS matter \cite{Christian2019,
  Montana2018, Ayriyan2019}. The first numerical-relativity simulations
of BNS mergers in which matter undergoes a hadron-quark phase transition
have been performed recently \cite{Most2018b, Bauswein2019} (see
Sect. \ref{postmerger phase transitions}).

Some hadronic EoSs that have been found not to satisfy the GW170817
constraints on the tidal deformability \cite{Abbott2017, Annala2017,
  AlvarezCastillo2018, Abbott2018a, Abbott2018b, Vuorinen2019}
and/or the lower limit for the maximum NS mass
\cite{Antoniadis_fulllist:2013, Demorest2010, Cromartie2019} can be
salvaged from the closed files of theories\footnote{In the next
  paragraphs I will write about some ways to {\it salvage} EoSs, but
  other physical effects have been found that may make the tidal
  deformability obtained with a given EoS higher, and therefore more
  easily excluded by GW observations of BNS systems. In particular,
  Ref. \cite{Char2018} found that, at least for the EoSs tried,
  considering superfluidity may increase the tidal deformability of $\sim
  5 - 10\%$.}, by proposing that they are the lower-density part of an EoS
with such a hadron-quark phase transition, since hybrid stars built in
that way have been shown to be compatible with the constraints imposed by
GW170817 \cite{Paschalidis2018a, Nandi2018, Nandi2018a}.

Stiff EoSs excluded at a first investigation may also be salvaged by
considering the additional effects that vector-meson interactions can
have on EoS models that do not consider them, since these interactions
were found to decrease the tidal deformability of a $1.4 \Msun$ star by
$\sim 30\%$ \cite{Dexheimer2018}, or by assuming that dark matter
interacts with nucleonic matter inside the NS, since this softens the EoS
and lowers the value of tidal deformability \cite{Ellis2018a, Ellis2018b,
  Nelson2018, Das2019, Quddus2019} (see Sect. \ref{dark matter}).

Before continuing the topic of hybrid stars, I mention here for
completeness that a third way to salvage EoSs that seem ruled out at
first sight by recent observations consists in considering anisotropic
compact stars \cite{Biswas2019}, which have been studied in many works
since the seminal one of Ref. \cite{Bowers1974} (see
Refs. \cite{Doneva2012, Silva2015, Yagi2015, Yagi2016, Raposo2018}, for
some of the most recent references). Pressure in stars in equilibrium is
usually assumed to be isotropic, but the possibility that the radial
pressure and the transverse pressure are not equal has been proposed in
several works \cite{Herrera1995, Kippenhahn2012, Sulaksono2015,
  Chakravarti2019a, Chakravarti2019b}. Basically, pressure anisotropy can
arise whenever the velocity distribution of particles in a fluid is
anisotropic, which in turn can be caused by magnetic fields, turbulence,
convection \cite{Herrera1995}, phase transitions \cite{Kippenhahn2012},
superfluidity or solid cores \cite{Herrera1995,
  Kippenhahn2012}. Additionally, some braneworld models of gravity, with
extra spatial dimensions, predict an effective four-dimensional
stress-energy tensor with anisotropic pressure\footnote{Conversely, with
  accurate enough measurements of stellar properties, one could
  constrain the {\it brane tension} of those alternative models of
  gravity \cite{Chakravarti2019a, Chakravarti2019b}.}
\cite{Chakravarti2019a, Chakravarti2019b}.

Using the formula for pressure anisotropy proposed (for facility of use,
rather than for physical reasons) in Ref. \cite{Bowers1974},
Ref. \cite{Biswas2019} computed the tidal deformability of anisoptropic
stars and found that, for a wide range of values for the parameters of
the anisotropy ansatz and for stars with mass in the range of interest
$[1, 2]\Msun$, the tidal Love number for an anisotropic star is smaller
than that for a spherically symmetric star with the same mass. This
leads to the conclusion that EoSs that are apparently ruled out by
GW170817 because of their stiffness might still be viable if the stars in
that BNS system were supported by a sufficiently anisotropic
pressure. Additionally, it has been noted that there is then a degeneracy
between EoS and anisotropy in the interpretation of the tidal
deformability measured from BNS inspirals. This degeneracy may be solved
if the stellar radius is measured independently, since anisotropic stars
usually have a larger radius for a given mass \cite{Biswas2019}.

\begin{figure}[tb]
\begin{center}
\begin{minipage}[t]{12 cm}
   \includegraphics[width=1.0\textwidth]{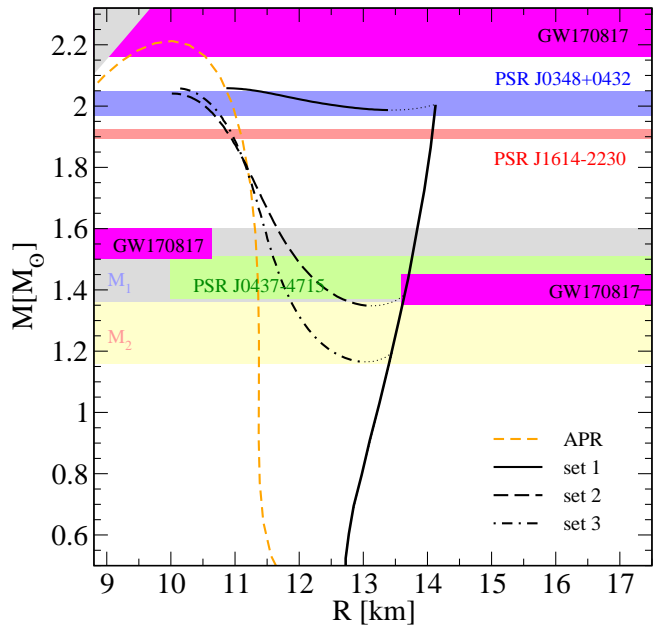}
\end{minipage}
\begin{minipage}[t]{16.5 cm}
  \caption{Mass vs radius for sequences of compact stars with the hybrid
    EoS of Ref. \cite{AlvarezCastillo2018}. {\it Sets} 1, 2 and 3
    correspond to different onset masses for the deconfinement
    transition. The dotted lines denote unstable configurations and
    connect two stable branches. For comparison, the mass-radius sequence
    for the APR EoS \cite{Akmal1998a} (a standard EoS for nuclear
    astrophysics applications) is shown with a dashed line. The
    horizontal bands denote: the mass measurement for PSR J0348+432
    \cite{Antoniadis_fulllist:2013}, PSR J1614-2230 \cite{Arzoumanian2018}, and PSR
    J0437-4715 \cite{Johnston1993}; the mass ranges for the compact stars
    in GW170817 (labelled $M_1$ and $M_2$); some exclusion regions from
    the observation of GW170817 (in particular, Ref. \cite{Bauswein2017b}
    excluded radii of smaller than $10.68\, \km$ for stars of $1.6 \Msun$,
    Ref. \cite{Annala2017} excluded radii exceeding $13.6\, \km$ for stars
    of $1.4 \Msun$, and Ref. \cite{Rezzolla2017} excluded masses higher
    than $2.16 \Msun$). (From
    Ref. \cite{AlvarezCastillo2018}) \label{fig:AlvarezCastillo2018-Fig10}}
\end{minipage}
\end{center}
\end{figure}

After digressing about the effects of anisotropy, I now resume the main
topic of this Section, saying that the interest in hybrid EoSs goes even
further than what written above, since some choices of phase transition
predict the existence of {\it twin stars} \cite{Glendenning2000,
  Schaffner-Bielich2002, Fraga2002}: NSs and hybrid stars having the same
mass but different radii. The appearance of a second stable branch in the
mass-radius relation of compact stars (see \eg~Fig. \ref{fig:AlvarezCastillo2018-Fig10})
is thought to arise from the occurrence of a phase transition. The
lower-mass star in the twin is made solely of nucleons and has larger
radius and larger tidal deformability, while the more massive star is a
hybrid star with a quark content and smaller radius and tidal
deformability\footnote{There are exceptions: The so-called {\it rising
    twins} are twin star combinations where the more massive one has a
  larger radius \cite{Schertler2000, Sieniawska2019}.}. See
Refs. \cite{Alford2013, Christian2018} for studies of classifications of
twin-star scenarios, Ref. \cite{Montana2018} for a recent extensive
analysis of the features of the phase transitions that lead to twin-star
configurations and Refs. \cite{Alvarez2017, Kaltenborn2017, Ayriyan2018}
for recent works on viable EoSs that allow for twin stars. It has been
recently pointed out that strong magnetic fields can influence noticeably
twin-star scenarios \cite{Gomes2019}. Even if, given an EoS,
non-magnetized or weakly magnetized twin stars are found to exist, it is
possible that strongly magnetized twin stars with that same EoS do not
exist. Conversely, even if, given an EoS, non-magnetized or weakly
magnetized twin stars are found not to exist, it is possible that
strongly magnetized twin stars with that same EoS do exist
\cite{Gomes2019}.

As a somewhat different idea, the {\it two-family} scenario
\cite{Drago2018} has been proposed, in which hadronic stars exist up to
$1.5 - 1.6 \Msun$ (their radius and tidal deformability decreasing with
increasing mass) and more massive stars are strange quark stars, with
larger radii and tidal deformabilities (note that the relations between
mass and radii or tidal deformabilities are qualitatively different from
those of hybrid stars in the twin-star scenario). A complete transition
may occur in finite time during the merger of the binary (or the collapse
of a star, \eg~in core-collapse supernovae) \cite{Drago2018} and the
process of conversion may have two different stages: a turbulent
combustion, which, in a timescale of the order of a few ms, converts most
of the star, and a diffusive combustion, which converts the unburnt
hadronic layer in a timescale of the order of $10$ s
\cite{Drago2016}. Strictly speaking, then, in the two-family scenario
hybrid stars exist only briefly during the conversion of hadronic matter
into quark matter.

In addition to influencing the computation of tidal deformabilities, the
two-family scenario predicts different rates for prompt collapses after
the merger with respect to other EoSs, because the threshold mass
\cite{Bauswein2017} against prompt collapse to black hole after the
merger is different. Related to this, it has also been found that
GW170817 cannot be a binary of hadronic stars with the hadronic EoS
adopted in the two-family scenario (containing hyperons and $\Delta$
resonances), because it would have undergone prompt collapse
\cite{Drago2018a}, while this is not believed to be the case
\cite{Murguia-Berthier2017, Shibata2017c, Ma2017, Pooley2018, Ai2018},
but it could be a binary of one hadronic and one quark star: The prompt
collapse would be avoided by the formation of a hypermassive hybrid
configuration, whose ejecta may give rise to the macronova
\cite{Drago2018a}. The first numerical-relativity simulations of BNS
mergers with the hadronic EoS of the two-family scenario have been
recently performed \cite{DePietri2019}.

Hybrid stars that give rise to the twin-stars or the two-family scenarios
allow for both stars that have a tidal deformability large enough to fit
with the lower-bound from GW170817 and stars that have radii small
enough\footnote{By using a set of relativistic polytropes connected to a
  quark bag model for the phase transition with a Maxwell construction
  (see note on page \pageref{Maxwell-Gibbs}) and with a realistic
  description of the NS crust (SLy4 \cite{Haensel94, Douchin01} up to the
  nuclear saturation density), Ref. \cite{Sieniawska2019} studied the
  radii and tidal deformabilities of compact stars compatible with
  current observational constraints and found that the minimum radius
  that can be produced on a twin branch lies between $9.5$ and $10.5 \,
  \km$, while the minimum radius for lighter stars that do not undergo
  the phase transition is about $12 \km$.} to fit with data from x-ray
observations of isolated NSs.
If both such observational constraints are confirmed and become tighter
in the future, the twin-stars or the two-family scenarios may become the
only viable ones \cite{Burgio2018}.

One way to show the qualitative difference between scenarios that allow
for twin stars and those that do not is to plot the dimensionless tidal
deformabilities of the stars in the binary against each other in a
$\Lambda_A - \Lambda_B$ plane. This is often done when examining the
constraints imposed by GW observations of BNS on the EoS. Since the tidal
deformabilities are related to each other through the total mass (or,
more practically, the chirp mass) of the system, in the absence of twin
stars, such a plot results in a broad band, whose position and shape are
governed by the EoS and whose width by the error of the measurement of
the total mass. Since a twin-star pair has two significantly different
values of the tidal deformability, more distinct bands may appear in the
$\Lambda_A - \Lambda_B$ plot, one for the NS--NS case, one for the
NS-hybrid-star case and one for the hybrid-star--hybrid-star case
\cite{Christian2019, Montana2018}.

Before going into some details about work on hybrid stars, I would like
to stress again two general points. The first is that, as already noted
in Sect. \ref{quark stars}, the constraints on the tidal deformability
set by the analysis of GW170817 performed by the LIGO-Virgo
Collaborations \cite{Abbott2017, Abbott2018a, Abbott2018b} and by similar
works were obtained by expanding the tidal deformability as a function of
mass $\Lambda(M)$ linearly about $M=1.4\,\Msun$. However, such a linear
expansion is valid only for EoSs without phase transitions, as can be
seen, \eg, in Fig. \ref{fig:AlvarezCastillo2018-Fig10}. In fact, using
this approximation to estimate the tidal deformability of a $1.4 \Msun$
star should be avoided because it excludes the possibility of testing for
hybrid stars. In particular, it has been shown that if one does not make
that linearity assumption the upper bound on $\Lambda_{1.4}$ could be
smaller \cite{Most2018, Paschalidis2018a, Tews2018, Burgio2018, Han2018,
  Montana2018}.

The second general note on hybrid stars is that up to the time of the
writing of this review there are actually no numerical-relativity
simulations in full general relativity of the merger of binaries composed
of one hybrid star and one NS or of two hybrid stars. Such simulations
are necessary for further comparison with mergers of stars described by
purely nucleonic EoSs.


For the rest of this Section, I will give more details on specific works
involving hybrid stars. Ref. \cite{Paschalidis2018a} was the first to
investigate how GW170817 can constrain the properties of hybrid
hadron-quark compact stars. Ref. \cite{Nandi2018} later did a very
similar work. They constructed (with the Maxwell
method\label{Maxwell-Gibbs}\footnote{The Maxwell method, or Maxwell {\it construction},
  describes a constant-pressure phase transition in which density is
  discontinuous at the interface between hadronic and quark matter,
  without a phase in which the two are mixed; the Gibbs method (or
  construction), instead, allows for a mixed phase in which pressure and
  density vary monotonically. More in general, a hadron-quark phase
  transition is described by the surface tension between the two
  phases. Infinite surface tension corresponds to the Maxwell
  construction and a zero surface tension to a Gibbs construction. It is
  largely unknown what may be realistic values for the surface tension
  and different values for it (\ie~different
  constructions) lead to stars with different properties, especially
  tidal deformability and radius \cite{Xia2019}. Particular care should
  be therefore taken when using results obtained from such
  constructions.}  \cite{Paschalidis2018a} or the Gibbs method
\cite{Nandi2018}) parameterized hybrid hadron-quark EoSs that (i) consist
of zero-temperature nuclear matter in $\beta$-equilibrium with a
low-density phase of nucleonic matter and a high-density single-phase
core of quark matter (parameterized with piecewise polytropes and other
methods), with a first-order phase transition at their interface, (ii)
are supplemented with a low-density EoSs for crustal matter, and (iii)
are consistent with the existence of $\approx 2 \Msun$ pulsars. Stars
built with such EoSs were found to have the twin-star property at
$\approx 1.5 \Msun$.

These articles \cite{Paschalidis2018a, Nandi2018} were then the first to
stress that the tidal deformation observed from GW170817 is consistent
with the coalescence of a binary composed of one hybrid star and one
NS, that certain hadronic EoSs that do not satisfy the GW170817
constraints on the tidal deformation become compatible with GW170817 if
a first-order phase transition occurs in one of the stars, and that,
while for purely hadronic EoSs the dimensionless tidal deformability can
be approximated as a linear function of the gravitational mass for masses
in the vicinity of $1.4 \Msun$, this is not true for hybrid hadron-quark
EoSs with low-mass twins \cite{Paschalidis2018a}. This issue was later
elaborated in detail in Ref. \cite{Sieniawska2019}.

Similarly, Ref. \cite{AlvarezCastillo2018}, adopting an EoS that predicts
stars possibly composed of a core of two-flavour quark matter and a shell
of hadronic matter and that allows for twin stars, interpreted the
GW170817 event as the merger of either such a hybrid star and a NS or a
merger of two such hybrid stars, while the BNS scenario was deemed
disfavoured mostly because the stiffness of the hadronic EoS employed
made a BNS merger incompatible with the compactness
expected from GW170817 (see also
Fig. \ref{fig:AlvarezCastillo2018-Fig10}).

Also Refs. \cite{Christian2019, Montana2018} interpreted the GW170817
event as the merger scenario of either a NS-NS, a NS--hybrid-star, or a
hybrid-star--hybrid-star binary. In particular, Ref. \cite{Montana2018}
performed an extensive analysis of the merger scenarios in which the
parameters characterizing the phase transition have also been varied and
both Maxwell and Gibbs constructions for the phase transition were
considered. For the lower-density region of the inner core the hadronic
EoSs of Refs. \cite{Tolos2017a, Tolos2017b}, for the inner and outer
crust the EoS of Ref. \cite{Sharma2015}, and for the quark phase a
constant-speed-of-sound parameterization \cite{Chamel2013, Zdunik2013,
  Alford2013} were used. They also predicted that a phase transition
would be revealed in the inspiral waveform if GW detectors measured
values of the chirp mass smaller than $1.2 \Msun$ together with tidal
deformabilities of $\Lambda_{1.4}\lesssim 400$.

Ref. \cite{Ayriyan2019} performed a Bayesian analysis that included some
models for the hadronic and quark matter phases that were said to be
realistic and that give rise to twin stars and found that their
analysis favours models with a strong mixed phase.

Ref. \cite{Han2018} has explored the sensitivity of the tidal
deformability to the properties of a phase transitions or of two
sequential phase transitions (the first from nuclear matter to quark
matter, the second, at a higher density, to a different quark-matter
phase\footnote{for example, from the two-flavour colour-superconducting to
  the colour-flavor-locked phase \cite{Alford2017a}; see, \eg,
  Ref. \cite{Alford2008} for a review.}), not necessarily producing twin
stars, using the constant-speed-of-sound parameterization
\cite{Chamel2013, Zdunik2013, Alford2013} \cite{Blaschke2009,
  Alford2017a, Alford2008}. In their calculations, they found that the
tidal deformability is smaller than that of purely hadronic stars and
small enough to be distinguishable in future GW observations of BNS
merger events.

Refs. \cite{Maslov2018, Xia2019}, on the basis of an idea of
Ref. \cite{Voskresensky2002}, investigated the effects of the surface
tension\footnote{See also the footnote on the Maxwell and Gibbs
  constructions, on page \pageref{Maxwell-Gibbs}.} between the hadron
and quark phases (its value is largely unknown) and found that assuming
different values of the surface tensions leads to variations in the
maximum mass of the hybrid star of $\approx 0.02 \Msun$ (which is not
particularly significant), in its radius of $\approx 0.6 \km$ and in its
tidal deformability of $\Delta\Lambda/\Lambda\approx 50\%$ (which are
both very significant). Conversely, with future, more accurate
measurements of tidal deformabilities and radii of hybrid stars it may be
possible to constrain the surface tension. The current measurements from
GW170817 were found to be unconstraining \cite{Xia2019}.

Ref. \cite{Csaki2018} pointed out that GW measurements from BNS mergers
may offer for the first time access to the structure of objects that
might have a non-negligible contribution from vacuum energy to their
total mass. The presence of such vacuum energy in the inner cores of
NSs would occur in new QCD phases at large densities
\cite{Pacini1966, Ellis1991}. If such phase transitions, which are
different from those mentioned above, occur, this would lead to a change
in the internal structure of NSs and thus possibly influence
their tidal deformabilities. In their computations, performed with a
piecewise-polytropic representation of some commonly used EoSs augmented
with a phase transition involving vacuum energy, they found that the
effects of vacuum energy could be measurable (in future detectors and
with enough events) for mergers with high chirp masses ($\approx 1.9
\Msun$), while for smaller chirp masses like that of GW170817 the
deviations of the tidal deformability are below detection capabilities
\cite{Csaki2018}.
Measurements of this sort may be the only possibility for testing the
gravitational properties of vacuum energy independently from the
acceleration of the Universe.

Ref. \cite{Tews2018} extended the work of the LIGO-Virgo Collaborations
\cite{Abbott2017, Abbott2018a, Abbott2018b} by using a different general
representation of EoSs that allows for a phase transition at high densities
through a parameterization of the speed of sound \cite{Tews2018a,
  Tews2018}, which was developed starting from ideas of
Refs. \cite{Alford2013, Bedaque2015}. This is different from the
constant-speed-of-sound parameterization \cite{Chamel2013, Zdunik2013,
  Alford2013} mentioned in Sect. \ref{eos-parameterization}, but
analogously offers the advantage that the speed of sound is continuous
everywhere except when first-order phase transitions are explicitly
inserted. Six values of the density are randomly selected in the interval
between the saturation density and twelve times the saturation density
and the values of the speed of sound at those densities are treated as
reference points, while all other points are computed by connecting the
reference points through linear segments. It is claimed that this model
parameterizes the widest possible domain of EoSs \cite{Tews2018,
  Tews2019}. With this setup, the authors computed probability contours
smaller than those from the LIGO-Virgo Collaborations \cite{Abbott2018a,
  Abbott2018b} and obtained $80 < \tilde{\Lambda}< 570$.
In a follow-up work \cite{Tews2019}, it was estimated that, in order to
test the predictions of nuclear theory (in the form of chiral effective
field theory) in the range between one and two times the saturation
density, the uncertainty in the dimensionless tidal deformability needs
to be less than about $300$, while in order to test the existence of
phase transitions in denser matter it needs to be below $100$.

As discussed in Sects. \ref{eos-parameterization} and \ref{sec:inspiral},
the choice of parameterization can have a significant effect on the
global mass-radius relation and the EoS constraints inferred from GW
observations of BNS systems. This was shown also for the case of EoSs
allowing for first-order phase transitions in Ref. \cite{Greif2019}, that
compared a piecewise-polytropic parameterization to their own
constant-speed-of-sound parameterization, similar but different from that
of Ref. \cite{Tews2018}. It was also mentioned above that, in order to
overcome the drawbacks of parameterized models, non-parametric models
have been proposed. Ref. \cite{Landry2018} was the first to introduce a
non-parametric method for directly inferring the NS EoS from GW data. It
used Gaussian process regression (see, \eg, Ref.  \cite{Rasmussen2006})
to generate a large number of possible dependencies of density on
pressure, which span a very wide range of stiffnesses and core pressures,
while being consistent with thermodynamic stability, causality, observed
astrophysical data, and candidate models from nuclear theory. In their
work, they train what they call their {\it synthetic EoSs} on seven (of
which five purely hadronic and two containing quarks or hyperons)
fiducial tabulated candidate EoSs selected from well-established
nuclear-theory models
that span a wide range of stiffnesses and support a $1.93 \Msun$ star.
Two models were considered: one built with a very uninformative EoS prior
(tabulated EoSs provide only weak, general guidance to the form of the
synthetic EoSs) and the other more conformed to the chosen fiducial EoSs
(in practice the synthetic EoSs depart relatively little from the
average of the tabulated EoSs) \cite{Landry2018}.
Applying the method to GW170817 gave
tidal deformabilities and their error bands that for both models agree
with the results of the LIGO-Virgo Collaborations \cite{Abbott2018a,
  Abbott2018b}. Perhaps the most important claim of
Ref. \cite{Landry2018} is that, since a Bayesian analysis of the two
models (the one weakly influenced by the tabulated EoSs and the one
strongly influenced by them) finds that neither is strongly preferred by
the data (Bayes factor of about $1.12$), constraints on the EoS from
GW170817 alone may be relatively prior-dominated and thus should be
interpreted with care.

\subsection{Possible influence of dark matter}
\label{dark matter}

\begin{figure}[tb]
\begin{center}
\begin{minipage}[t]{9 cm}
  \includegraphics[width=1.0\textwidth]{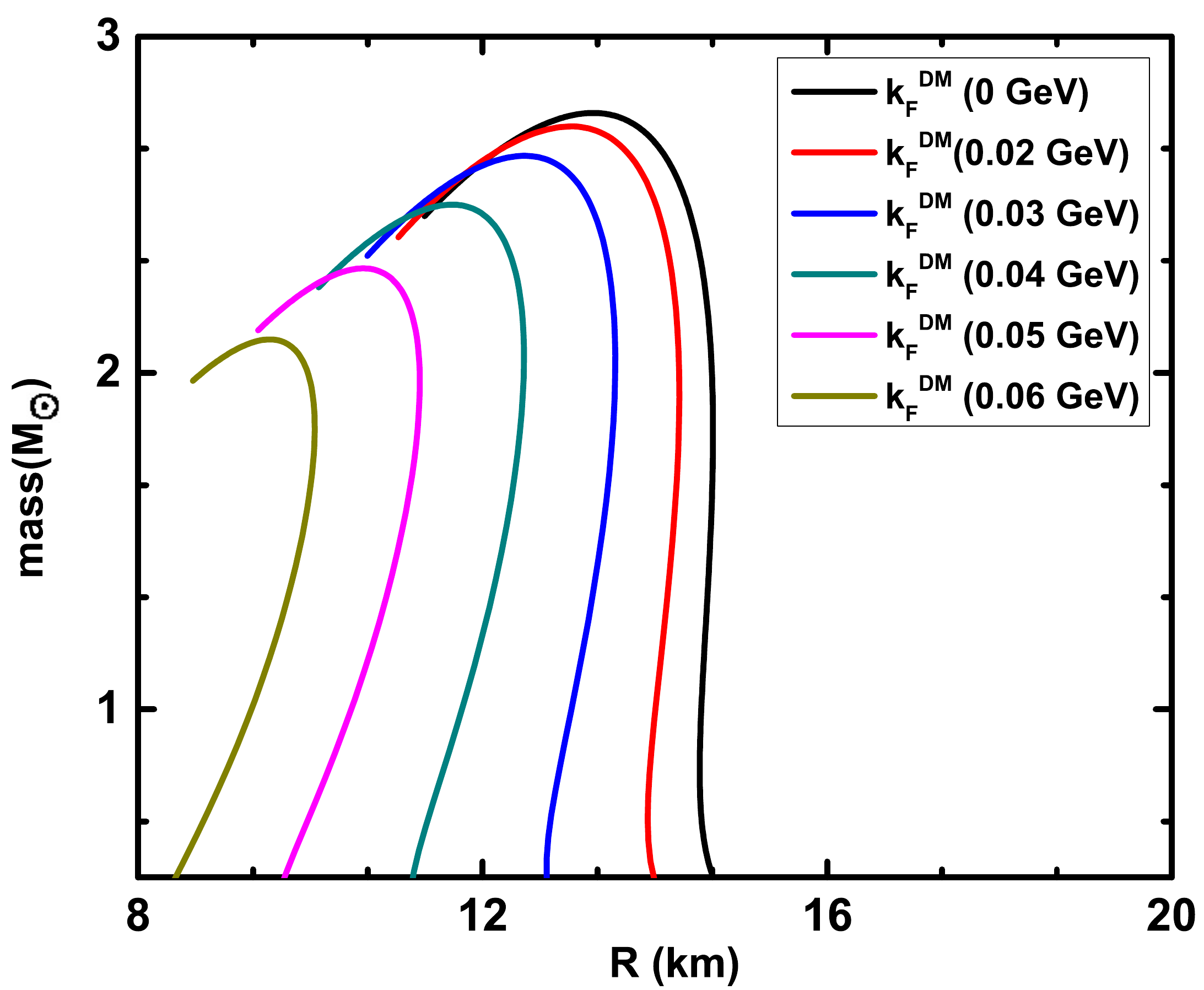}
\end{minipage}
\begin{minipage}[t]{16.5 cm}
\caption{Mass-radius relation of NSs for EoSs with different dark-matter
  Fermi momentum. (From
  Ref. \cite{Das2019}) \label{fig:Das2019_PRD99_Fig2}}
\end{minipage}
\end{center}
\end{figure}

A few works have been published on more exotic possibilities of what
could also be inside compact stars and how this would influence the
interpretation of GW data, starting from those of GW170817. Some works
have studied the influence that dark matter may have on NSs (\eg, their
maximum possible mass and the mass-radius relation
\cite{Panotopoulos2017, Ellis2018b}) and their tidal deformability
\cite{Sandin2009, Ciarcelluti2011, Tolos2015}, in particular in view of
the data of GW170817 \cite{Ellis2018a, Ellis2018b, Nelson2018, Das2019,
  Quddus2019}. In general, these works have determined that the tidal
deformability and the maximum possible mass for NSs are smaller in the
presence of a dark-matter core \cite{Ellis2018b, Das2019, Quddus2019}
(see Fig. \ref{fig:Das2019_PRD99_Fig2}). As already mentioned in
Sects. \ref{postmerger} and \ref{exotic binaries}, it has been found that
the presence of sizeable dark matter in merging NSs may cause the
appearance of one or two additional peaks in the post-merger frequency
spectrum \cite{Ellis2018a, Bezares2019} and that it can affect the
one-armed deformations of the stellar object formed after the merger
\cite{Bezares2019}.

However, these effects of dark matter are non-negligible only if enough
dark matter accumulates in (or in and around) NSs and this does not
happen during what is considered the usual evolution of NSs. Namely, via
gravitational accretion a NS cannot accumulate dark matter in quantities
sufficient to form substantial dark-matter cores or halos. The formation
of substantial dark-matter structures would require additional processes
to occur, like {\it dark conversion} of neutrons to scalar dark matter
\cite{Ellis2018a, Ellis2018b},
production of dark-matter particles via bremsstrahlung in neutron-neutron
scatterings \cite{Nelson2018}, or a nonstandard dark-matter sector with a
subdominant dissipative component \cite{Foot2004, Panotopoulos2017,
  Sandin2009, Leung2011a, Fan2013, Pollack2015}.

Ref. \cite{Ellis2018b} considered eleven representative nucleonic EoSs
from the literature and added to NSs built with them self-interacting
bosonic dark matter (they also argued that asymmetric dark matter
\cite{Petraki2013} with or without self-interactions would yield similar
results). It was found that, if enough dark matter accumulates in NSs,
the presence of a dark-matter core with a mass of about $5\%$ of the NS
mass could make the NS more compact (and so decrease its tidal
deformability and the maximum mass attainable) to the point that some EoS
currently compatible with the existence of a $\approx 2 \Msun$ NS would
instead become incompatible and that some EoSs that lie outside the
confidence region of tidal deformability determined from GW170817
\cite{Abbott2018a, Abbott2018b} could instead no longer be excluded with
the current observation. These results are nearly independent from the
strength of the dark-matter self-interaction.
Ref. \cite{Das2019}, then, arrived at similar conclusions, considering
fermionic dark matter interacting with nucleonic matter (described with
the Walecka model \cite{Walecka1974, Serot1997, Panotopoulos2017}) inside
the NS.

Refs. \cite{Ellis2018b, Nelson2018} considered not only cores of dark
matter, but also halos extending outside the NS. In this case,
NSs with dark matter would have a larger tidal deformability.

In conclusion to this Section, note that the dark-matter mass fraction in a
NS may not be the same in all stars, since it may depend on the
NS age, initial temperature, or the environment in which it was
formed. Such possibilities would further complicate the
interpretations of measurements of NS properties.

\subsection{Summary of radius measurements}
\label{radius measurements}

A summary of radius estimates obtained from GW170817 (and some other
observations) from various works is presented in Table \ref{table:radii}
(see also Ref. \cite{Montana2018} for a similar table). The Table is not
meant to be comprehensive of all published results, but only to give the
reader an image of the radius intervals and error bars.

In addition to the estimates obtained from GW170817, radii of NSs have
been and are being measured with several other approaches, mostly based
on x-ray, optical and radio observations. Comments on these observations
and results are outside the scope of this review. Several other recent
reviews are available on the topic \cite{Ozel2016, Watts2016,
  Miller2016a, Degenaar2018}.  However, let me note that one of the most
fruitful methods is the analysis of thermal emissions from low-mass x-ray
binaries in their quiescent stage and photospheric-radius expansion of
type I x-ray bursters \cite{Steiner2010, Ozel2016, Nattila2016,
  Suleimanov2017, Nattila2017, Steiner2018}. Furthermore, it needs to be
said that the systematic errors of these methods are not completely known
(some of the data are more controversial than others; see, \eg,
discussions in Refs. \cite{Kajava2014, Miller2016a, Degenaar2018}) and
different methods yield different values for the estimates of the radius
$R_{1.4}$ of a $1.4 \Msun$ NS: with the so-called {\it touchdown} method
$R_{1.4}$ is in the range $10-11 \km$ \cite{Ozel2016} and with the
so-called {\it cooling tail} method it is in the range $11-13 \km$
\cite{Nattila2017}.

Future observations by the NICER (Neutron star Interior Composition
Explorer) mission \cite{Arzoumanian2014, Gendreau2016}, expected to
provide first results within this year, and the planned eXTP (enhanced
x-ray Timing and Polarimetry) Mission \cite{Watts2019}, LOFT (Large
Observatory For x-ray Timing) satellite \cite{WilsonHodge2016}, and
ATHENA (Advanced Telescope for High Energy Astrophysics) \cite{Motch2013}
are seen to have completely different systematic errors, and thus, by
combining many measurements with differing systematics one can hope to
significantly narrow the range for $R_{1.4}$. These endeavours are based
on the idea of measuring pulsations in the x-rays emitted from the hot
polar cap of a pulsar due to its rotation. These x-ray waveforms depend
on the relativistic gravitational properties of the star because they
initially propagate through the highly curved NS space-time. Ray-tracing
techniques then can yield the radius of the NS or be used directly in
Bayesian analyses of EoS parameters (see, \eg, Refs. \cite{Nattila2018,
  Salmi2018} and references therein for the latest results).

In the future, accurate estimates of the NS radius may also come from
simultaneous measurements of the mass and the moment of inertia of
NSs. These can be obtained mainly in two ways: (i) exploiting geodetic
precession, namely the precession of the rotation and orbital axes (if
these are not aligned) around the direction of the total angular
momentum;
(ii) exploiting the advance of the pericentre of the system caused by
spin-orbit coupling. Refs. \cite{Lattimer2005b, Kramer2009} estimated
that the moment of inertia will eventually be determined to a precision
of about $10\%$.

\begin{table}[ht]
\begin{center}
  \caption{Constraints on the radius of NSs from GW170817 (and some other
    observations) from works that report in their text estimates for the
    radius $R_{1.4}$ of a $1.4 \Msun$ star or the radius of the NSs in
    GW170817, considering EoSs without and with a phase transition and
    from multimessenger analyses. In each category, the entries are in
    order of publication date. 
    See also Ref. \cite{Montana2018}.}
  \label{table:radii}
  \setlength{\tabcolsep}{-0.03em}
\vspace{0.5 cm}
\begin{tabular}{lllrcll}
  \hline
\multicolumn{2}{l}{Reference} & \phantom{MTHR} & & $R_{i}\,{\rm [km]}$ & & Notes\\ \hline

\multicolumn{2}{l}{\textit{Without a phase transition}}     & &             &          & &\\
\multicolumn{2}{l}{Bauswein et al. \cite{Bauswein2017b}} & & $10.68^{+0.15}_{-0.03}\leq$ & $R_{1.6}$ && $\quad^{\circ}$ \\
\multicolumn{2}{l}{Fattoyev et al. \cite{Fattoyev2017}} & &  & $R_{1.4}$ & $\leq 13.76$ &\\
\multicolumn{2}{l}{Most et al.     \cite{Most2018}}     & & $12.00\leq$ & $R_{1.4}$ & $\leq 13.45$&$\quad^{\times}$\\
\multicolumn{2}{l}{Lim Holt      \cite{Lim2018}}   & & $10.36\leq$  & $R_{1.4}$ & $\leq 12.87$&$\quad^{\times}$\\
\multicolumn{2}{l}{De et al.       \cite{De2018}}       & & $8.9\leq$   & $R_{1.4}$ & $\leq 13.2$&$\quad^{\div}$\\
\multicolumn{2}{l}{Malik et al.      \cite{Malik2018}}   & & $11.82\leq$  & $R_{1.4}$ & $\leq 13.72$&$\quad^+$ \\ 
\multicolumn{2}{l}{LIGO/Virgo      \cite{Abbott2018b}}   & & $10.5\leq$  & $R_{{}_{\rm{{GW170817}}}}$ & $\leq 13.3$& $\quad^*$\\ 
\multicolumn{2}{l}{Tews et al.     \cite{Tews2018}}     & & $11.3\leq$  & $R_{1.4}$ & $\leq 12.1$& $\quad^\star$\\ 
\multicolumn{2}{l}{K\"{o}ppel et al.     \cite{Koeppel2019}}     & & $10.92\leq$  & $R_{1.4}$ & &\\ 
\multicolumn{2}{l}{Raithel      \cite{Raithel2019}}   & & $9.8\leq$  & $R_{{}_{\rm{{GW170817}}}}$ & $\leq 13.2$& $\quad^*$\\

\hline                                                          
\multicolumn{2}{l}{\textit{With a phase transition}}        & &             &          & \\
\multicolumn{2}{l}{Annala et al.   \cite{Annala2017}}   & & $9.9 \leq$  & $R_{1.4}$ & $\leq 13.6$ &\\ 
\multicolumn{2}{l}{Most et al.     \cite{Most2018}}     & & $8.53\leq$  & $R_{1.4}$ & $\leq 13.74$ &$\quad^{\times}$\\ 
\multicolumn{2}{l}{Tews et al.     \cite{Tews2018}}     & & $9.2\leq$   & $R_{1.4}$ & $\leq 12.5$&$\quad^\star$\\ 
\multicolumn{2}{l}{Montana et al.      \cite{Montana2018}}   & & $10.1\leq$  & $R_{1.4}$ & $\leq 13.11$&$\quad^{\times}$ \\ 

\hline                                                          
\multicolumn{2}{l}{\textit{From multimessanger analyses}}        & &             &          & \\
\multicolumn{2}{l}{Radice Dai   \cite{Radice2018c}}   & &    $11.4\leq$  & $R_{1.4}$ & $\leq 13.2$&$\quad^{*\,\dagger}$ \\ 
\multicolumn{2}{l}{Coughlin et al. \cite{Coughlin2018a}}   & & $11.1\leq$& $R_{1.4}$ & $\leq 13.4$&$\quad^{*\,\dagger}$\\ 
\multicolumn{2}{l}{Kumar Landry    \cite{Kumar2019}}   & & $9.4\leq$& $R_{1.4}$ & $\leq 12.8$&$\quad^{*\,\ddagger}$\\

\hline
\end{tabular}
\vspace{-0.3 cm}
\end{center}
\begin{spacing}{0.6}
{\tiny
  \noindent  { }\\
  $\quad^{\circ}$ This work gives only an estimate for the radius
  $R_{1.6}$ of an $1.6 \Msun$ star. However, I still include it in this
  table because it was probably the first work to give an estimate of NS
  radii based on the observation of GW170817.\\
  $\quad^{\times}$ At $2\,\sigma$ confidence level.\\
  $\quad^{\div}$  At $90\%$ confidence level and including an estimate of systematic error.\\
  $\quad^+$ The lower limit is based on results from Ref. \cite{Radice2017b}, related to macronova emission.\\
  $\quad^*\,$ At $90\%$ confidence level.\\
  $\quad^\star$ Total envelopes with GW input at $90\%$ confidence level.\\
  $\quad^\dagger\,$ These authors suggest an additional systematic uncertainty of $0.2 \km$ on these values.\\
$\quad^{\ddagger}$ This work combined the GW data with electromagnetic observations of other NSs in
  accreting low-mass x-ray binaries (from Ref. \cite{Ozel2015}).}
\end{spacing}

\end{table}

As described in the previous sections and in Table \ref{table:radii},
various works extracted estimates of the NS radius from data obtained
from GW170817 and the overall picture that appears from such estimates is
relatively robust. Nuclear theory (see, \eg, Refs. \cite{Gandolfi2012,
  Lynn2016, Drischler2016a, Tews2017}) and experiments currently predict
approximately the same radius intervals (see, \eg,
Ref. \cite{Lattimer2016} for a review) as estimated from GW170817.
Results from GW170817 are also consistent with the above-mentioned
measurements from electromagnetic observations.

\section{Conclusion}

Even if not all agree, most people in the community are saying that, just
from the one detection of GWs from BNS mergers that has been analysed, it has
already been possible to set new constraints on the EoS of very
high-density matter. It may be not so useful to debate on whether such a
statement is true or not, since the new observation run of the LIGO-Virgo
Collaborations has already started to give us new datasets on BNS
mergers, which will definitively tell us more about the EoS of compact
stars, including, perhaps, about possible phase transitions at
supranuclear density. While the GW170817 event was very fortunate because
of its proximity to us, it can be expected that future observations
will not all have such a high signal-to-noise ratio, but the increasing
sensitivity of advanced GW detectors over the next years will surely lead
to a large number of detections of merging BNS systems (if we are lucky,
we may also observe the merger of an eccentric NS binary)
\cite{Abbott2016h, Abbott2018} and, together with accurate
electromagnetic observations related and unrelated to GW events, will
allow for an even more precise measurement of the supranuclear EoSs. It
has been estimated that a number of detections (at $100 \Mpc$) of order $10$
will set percent limits on NS radii and $10\%$ limits on the EoS
\cite{Margalit2019, Miller2019, Tews2019}.

The techniques reviewed in this article will continue to be used in the
data analyses and will of course undergo further developments. Numerical
simulations, in addition to delivering waveforms with higher and higher
accuracy for constructing improved approximants, may also at some point
finally provide us with more realistic hints on the evolution of magnetic
fields and radiation after the merger. Semianalytic techniques that
provide the actual templates for the data analyses are also continuously
improving.

This is all really exciting.

\section{Acknowledgements}

I would like to thank all the colleagues who gave me valuable
suggestions on how to improve this review. Partial support has come from
JSPS Grant-in-Aid for Scientific Research (C) No. T18K036220.

\bibliographystyle{elsarticle-num} 
\bibliography{aeireferences}

\end{document}